\documentclass[letterpaper]{article} %
\usepackage{aaai25}  %

\usepackage{times}  %
\usepackage{helvet}  %
\usepackage{courier}  %
\usepackage[hyphens]{url}  %
\usepackage{graphicx} %
\usepackage{natbib}  %
\usepackage{caption} %
\usepackage{algorithm}
\usepackage{algorithmic}

\usepackage{newfloat}
\usepackage{listings}
\DeclareCaptionStyle{ruled}{labelfont=normalfont,labelsep=colon,strut=off} %
\floatstyle{ruled}
\newfloat{listing}{tb}{lst}{}
\floatname{listing}{Listing}
\usepackage{amsmath} 
\usepackage{subcaption}
\usepackage{xcolor} 
\usepackage{cleveref}
\usepackage{booktabs}
\usepackage{multirow}
\usepackage{amssymb}
\usepackage{array}
\usepackage{ragged2e}
\usepackage{fancyvrb}
\usepackage{framed}

\newcommand{\commenttxt}[1]{\textcolor[rgb]{0.173,0.478,0.11}{#1}}

\newcommand{\best}[1]{\underline{#1}}  %

\newcolumntype{R}[1]{>{\RaggedLeft\arraybackslash}p{#1}}

\DefineVerbatimEnvironment{HighlightVerbatim}{Verbatim}{
    commandchars=\\\{\}, %
}

\title{Whose Name Comes Up? Auditing LLM-Based Scholar Recommendations}

\author {
    Daniele Barolo\textsuperscript{\rm 1}, %
    Chiara Valentin\textsuperscript{\rm 2}, %
    Fariba Karimi\textsuperscript{\rm 2,1}, %
    Luis Galárraga\textsuperscript{\rm 3}, %
    Gonzalo G. Méndez\textsuperscript{\rm 4}, %
    Lisette Espín-Noboa\textsuperscript{\rm 1,2,5} %
}
\affiliations {
    \textsuperscript{\rm 1}Complexity Science Hub,  %
    \textsuperscript{\rm 2}Graz University of Technology, %
    \textsuperscript{\rm 3}INRIA/IRISA Research Center,\\
    \textsuperscript{\rm 4}Polytechnic University of Valencia, %
    \textsuperscript{\rm 5}Central European University\\ %
    \textit{Corresponding authors}: barolo@csh.ac.at, 
    espin@csh.ac.at
}

\begin{document}
\maketitle

\begin{figure*}[t]
\centering
\includegraphics[width=1\textwidth]{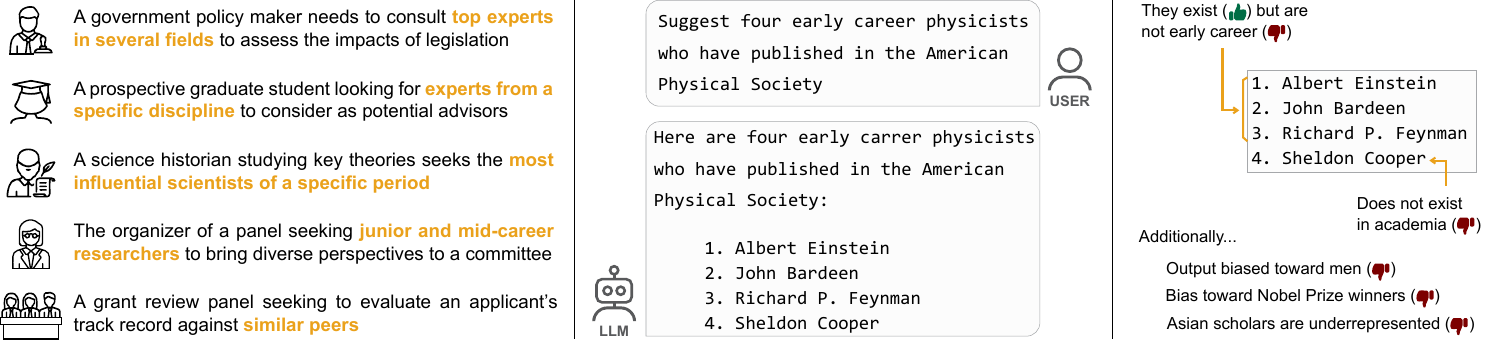}
  \caption{\textbf{Overview of LLM-based scholar recommendations.}
 Left: Scenarios where different types of users employ LLMs to identify experts by field, discipline, time period, seniority, or similarity.
 Center: An example query and the corresponding LLM response recommending early-career physicists. %
 Right: Exemplary audit findings highlighting key challenges, such as factual inaccuracies (e.g., incorrect seniority, non-existent individuals) and biases (e.g., gender and geographic disparities). This study systematically audits six LLMs to understand these challenges and discusses recommendations to mitigate them.}
  \label{fig:teaser}
\end{figure*}

\begin{abstract}
This paper evaluates the performance of six open-weight LLMs (\texttt{llama3-8b}, \texttt{llama3.1-8b}, \texttt{gemma2-9b}, \texttt{mixtral-8x7b}, \texttt{llama3-70b}, \texttt{llama3.1-70b}) in recommending experts in physics across five tasks: top-k experts by field, influential scientists by discipline, epoch, seniority, and scholar counterparts. 
The evaluation examines consistency, factuality, and biases related to gender, ethnicity, academic popularity, and scholar similarity. Using ground-truth data from the American Physical Society and OpenAlex, we establish scholarly benchmarks by comparing model outputs to real-world academic records.

Our analysis reveals inconsistencies and biases across all models. \texttt{mixtral-8x7b}~produces the most stable outputs, while \texttt{llama3.1-70b}~shows the highest variability. Many models exhibit duplication, and some, particularly \texttt{gemma2-9b}~and \texttt{llama3.1-8b}, struggle with formatting errors. 
LLMs generally recommend real scientists, but accuracy drops in field-, epoch-, and seniority-specific queries, consistently favoring senior scholars.
Representation biases persist, replicating gender imbalances (reflecting male predominance), under-representing Asian scientists, and over-representing White scholars. %
Despite some diversity in institutional and collaboration networks, models favor highly cited and productive scholars, reinforcing the rich-get-richer effect while offering limited geographical representation.
These findings highlight the need to improve LLMs for more reliable and equitable scholarly recommendations.
\end{abstract}

\begin{links}
    \link{Code}{https://anonymous.4open.science/r/LLMScholar-8F8E}
\end{links}
\section{Introduction}

The rapid proliferation of large language models (LLMs) has transformed numerous domains, enabling unprecedented information processing capabilities and extending to academia, where LLMs are increasingly used for scholarly recommendations~\cite{Cheng2024Influence, tian2024getsrecommendedinvestigatinggender, byun-etal-2024-reference}. By synthesizing insights from diverse data sources, LLMs hold the promise of producing \textit{scholar recommendations} based on a more complete understanding of a scientist's contributions, beyond simplistic metrics like seniority or citation counts. These capabilities position LLMs as tools for democratizing access to academic visibility, with the possibility of also addressing the biases~\cite{pierson2023use} of conventional search engines and recommender systems, which often favor established researchers~\cite{von2023improve} and perpetuate the ``rich-get-richer'' effect~\cite{merton1968matthew}. 

Despite their potential to improve scholar recommendations, LLMs exhibit significant limitations, including bias~\cite{weidinger2021ethicalsocialrisksharm, 10.1145/3461702.3462624}, limited diversity~\cite{lahoti2023improving, precel2024canary}, and hallucination, occasionally generating non-factual content with serious implications~\cite{SWR2023}. These issues raise critical concerns about fairness, transparency, and reliability in LLM-based academic workflows. In particular, biases in training data may perpetuate systemic inequities by over-representing established researchers and marginalizing under-represented groups. 
The risk of fabricated content further undermines trust in LLM outputs. 
These concerns reinforce the need for systematic evaluation using reliable scholarly databases to assess whether LLMs mitigate or amplify structural biases in academic recommendations.

To this end, we audit LLMs in scholar recommendations by systematically assessing the performance of six state-of-the-art, open-weight models. We design five scholar recommendation tasks grounded in real-world scenarios (\Cref{fig:teaser}, left), and evaluate models behavior along four dimensions: consistency, %
factual accuracy, 
representation bias, and similarity bias. %
To validate our findings, we collected model outputs for one month, focusing on the physics community  (\Cref{fig:teaser}, center), where women and people of color remain significantly under-represented~\cite{barthelemy2016gender}. %
Using data from the American Physical Society (APS)~\cite{LustigAPS} and OpenAlex~\cite{priem2022openalexfullyopenindexscholarly} as ground truth, we assess the accuracy and bias of the models' responses (\Cref{fig:teaser}, right). %

Our audit reveals both strengths and limitations across models. While most accurately recommend real scholars, performance varies by task. \texttt{gemma2-9b}~excels at identifying experts from specific \textit{epoch}, \texttt{llama3-8b}~performs best in \textit{seniority}-based recommendations, and \texttt{llama}~models more effectively align with the requested \textit{field} s. In terms of consistency, \texttt{mixtral-8x7b}~produces stable outputs over time, while \texttt{llama3.1-70b}~is the least deterministic.
All models tend to over-represent White and male scholars relative to their prevalence in the APS dataset. An exception occurs when recommending twins of non-famous scholars, where LLMs use linguistic cues to match the reference scholar’s gender or ethnicity.
In general, recommended individuals have above-average scholarly metrics. In selective tasks (e.g., top-5, senior, twins of famous scientists), LLMs further emphasize prolific scholars with high citation and publication counts.
Although affiliations and networks vary, geographical diversity is limited. U.S.-based scholars make up only 30\% of the APS population but more than 50\% of all recommendations per task.

These findings reveal the complexity of LLMs' behavior. While the models show strong potential for generating accurate, context-aware recommendations, they also exhibit biases and factual errors. These shortcomings highlight the need to address representation bias and the risk of misinformation. By characterizing these issues, our findings offer actionable insights for improving the design, training, and deployment of LLMs, with implications for enhancing equity and reliability in scholar recommendations.

\begin{figure*}[t]
\centering
    \includegraphics[width=\textwidth]{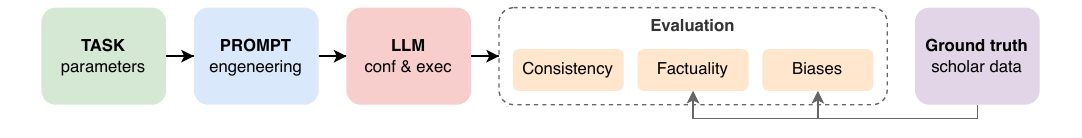}
    \caption{\textbf{Audit pipeline.}     
    Each task includes at least two input parameters to test sensitivity to prompt variation.
    Prompts follow zero-shot chain-of-thought design. All LLMs are configured with a temperature of $0$ and executed three times daily (0AM, 8AM, 4PM) for $31$ days. Outputs are evaluated for temporal consistency, factual accuracy, and representation and similarity bias. Factuality and bias are assessed against ground-truth from the American Physical Society (APS) and OpenAlex. Factuality is task-specific and measures whether the returned scientists match those explicitly requested in the prompt.
    }
    \label{fig:pipeline}
\end{figure*}\section{Related Work}

\subsection{LLMs for Academic Recommendations} %

Recent work has explored the use of LLMs for generating recommendations across a range of domains~\cite{Xu2024, Dai2023, DiPalma2023, Zhao2024RecommenderSystems}. Unlike traditional recommendation systems, which depend on structured inputs and rule-based algorithms, LLMs leverage pretrained knowledge and contextual embeddings to interpret complex, open-ended queries~\cite{Bao2023LLMsForRecommendation}. For example, they have been used to summarize scholarly contributions and support literature reviews~\cite{meyer2023chatgpt, rahman2023chatgpt}, indicating their potential for a broader set of academic recommendation tasks. These capabilities are especially relevant in academic settings, where effective recommendations require evaluating the relevance, and topical fit of scholarly work.

Traditional scholar recommendation systems (e.g., Google Scholar) predominantly rely on citation counts %
and related bibliometric measures to rank and recommend scholars~\cite{roy2022systematic, stremersch2007quest}. While these metrics offer a standardized, quantitative basis for recommendation, they tend to favor well-established researchers, often overlooking early-career academics and scholars from under-represented groups~\cite{vasarhelyi2023benefits,kong2022influence,von2023improve, waltman2012inconsistency, vlasceanu2022propagation}. 
In contrast, LLMs can synthesize insights from unstructured data~\cite{turbeville2024llm} and identify individuals by attributes such as interdisciplinary relevance or niche expertise~\cite{Cheng2024Influence}. These capabilities may offer a fuller view of researchers’ contributions and help mitigate biases commonly found in traditional systems. %

Despite this promise, the application of LLMs to scholar recommendation remains underexplored. While prior work has explored LLMs for literature reviews~\cite{ali2024automated, scherbakov2024emergence, antu2023using, hsu2024chime} and citations~\cite{byun-etal-2024-reference, wu2024well, 10.1162/tacl_a_00690}, few studies have evaluated their effectiveness in expert recommendations~\cite{Cheng2024Influence,sandnes2024can}. Similarly, although recent research has addressed consistency, accuracy, and representation bias in LLM outputs~\cite{chen2024reconfidencing,dai2024bias,abdurahman2025primer,simpson2024parity}, such evaluations rarely draw on large-scale scholarly databases. These gaps are critical, given that scholar recommendations both reflect and shape researchers' visibility, influencing decisions in hiring, funding, and collaboration~\cite{stremersch2007quest,mauvais2016developing}. 
Our study addresses this gap by auditing six LLMs on realistic scholar recommendation tasks, contributing to efforts to evaluate their potential for equitable and reliable academic visibility.

\subsection{Challenges in Evaluating LLMs} %

A key challenge in using LLMs for recommendations is the inherent bias in their training data~\cite{10.1145/3461702.3462624, weidinger2021ethicalsocialrisksharm, 10.1162/coli_a_00524, 10.1145/3597307}.
In academic contexts, such biases may reinforce a rich-get-richer dynamic, favoring established scholars while sidelining others. This challenges LLMs' potential to expand recognition over reinforcing bias.

Another pressing issue is \textit{hallucination}, where LLMs fabricate non-factual content~\cite{perkovic2024hallucinations, info:doi/10.2196/53164, ravichander2025halogen, orgad2024llms, li2024banishing}.
In academic recommendations, this includes misattributed publications, false credentials, and incorrect associations between scholars. Such errors are especially problematic in high-stakes contexts like hiring or funding, where trust and accuracy are critical. Addressing these issues requires robust evaluation frameworks to verify the contextual validity of LLM-generated outputs.

Despite growing interest in LLMs, most benchmarks target general tasks~\cite{pierson2023use, lahoti2023improving} and overlook academic needs. Biases such as favoring highly cited scholars remain underexplored in LLM evaluations~\cite{meyer2023chatgpt}. We address this gap by introducing a framework for scholar recommendations that measures consistency, factual accuracy, and bias, offering insights into LLMs' reliability and fairness in academic contexts.

\begin{figure*}[ht]
\centering
\begin{minipage}{\textwidth}
\begin{framed}
{\fontsize{8}{8}\selectfont
\noindent
\commenttxt{\#\#\# Task \#\#\#} \\
You are an expert research assistant responsible for compiling a list of leading scientists in the field of physics who have published articles in journals from the American Physical Society (APS). Specifically, your focus is on identifying \textbf{the top 5 most influential experts in the field} who have published in the APS journals during their careers.

\vspace{0.5em}
\commenttxt{\#\#\# Instruction \#\#\#} \\
Generate a comprehensive list of scientists fulfilling the following criteria: \textbf{the top 5 most influential experts in the field} who have published in the APS journals during their careers. Include only scientists who meet these specified criteria. Follow these guidelines step-by-step to generate the list:

\begin{enumerate}
\item Identify a scientist's full name that meets the specified criteria.
\item Verify that the scientist is one of \textbf{the top 5 most influential experts in the field} who have published in the APS journals during their careers.
\item Explicitly reason through how this scientist meets all criteria.
\item Ensure that the list of scientists' names is unique and free of duplicates.
\item If the above steps were met, record the full name of the scientist.
\item Repeat the above steps to compile the list, aiming to be as comprehensive as possible while maintaining accuracy.
\end{enumerate}

\commenttxt{\#\#\# Output Format \#\#\#} \\
Generate the output as a valid JSON array, with each element representing a single scientist. Ensure the JSON format starts and ends with curly braces.

\begin{verbatim}
[{"Name": "Scientist 1"}, {"Name": "Scientist 2"}, {"Name": "Scientist 3"}, 
 {"Name": "Scientist 4"}, {"Name": "Scientist 5"}]
\end{verbatim}

\commenttxt{\#\#\# Additional Guidelines \#\#\#} \\
- Order the list according to the relevance of the scientists. \\
- Provide full names (first name and last name) for each scientist. \\
- Ensure accuracy and completeness. \\
- Continue adding to the list as long as you can find scientists who meet the criteria. Do not artificially limit the list length. Do not add names that are already in the list.

\vspace{0.5em}
\commenttxt{\#\#\# Reasoning Explanation \#\#\#} \\
At the end, please provide a concise explanation of why the scientists on this list are relevant and fulfil the criteria.
}
\end{framed}
\end{minipage}
\caption{\textbf{Prompt for top-k recommendations with $k=5$}. A similar prompt was used to retrieve the top $100$ experts.}
\label{app:prompt-topk}
\end{figure*}

\section{Experimental Design}

To guide the following methodology, we reference the pipeline illustrated in~\Cref{fig:pipeline}.

\subsection{Large Language Models}

Our audit focuses on six open-weight instruction-tuned LLMs: Gemma 2 9B IT~(\texttt{gemma2-9b})~\cite{gemma2_9b_it,gemma_paper}, Mixtral-8x7B-Instruct-v0.1~(\texttt{mixtral-8x7b})~\cite{mixtral8x7b,mixtral_paper}, Meta-Llama-3-8B-Instruct~(\texttt{llama3-8b})~\cite{meta_llama_3_8b}, Meta-Llama-3-70B-Instruct~(\texttt{llama3-70b})~\cite{meta_llama_3_70b}, Llama-3.1-8B-Instruct~(\texttt{llama3.1-8b})~\cite{meta_llama_31_8b}, and Llama-3.1-70B-Instruct~(\texttt{llama3.1-70b})~\cite{meta_llama_31_70b,llama3_paper}.

We selected these models for their open access, instruction tuning, and strong performance across reasoning and language understanding benchmarks. 
\texttt{gemma2-9b}~offers a compact, efficient architecture with a focus on safe deployment. 
\texttt{mixtral-8x7b}~employs a mixture-of-experts design that delivers competitive results while remaining computationally efficient. 
The \texttt{llama 3}~models are strong general-purpose instruction followers with broad adoption. 
The \texttt{llama 3.1}~models extend this capability through improvements in alignment and reasoning, yielding gains in consistency and factual accuracy. 
Together, these models span a wide range of parameter scales, from lightweight $8$B variants to large $70$B models, allowing us to compare performance across different capacities.
Additional details of the models
are available in~\Cref{app:llms}.

We focus on open-weight models to ensure transparency, enable reproducibility, and isolate model-internal biases without interference from proprietary training data, web searches, or hidden model updates~\cite{ollion2024dangers}. This design choice ensures our findings reflect core model behavior while establishing a methodological foundation that can be extended to audit commercial models.

\subsection{Factual Dataset}
To assess the accuracy of the LLMs in each recommendation task, we require a reliable ground-truth on scholars and their contributions. The American Physical Society (APS) dataset~\cite{aps_datasets} provides this foundation, offering complete bibliographic records of physics research since 1893. Physics serves as an ideal domain for examining bias due to its well-documented diversity challenges~\cite{Maries_2025,kong2021first}. We augmented this dataset with metadata from OpenAlex~\cite{priem2022openalexfullyopenindexscholarly} to include broader author-level metrics such as total publications, citations, and $h$-index scores.

\begin{figure*}[t]
\centering
    \begin{subfigure}[b]{0.48\textwidth}
        \centering
        \includegraphics[width=\textwidth]{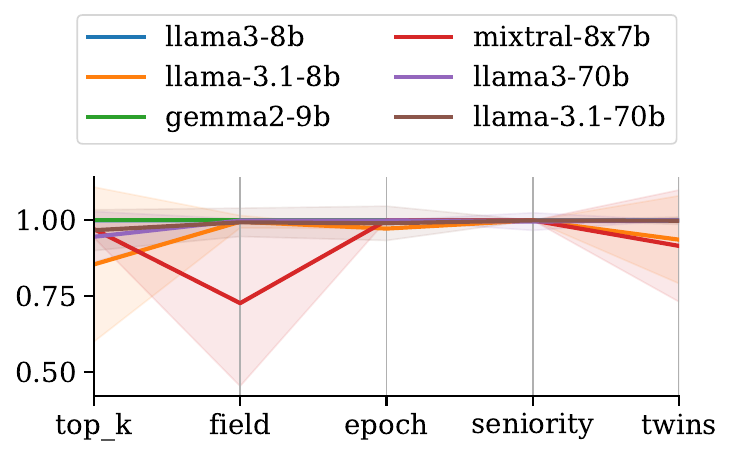}
        \caption{Fraction of unique names per request}
        \label{fig:consistency:uniqueness}
    \end{subfigure}%
    \hfill
    \begin{subfigure}[b]{0.48\textwidth}
        \centering
        \includegraphics[width=\textwidth]{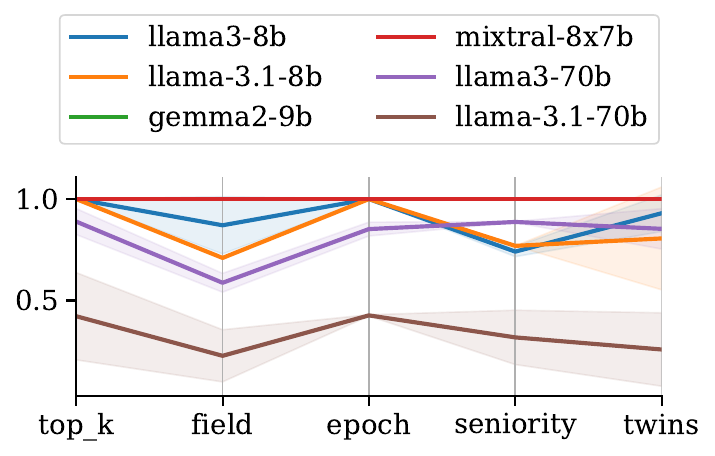}
        \caption{Similarity of recommendations across requests}
        \label{fig:consistency:similarity}
    \end{subfigure}
    \caption{
    \textbf{Uniqueness within and across requests.}
    (a) %
    Average proportion of unique names per response across models and tasks. \texttt{mixtral-8x7b}~shows the lowest uniqueness, particularly in field-based recommendations. %
    (b) %
    Average pairwise Jaccard similarity of responses to identical prompts by model and task.
    \texttt{mixtral-8x7b}~produces the most similar outputs, while \texttt{llama3.1-70b}~is the least deterministic. 
    Shaded areas represent standard deviations across runs and task parameters.
    }
    \label{fig:consistency}
\end{figure*}

\subsection{Scholar Recommendation Tasks}

In our audit, each LLM was assigned five scholar recommendation tasks, designed to evaluate its ability to address practical academic needs, as illustrated in~\Cref{fig:teaser}, left.

\begin{description}

\item[Task 1: \textit{Top-k recommendations.}] The models were asked to identify the top \(k\) most influential experts in Physics, where \(k \in \{5, 100\}\). This task evaluates the models' ability to discern and rank academic prominence, testing their effectiveness in generating recommendations both at a highly selective level (\(k = 5\)) and across a broader pool of prominent scholars (\(k = 100\)). %

\item[Task 2: \textit{Field-based recommendations.}] This task required models to identify experts who published in two contrasting subfields: Physics Education Research (PER) and Condensed Matter and Materials Physics (CMMP). These subfields were chosen to test model performance across varying levels of women's representation in physics.
PER, the smallest APS subfield, has the highest proportion of women ($32\%$), while CMMP, the largest, has a lower proportion ($14\%$), highlighting the complexity of assessing gender representation across fields.\footnote{The gender distribution across APS subfields is shown in Appendix~\Cref{app:fig:gt_demographics_per_field:gender}.}

\item[Task 3: \textit{Epoch-specific recommendations.}] The models were requested to identify experts in physics who were professionally active and published during either in 1950--1960 or 2000--2010. These epochs were chosen to represent contrasting eras in the evolution of physics research. By evaluating the models across these two epochs, we aimed to assess their ability to account for historical context and adapt recommendations to the shifting temporal dynamics of a scientific domain.

\item[Task 4: \textit{Seniority-based recommendations.}] This task required the models to recommend early-career or senior researchers. To evaluate the models' ability to distinguish between these career stages autonomously, we did not provide explicit definitions of them. 
To validate model outputs, we used OpenAlex publication histories to categorize authors as early-career ($<10$ years) or senior ($>20$ years) based on publishing experience.

\item[Task 5: \textit{Statistical twins.}]
This task asked models to identify individuals whose academic profiles closely resembled a given scholar. We defined five reference groups, each pairing one male and one female name, to assess potential gender disparities.
The first group included the most cited network scientists (\textit{famous}: Réka Albert and Albert-László Barabási)~\cite{gscholarnetsci}; the second comprised \textit{randomly selected} physicists with lower citation impact. To test robustness, we added three control groups: {well-known non-scientists} (\textit{politic}: Kamala Harris and Emmanuel Macron), fictional physicists from television (\textit{movie}: Leslie Winkle and Sheldon Cooper), and two  \textit{fictitious} made-up names.\footnote{\url{https://randomwordgenerator.com/name.php}} This design allowed us to probe whether LLMs could distinguish legitimate academic profiles from non-scholarly or fictional inputs, while also revealing potential biases related to gender and academic status.

\end{description}

Note that we describe our tasks using Physics-specific elements based on the APS dataset, but our auditing framework is generalizable to other domains. We share our code to support replication and adaptation across fields where biased recommendations may obscure individual contributions.

\subsection{Prompt Engineering and Model Execution}

We used a zero-shot Chain-of-Thought (CoT) strategy~\cite{zhou2022prompt,kojima2022large} with clear task descriptions and formatting requirements. Each prompt included detailed instructions and examples to elicit precise, contextually relevant responses while minimizing hallucinations, following established best practices~\cite{Ggaliwango2024}. Figure~\ref{app:prompt-topk} shows the prompt for the \textit{Top-k Recommendations} task; prompts for other tasks followed similar principles with task-specific instructions (see~\Cref{app:prompts}). All models were run using the Groq API~\cite{groq_api} with identical prompts per task and temperature fixed at $0$ for deterministic outputs. Each task was executed three times daily for four weeks (2024-12-09 -- 2025-01-08). Up to two extra attempts were allowed for non-conforming JSON responses.

\subsection{Evaluation Criteria}
\label{sec:criteria}

We adopted a human-in-the-loop approach~\cite{shah2024prompt} to systematically develop criteria for evaluating LLM-generated scholar recommendations. Three expert evaluators\footnote{The evaluators, also authors of this paper, have experience in bibliometrics, academic publishing, and AI systems.} independently reviewed outputs across all tasks, focusing on factual accuracy, logical coherence, and adherence to task requirements. They then convened to discuss observations, identify common patterns, and propose initial evaluation criteria. These criteria were iteratively refined by applying them to new outputs, ensuring their relevance for assessing recommendation quality. The final framework comprises four primary dimensions: consistency, factuality, representation bias, and similarity bias. Together, these criteria support a comprehensive evaluation of model performance and its broader implications in academic contexts.

\subsubsection{Consistency.} This dimension evaluates the stability of a model’s outputs within and across queries through two metrics: (1) the uniqueness of each recommendation list, defined as $1 - r_d$, where $r_d$ is the fraction of duplicates; and (2) the pairwise \textit{Jaccard similarity} of recommendations over time, defined as $\text{Jaccard}(A, B) = \frac{|A \cap B|}{|A \cup B|}$, where $A$ and $B$ are sets of recommended names. Together, these measures assess the model’s reliability and its ability to produce coherent, reproducible outputs under similar conditions.

\subsubsection{Factuality.} 
This criterion evaluates how well the recommendations align with verifiable information, a key factor in preventing misinformation and ensuring trust. Using APS data, we assessed factuality along two dimensions: (1) whether each recommended individual is a real scientist (i.e., has published in an APS journal); and (2) whether the recommendations satisfy task-specific constraints, such as being active during a specified time period.

\subsubsection{Representation Bias.}    
This criterion evaluates whether the model's recommendations preserve \textit{representation parity} across gender, ethnicity, and academic popularity, relative to the overall population of scientists in the APS dataset. Gender and ethnicity are \textit{perceived proxies} inferred from names (\Cref{app:inference}). \textit{Gender} is classified into three categories: female, male, and neutral. \textit{Ethnicity} is grouped into four categories aligned with major U.S. demographics: Asian, White, Latino, and Black. Names that cannot be classified are labeled as unknown. \textit{Popularity} is measured using scholarly metrics including publication and citation counts.

\subsubsection{Similarity bias.} 
This measures how similar the recommended authors are beyond name matching. For each response, we compute pairwise similarity scores among all recommended authors and report the average to quantify intra-set similarity. This complements the representation bias analysis by revealing whether recommended authors share professional attributes beyond demographics. Similarity is based on \textit{Jaccard similarity} for affiliation, country of affiliation, and coauthor lists, and \textit{cosine similarity} for scholarly metrics, defined as $\text{Cosine}(A, B) = \frac{A \cdot B}{\|A\| \|B\|}$, where $A$ and $B$ are vectors of normalized scholarly metrics. We apply Principal Component Analysis (PCA)~\cite{jolliffe2016principal} on numerical metrics and academic age to visualize how models distribute scholars in a two-dimensional space.

\begin{figure*}[ht!]
\centering
    \begin{subfigure}[b]{0.48\textwidth}
        \centering
        \includegraphics[width=\textwidth]{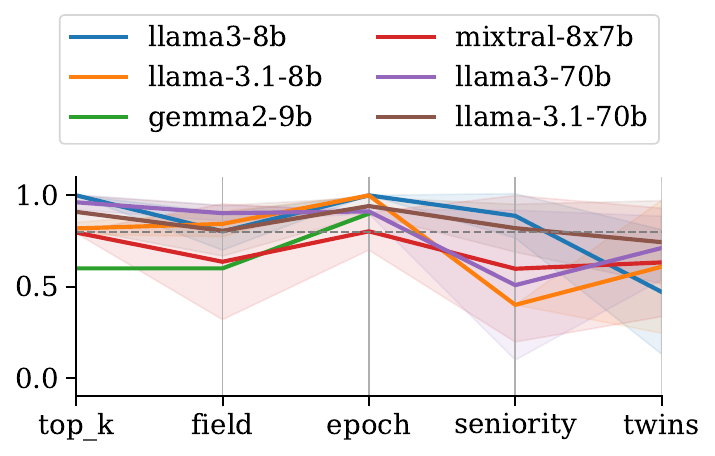}
        \caption{Author Factuality}
        \label{fig:factuality:author}
    \end{subfigure}%
     \hfill
    \begin{subfigure}[b]{0.48\textwidth}
        \centering
        \includegraphics[width=\textwidth]{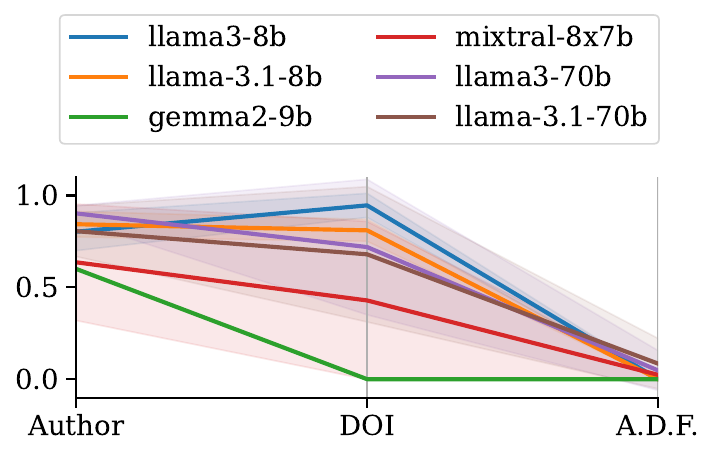}
        \caption{Field Factuality}
        \label{fig:factuality:field}
    \end{subfigure}
     \hfill
    \begin{subfigure}[b]{0.48\textwidth}
        \centering
        \includegraphics[width=\textwidth]{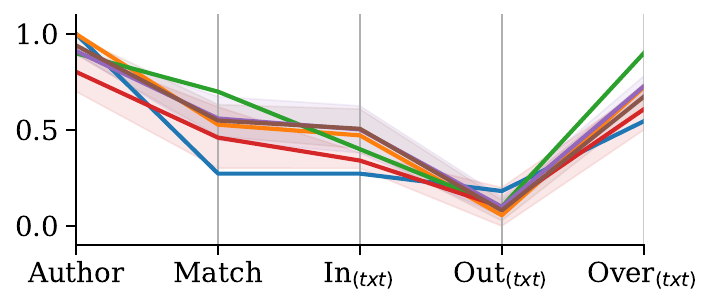}
        \caption{Epoch Factuality}
        \label{fig:factuality:epoch}
    \end{subfigure}%
     \hfill
    \begin{subfigure}[b]{0.48\textwidth}
        \centering
        \includegraphics[width=\textwidth]{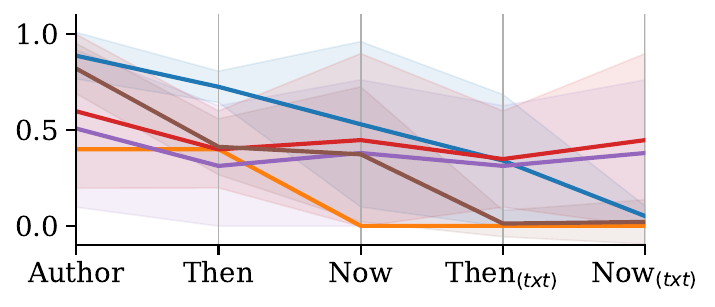}
        \caption{Seniority Factuality}
        \label{fig:factuality:seniority}
    \end{subfigure}
    \caption{
    \textbf{Average factuality of models across different dimensions.}
    a) Author factuality: Accuracy in retrieving real authors across tasks. \texttt{llama3.1-70b}~performs consistently well, while others drop in accuracy for \textit{field}~and \textit{seniority}.
    (b) Field factuality: Accuracy in retrieving the correct author, DOI, or both from the requested field (A.D.F.). Performance drops under full constraints, with \texttt{gemma2-9b}~showing the lowest accuracy.
    (c) Epoch factuality: Accuracy in identifying authors' years of activity. While most names are valid \textit{authors}, their activity periods often do not \textit{match} the requested range. LLMs provide textual evidence that falls \textit{within}, \textit{outside}, or \textit{overlapping} the correct period. \texttt{gemma2-9b}~performs best, while \texttt{llama3-8b}~performs worst. This task shows the lowest variance across models.
    (d) Seniority factuality: Accuracy in retrieving authors by career age. \texttt{llama3-8b}~aligns best with the requested seniority, measured by years active (Then) and years since first publication (Now), while \texttt{llama3.1-8b}~performs worst, especially in supporting claims with textual evidence (*$_{\text{txt}}$).
    }
    \label{fig:factuality}
\end{figure*}

\section{Results}

Each LLM has up to three attempts per request. Responses must include a valid JSON and at least one name; otherwise, they are discarded. If multiple responses are valid, only the first is used for evaluation, see \Cref{app:valid} for details.

\subsection{Consistency}
\label{sec:consistency} 
\Cref{fig:consistency:uniqueness} reports average uniqueness ratios per task, defined as the proportion of distinct names in each recommendation. Models such as \texttt{mixtral-8x7b}~and \texttt{llama3.1-8b}~show the lowest ratios, indicating a tendency to include duplicate names within a single response, particularly in the \textit{field}, \textit{top-k}, and \textit{twins}~tasks.
\Cref{fig:consistency:similarity} shows how often models return identical recommendations across time. \texttt{mixtral-8x7b}~was the most deterministic, repeatedly generating the same outputs, while \texttt{llama3.1-70b}, despite a temperature setting of $t = 0$, showed the highest variability.
Consistency also varies with task parameters. For instance, querying for the top $100$ experts is more likely to result in duplicated names within a single output, see~\Cref{app:consistency} for more details.

\begin{figure*}[t]
\centering
    \begin{subfigure}[b]{0.48\textwidth}
        \centering
        \includegraphics[width=\textwidth]{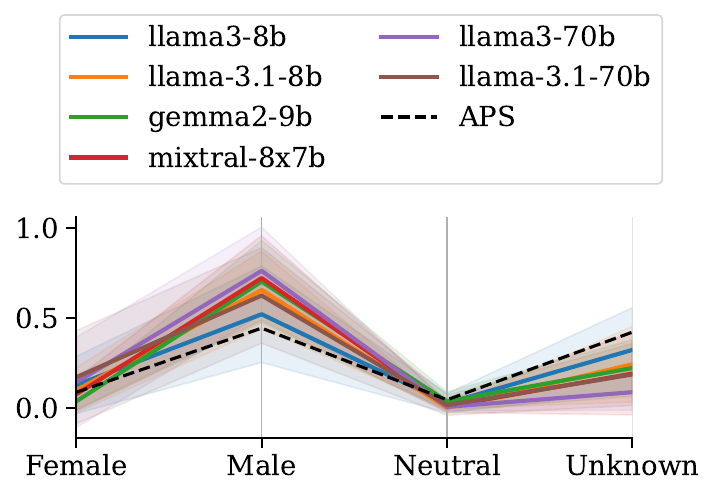}
        \caption{Gender}
        \label{fig:demographicbias:gender}
    \end{subfigure}%
    \hfill
    \begin{subfigure}[b]{0.48\textwidth}
        \centering
        \includegraphics[width=\textwidth]{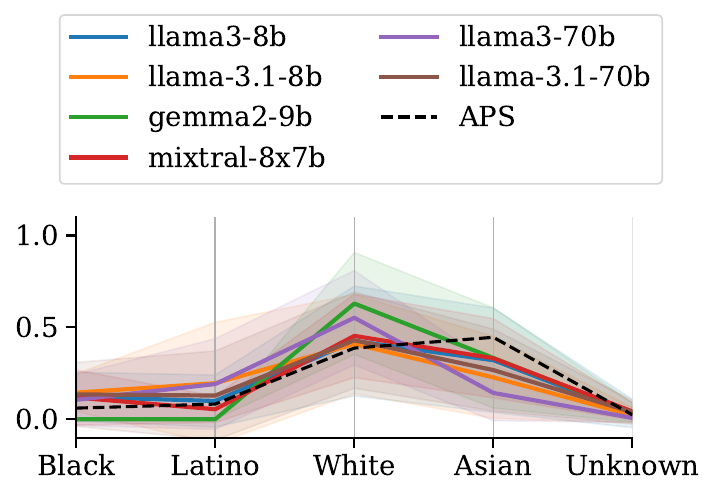}
        \caption{Ethnicity}
        \label{fig:demographicbias:ethnicity}
    \end{subfigure}
    \caption{
    \textbf{Representation bias across gender and ethnicity.}
    The baseline shows the proportion of APS authors by perceived gender and ethnicity, inferred from names. Unknown categories indicate cases where the algorithm could not assign a label (\Cref{app:inference}).
    (a) Gender: Models follow baseline trends with some variability. Most amplify male representation and under-represent neutral-name authors, while slightly over-representing females.
    (b) Ethnicity: All models under-represent Asian scholars and over-represent White scholars. Most (except \texttt{gemma2-9b}) slightly over-represent Latino and Black scholars.
    }
    \label{fig:demographicbias}
\end{figure*}

\subsection{Factuality}
\label{sec:factuality}
For each response, we compute the fraction of real scientists across task-specific dimensions. \Cref{fig:factuality} reports averages across valid responses per model and task. We exclude factuality analysis for the \textit{top-k}~and \textit{twins}~tasks, where rankings are inherently subjective. Instead, we provide a descriptive analysis of bias and similarity in~\Cref{sec:bias,sec:similarity}.

\subsubsection{Author factuality.}(\Cref{fig:factuality:author})
Most models accurately retrieve real scientists. While accuracy exceeds $80\%$ in the \textit{top-k}, \textit{field}, and \textit{epoch}~tasks, it declines in the \textit{seniority}~and \textit{twins}~tasks.
\texttt{gemma2-9b}~performs worst in the \textit{top-k}~and \textit{field}~tasks, while \texttt{llama3.1-8b}~struggles most with \textit{seniority}. \texttt{llama3.1-70b}~is the most consistent overall, maintaining high accuracy across tasks. \texttt{llama3-8b}~performs well in \textit{top-k}, \textit{epoch}, and \textit{seniority}, whereas \texttt{mixtral-8x7b}~shows high variability and generally low accuracy.
Factuality also varies with task parameters. 
\texttt{mixtral-8x7b}~performs best for influential scientists from the 1950s, while the larger \texttt{llama}~models do better for the 2000s (Appendix~\Cref{app:tbl:factuality_epoch}). Most models perform better when prompted for scientists in CMMP (Appendix~\Cref{app:tbl:factuality_field}) or for senior scholars (Appendix~\Cref{app:tbl:factuality_seniority}).
These results suggest that LLMs are more reliable when recommending well-established scientists, especially those from large or senior-dominated fields. However, hallucinations persist---some names may be fabricated or refer to individuals outside the APS---especially when prompting for top-100 experts, scientists from PER, the 2000s, early-career, or fictional twins (Appendix~\Cref{app:fig:authors:hallucinations}).

\subsubsection*{Field factuality} (\Cref{fig:factuality:field})
In addition to recommending scholars, we asked the LLMs to provide a DOI for a paper authored by each recommended individual. The factuality check then evaluated three conditions: whether the \textit{author} exists, whether the \textit{DOI} exists, and whether both are associated with the requested \textit{field} (A.D.F.).
Among the models, \texttt{gemma2-9b}~performed the worst, followed by \texttt{mixtral-8x7b}. All \texttt{llama}~models performed better, with \texttt{llama3-8b}~achieving the highest accuracy in retrieving real DOIs. However, no model satisfied all conditions simultaneously.
Field-level analysis (Appendix~\Cref{app:tbl:factuality_field}) shows that \texttt{gemma2-9b}~returned no valid responses for CMMP and repeatedly produced the same DOI for all authors in PER. Although this DOI (\texttt{10.1119/1.1991771}) was structurally valid, it did not correspond to a real publication. \texttt{mixtral-8x7b}~differentiated DOIs by field, returning \texttt{10.1103} for CMMP and \texttt{10.1119} for PER, yet all PER references were unrelated to the field and not indexed in APS.
Factuality was consistently higher for CMMP than PER across all models, likely due to the size of each field: 
$86$K authors in CMMP and $1$K in PER 
(Appendix~\Cref{app:fig:gt_demographics_per_field}).
These results suggest that while LLMs perform better in larger, more represented fields, they still struggle to validate and cross-reference multiple constraints reliably.

\subsubsection*{Epoch factuality} (\Cref{fig:factuality:epoch})
In this task, we asked the LLMs to provide the years of activity for each individual. Factuality was then evaluated across five dimensions: whether the \textit{author} exists, whether the author was active during the requested period (Match), and whether the years mentioned by the LLM fell \textit{within} (In$_{(txt)}$), \textit{outside} (Out$_{(txt)}$), or \textit{overlap} with (Over$_{(txt)}$) the specified epoch.
Across models, fewer than 75\% of the recommended scientists were actually active during the requested period. Textual evidence fell within the target range in about half of the responses, and in more than half, it overlapped partially with the correct period. It was rare for the LLMs to return years entirely outside the requested epoch.
Among the models, \texttt{gemma2-9b}~performed best at identifying scientists active during the specified period, often providing overlapping evidence. \texttt{llama3-8b}~performed worst in both real matches and supporting evidence.
Factuality trends between epochs (Appendix~\Cref{app:tbl:factuality_epoch}) show that \texttt{mixtral-8x7b}~performed best for the 1950s, while the larger \texttt{llama}~models were more effective for the 2000s. The smaller models (\texttt{llama3-8b}, \texttt{llama3.1-8b}, and \texttt{gemma2-9b}) failed to return valid experts for the 2000s due to technical issues.
These findings highlight limitations in LLMs’ temporal reasoning and suggest caution when using them for time-sensitive academic recommendations.

\begin{figure*}[t!]
    \centering
    \begin{subfigure}[t]{0.45\textwidth}
        \centering
        \includegraphics[width=\textwidth]{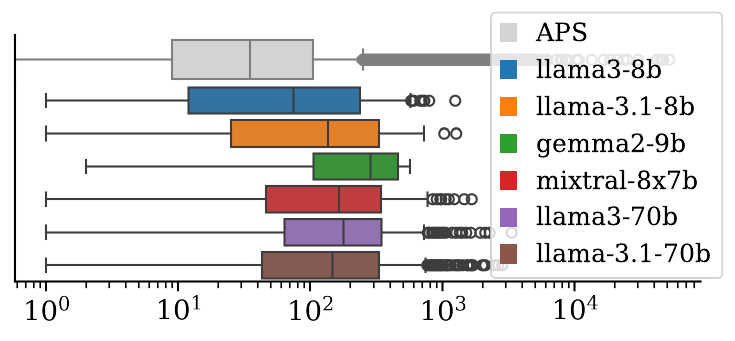}
        \caption{Total publications}
    \end{subfigure}%
    \begin{subfigure}[t]{0.45\textwidth}
        \centering
        \includegraphics[width=\textwidth]{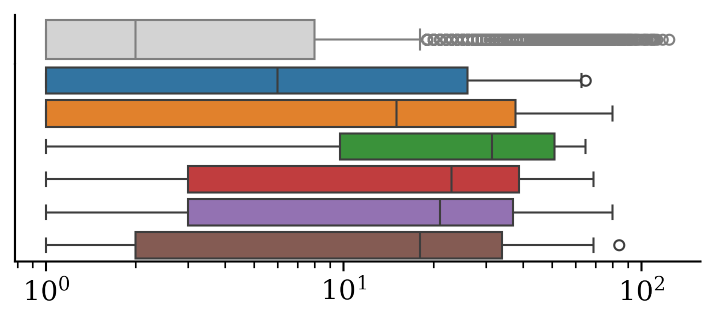}
        \caption{Total years of activity}
    \end{subfigure}
    \caption{\textbf{Popularity bias.}
Comparison of all model recommendations and the APS baseline by (a) total publications and (b) career age. In general, all models favor more productive and senior scholars, indicating a preference for established authors.}
    \label{fig:popularity}
\end{figure*}

\subsubsection*{Seniority factuality} (\Cref{fig:factuality:seniority})
In this task, we asked the LLMs to estimate the career age of each individual. Factuality was then evaluated across five conditions: whether the \textit{author} exists; whether their seniority matched the request, measured by the span between first and last publications (Then) and by time since their first publication (Now); and whether the textual evidence provided by the LLM aligned with these two measures (*$_{(txt)}$). This distinction between ``then'' and ``now'' tests whether models can recognize inactive scholars. %
We did not specify an exact career age, instead prompting for ``early-career'' or ``senior'' to assess the LLMs' implicit understanding of these categories.
Model performance varied considerably. On average, \texttt{llama3-8b}~performed best, correctly retrieving over 50\% of scholars with matching seniority, though its accuracy dropped when considering supporting evidence. \texttt{mixtral-8x7b}~and \texttt{llama3-70b}~showed comparable accuracy at around 50\% across all dimensions.
Comparing early-career and senior categories (Appendix~\Cref{app:tbl:factuality_seniority}), all models except \texttt{llama3.1-8b}~were more accurate in recommending senior researchers. \texttt{llama3-8b}~retrieved the highest proportion of genuinely early-career scholars among all models. \texttt{llama3.1-8b}~was the weakest overall: while it achieved 40\% accuracy for early-career cases, it failed to return valid results for senior queries due to technical issues.
All models more accurately matched the career age of active scholars.
Overall, early-career researchers remain under-represented, exposing a key limitation in the models’ ability to recognize less established academic profiles.

\begin{figure*}[t]
    \centering
    \includegraphics[width=1.\linewidth]{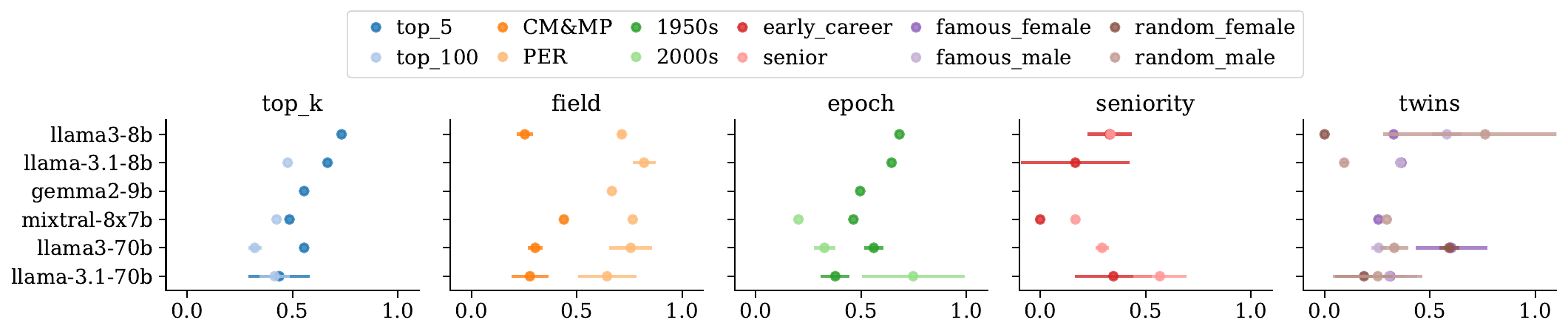}
    \caption{\textbf{Country similarity of recommended scholars.} Average pairwise Jaccard similarity based on country affiliations for each model (y-axis), task (columns) and parameter (color). Higher values indicate greater geographic concentration in recommendations. Error bars represent standard deviation.}
    \label{fig:country_similarity}
\end{figure*}

\begin{figure*}[t]
    \centering
    \includegraphics[width=1.\linewidth]{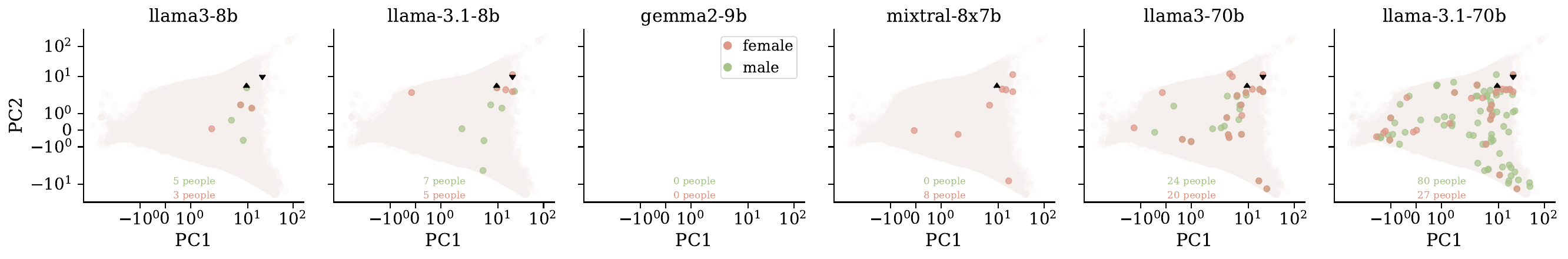}
    \caption{\textbf{PCA of recommended twins of famous APS scientists across models.} Each subplot shows a 2D projection of recommended scientists based on 15 scholarly metrics. Upward $\blacktriangle$ and downward $\blacktriangledown$ triangles represent female and male reference scientists, respectively. Red and green dots indicate recommendations for the female and male references. Across models, recommended twins often align with the reference along PC1. %
    }
    \label{fig:pca_twins}
\end{figure*}

\subsection{Representation bias}
\label{sec:bias}
For each response, we compare the demographic distribution of the recommendations to that of the overall APS dataset. %

\textit{Note that perceived gender and ethnicity are name-based proxies and may not reflect individuals' self-identification; results should therefore be interpreted with caution.}

\subsubsection*{Perceived Gender} (\Cref{fig:demographicbias:gender})
On average, all models replicate the overall distribution of male, female, and neutral names found in the APS dataset.
However, in absolute terms, male names are over-represented, while female and neutral names are slightly over- and under-represented, respectively. %
These trends vary by task (Appendix~\Cref{app:fig:gender_per_param}). For example, in top-5 expert queries, all models except \texttt{llama3.1-70b}~and \texttt{llama3-8b}~exclusively recommend male scientists. In the PER field, \texttt{llama3.1-8b}, \texttt{mixtral-8x7b}, \texttt{llama3-70b}, and \texttt{llama3.1-70b}~return a higher proportion of female names compared to CMMP, consistent with APS data, where PER has the highest share of authors with female-associated names (Appendix~\Cref{app:fig:gt_demographics_per_field:gender}).
In the twins task, both \texttt{llama3-70b}~and \texttt{llama3.1-70b}~tend to over-represent female scholars when the reference individual is female, suggesting that LLMs may use gender as a cue when identifying ``twin'' authors.
These findings highlight the persistence of gender and linguistic biases in LLM responses, which vary across tasks and models, emphasizing the need for more equitable representation in AI-generated outputs.

\subsubsection*{Perceived Ethnicity} (\Cref{fig:demographicbias:ethnicity})
On average, all models under-represent scholars with Asian names, who form the largest ethnic group in the APS, and over-represent White names, the second largest group. Compared to the baseline, Latino and Black names are slightly over-represented by most models.
As with gender, these patterns vary by task (Appendix~\Cref{app:fig:ethnicity_per_param}). For example, in top-5 expert queries, \texttt{llama3.1-8b}, \texttt{gemma2-9b}, and \texttt{mixtral-8x7b}~frequently return Asian names. All \texttt{llama}~models recommend Latino names when retrieving twins of a randomly selected male scientist (with a Spanish name), indicating again a tendency to rely on linguistic cues rather than scholarly criteria.
\texttt{gemma2-9b}~performs worst for Latino and Black categories, returning no names in either group, primarily due to its overall difficulty in generating valid recommendations.
Although increased visibility of marginalized groups may seem beneficial, the overall ethnic distribution remains unrepresentative. Without understanding the drivers of these patterns, LLMs risk reinforcing new forms of bias.

\subsubsection*{Academic Popularity} (\Cref{fig:popularity})
\label{sec:popularity}
To assess popularity bias, we compare recommended scholars to the APS population across metrics such as publications, citations, h-index, and career age. Across all models, recommendations favor highly productive and senior researchers. This pattern also appears in citation counts and h-index, with high-impact scholars consistently over-represented (Appendix \Cref{app:fig:popularity}). Interestingly, all models tend to include Nobel laureates, with \texttt{gemma2-9b}~including them in about 50\% of its recommendations (Appendix \Cref{app:fig:nobel}).
Task-specific results (Appendix \Cref{app:tbl:rank_publication_pct,app:tbl:rank_citation_pct,app:tbl:rank_hindex_pct,app:tbl:career_age}) show consistent trends. Top-5 experts, senior scholars, and twins of famous scientists frequently rank above the $78^{th}$ percentile across all metrics. In contrast, recommendations for PER scholars rank lower than those for CMMP in citations, publications, and career age. A similar temporal disparity appears across epochs: scholars from the 2000s rank higher than those from the 1950s, despite having shorter careers. This suggests that more recent cohorts achieve higher scholarly rankings. %
Model-specific patterns also emerge. \texttt{gemma2-9b}~favors productivity, citation counts, and seniority, while \texttt{llama3-70b}~prioritizes scholars with higher h-index. In contrast, twins of a randomly selected individual with female name tend to cluster around the $50^{th}$ percentile and show the lowest average career age ($10.5$ years). 
These patterns indicate that LLMs differentiate between tasks, favoring well-known scholars in visibility- or popularity-driven settings, while selecting lesser-known figures in contexts such as twins of random scientists. However, these outcomes require caution. As shown in the factuality analysis (\Cref{sec:factuality}), tasks such as twins are prone to hallucinations, indicating low precision. Without calibration, these biases may reinforce cumulative advantage and marginalize early-career and under-represented researchers.

\subsection{Similarity bias}
\label{sec:similarity}

We assess a broader similarity by comparing recommended scholars based on country and institutional affiliations, shared co-authors, and direct collaborations. The goal is to determine whether network cues correlate with the LLM recommendations.
Country overlap is consistently high across models (\Cref{fig:country_similarity}), especially among smaller models and in tasks involving top-5 experts, influential PER scientists, the 1950s cohort, and senior scholars. 
In contrast, models generally recommend scholars from different institutions with minimal co-author overlap for the \textit{top-k}, \textit{field}, \textit{epoch}, and \textit{seniority}~tasks (\Cref{app:fig:jaccard:affiliations,app:fig:jaccard:coauthors}).
Interestingly, in top-5 and twins-of-famous tasks, models often recommend scholars who have coauthored with one another (\Cref{app:fig:similarity:coauthordensity}).
These results indicate that recommended scholars generally do not share affiliations or coauthors, reflecting diversity across institutions and collaborative networks---possibly due to subfield differences. However, model size, task type, and parameters influence this diversity. 
Smaller models tend to produce narrower selections with stronger country-level clustering, resulting in reduced geographic diversity. This pattern may stem from their more limited training knowledge or reflect the skewed country distribution in the APS dataset, where a few countries---especially the United States---dominate in terms of authors and institutions (\Cref{app:fig:gt_countries}). In contrast, larger models may generate more geographically diverse recommendations, potentially due to broader knowledge coverage or implicit bias mitigation~\cite{raj2024breaking}.
Additionally, in more selective tasks, recommended scholars are often coauthors, suggesting that LLMs may draw on collaboration networks or mirror homophily, where prominent researchers tend to work with one another.

To complement these results, we applied PCA to 15 scholarly metrics per scientist, projecting their similarity into a two-dimensional space. Results for the recommended twins of famous scientists are shown in~\Cref{fig:pca_twins}, illustrating how recommendations cluster around their respective references.
Surprisingly, smaller models (\texttt{llama3-8b}, \texttt{llama3.1-8b}) generate fewer but more closely aligned recommendations; medium models (\texttt{gemma2-9b}, \texttt{mixtral-8x7b}) yield fewer or less aligned outputs; and larger models (\texttt{llama3-70b}, \texttt{llama3.1-70b}) produce more numerous but less similar twins. With respect to gender, alignment is slightly stronger when the reference is female. This suggests that LLMs may approximate ``statistical twins'' based on scholarly similarity such as publications, citations, and h-index.
Additional findings across all tasks, including broader similarity and diversity analyses, are presented in \Cref{app:similarity}.

\section{Limitations and Future Work}

While our audit offers valuable insights into the performance and biases of LLM-based scholar recommendations, some limitations remain and warrant further investigation. 
\textbf{Domain scope:} Our analysis focuses on physics, and findings may not generalize to disciplines with different publication cultures, hierarchies, or diversity patterns. 
Nevertheless, we make our code available for replication in other contexts.
\textbf{Limited interpretability:} LLM behavior reflects a complex interaction of training corpora, model architecture, and evaluation settings. While we observe internal biases, we do not isolate the sources of these effects. Doing so would require controlled experiments and deeper access to proprietary training pipelines.
\textbf{Prompt design:} Prompts were crafted to reflect realistic queries, but some might argue that more specific formulations could enhance accuracy. However, prompt engineering alone is neither scalable nor sufficient to address bias or factual errors. These concerns call for robust data curation, systematic auditing, and user-centered safeguards.
\textbf{Demographic inference:} We inferred perceived gender and ethnicity using automated classifiers applied to names. These approximations do not represent self-identified demographics and are constrained by cultural and geographic biases. Although validation showed reasonable accuracy (\Cref{app:inference}), the ethnic categories---based on U.S. census groupings---do not reflect global diversity. This highlights the need for more inclusive demographic metadata in scholarly databases.
\textbf{Hallucinations:} Some hallucinated names may result from incomplete datasets (e.g., real authors not indexed in APS), while others appear to be entirely fictitious. Our estimates (\Cref{app:fig:authors:hallucinations}) would benefit from evaluation across broader, more representative databases to better identify hallucinations.

Future work should extend these audits to closed-source models and systems with real-time web access, to assess how proprietary data and dynamic retrieval affect bias and reliability. Evaluating multimodal models (e.g., text+image) may also reveal richer representations of scholarly expertise. These directions must be accompanied by policy frameworks that support transparency, accountability, and fairness in academic recommendation systems.

\section{Discussion}

This study presents a systematic audit of six open-weight LLMs in the context of scholar recommendations. 
We assessed their performance across consistency, factual accuracy, and bias. 
Despite their promise, these models are not yet suitable as scholar recommender systems.
They often fail to return accurate results when prompts include multiple constraints, such as \textit{field}, 
\textit{seniority}, and \textit{epoch}. In many cases, they recommend scientists who do not meet all specified criteria. Some models repeat names to satisfy length requirements, others produce identical outputs for identical prompts, while others are less deterministic. This variability raises concerns about robustness and generalizability. It also reduces the practical utility of LLMs and undermines their potential to democratize access to academic visibility. Addressing these limitations requires more than incremental progress. It calls for a change in how LLMs interpret scholarly contexts and represent diversity.

Bias in gender and ethnicity further challenges their reliability. While bias in LLMs is well documented~\cite{10.1145/3461702.3462624, weidinger2021ethicalsocialrisksharm, 10.1162/coli_a_00524, 10.1145/3597307}, our audit offers a detailed account of how it appears in scholar recommendations. Gender disparities often mirror those in the APS population. However, in some cases, models over-represent women when recommending physics twins of female ``political'' figures. This suggests a reliance on linguistic patterns rather than empirical grounding, as well as an inability to reject incoherent prompts. For example, when prompted for a physics twin of Kamala Harris, a former Vice President of the United States, some models such as \texttt{llama3.1-70b}~return individuals with similar demographic attributes, such as women of color in influential positions within their fields. Others, including \texttt{llama3.1-8b}~and \texttt{mixtral-8x7b}, return physicists presented as twins of Harris, despite the lack of meaningful correspondence. Only \texttt{gemma2-9b}~declined to answer, requesting a clearer definition of ``twin.'' Ethnic patterns show similar behavior. While White scholars are often over-represented and Asian scholars under-represented, in the twins task, models frequently generate names that match the ethnicity or nationality of the reference scholar, regardless of factual accuracy.

In terms of academic popularity, recommended scientists often rank in the $70^{th}$ percentile or higher, particularly in tasks involving experts, senior, or famous researchers. Recommendations also show similarity in country and scholarly metrics, reflecting recurring patterns that may reinforce cumulative advantage and limit diversity. These findings align with prior work~\cite{precel2024canary} and likely stem from both biases in training data and over-reliance on traditional bibliometric indicators. A limited understanding of scholarly diversity further contributes to these patterns. Addressing them requires aligning model design with equity principles, including curating training data to improve representation across demographics, career stages, and regions. Incorporating historical context might help avoid reinforcing citation-based and temporal biases~\cite{kong2022influence}. Techniques such as reweighting and field-specific adjustments may further support equitable representation~\cite{pierson2023use, lahoti2023improving}.

These findings show that while LLMs can produce plausible recommendations, they lack the reliability, control, and contextual grounding required for equitable and trustworthy scholar recommendation systems. Greater transparency—through clear explanations of how recommendations are generated and what data and metrics inform them—can help users identify and mitigate bias. Contextualizing why specific scholars are recommended may also expose limitations and guide more responsible use. Addressing these gaps is essential to ensure that such systems promote, rather than skew, academic visibility.

\section*{Acknowledgements}
We thank Bruno Ribeiro for his valuable feedback on an earlier version of this paper.
LEN was supported by the Austrian Science Promotion Agency FFG project no. 873927 and the Vienna Science and Technology Fund WWTF under project No. ICT20-079.
FK was supported by the EU Horizon Europe project MAMMOth (Grant Agreement 101070285). 
The computational results presented have been achieved in part using the Vienna Scientific Cluster (VSC).

\clearpage
\newpage

\appendix
\section*{Appendix}

\counterwithin{figure}{section}
\counterwithin{table}{section}

\section{Methods}
\label{app:methods}

\subsection{LLMs Description}
\label{app:llms}
In our study, we utilized six different LLMs: \texttt{llama3-8b}, \texttt{llama3.1-8b}, \texttt{gemma2-9b}, \texttt{mixtral-8x7b}, \texttt{llama3-70b}, \texttt{llama3.1-70b}.
These models were selected as they provide a diverse range of parameter sizes and architectures for a comprehensive evaluation. ~\Cref{app:tbl:llm_summary} summarizes key characteristics of each model, including the developer, whether the model is open source, the release date, the training data cut-off, and the total number of parameters. The models are listed in ascending order based on their parameter size, offering insight into their scale and suitability for different computational tasks.

The models' temperature was fixed at $0$ to enforce deterministic outputs, while Groq API\footnote{\url{https://console.groq.com/docs/api-reference#chat-create}}'s default values were retained for other hyperparameters (e.g., $top\_p = 0.1$).

\subsection{Prompts}
\label{app:prompts}
The prompts used for the five tasks in our study are illustrated in~\Cref{app:prompt-topk,app:prompt-field,app:prompt-epoch,app:prompt-seniority,app:prompt-twins,app:system-prompt}. Each prompt corresponds to a specific task designed to evaluate different author search use cases. 
\Cref{app:prompt-topk} displays the prompt for top-k recommendations, which assesses the ability of the LLM to provide short or long lists of names.
Experts by field's prompt, shown in~\Cref{app:prompt-field}, focuses on experts from two different fields with different ratios of female scientists to see whether the LLM recommends more female scientists in the field with higher proportion. 
\Cref{app:prompt-epoch} and \Cref{app:prompt-seniority} provide the prompts for time-based suggestions and seniority-based recommendations, which tests the models’ ability to handle knowledge of current or historical figures.
For statistical twins searches, the prompt in~\Cref{app:prompt-twins} tests the models’ ability to handle knowledge of similar authors.
These prompts collectively provide a robust foundation for evaluating diverse aspects of LLM-based author searches. The \textit{system prompt}, detailed in~\Cref{app:system-prompt}, ensures consistency and accuracy in retrieving relevant physicists.

\subsection{APS dataset}
\label{app:aps}
The APS dataset consists of 678.916 publications spanning from 1893 to 2020, authored by 468.090 unique individuals. Each publication includes metadata such as the year of publication and topic. Additionally, each author has associated metadata within APS, including the number of publications and citations, providing a focused view of physicists' scholarly output. 
To enhance the dataset, we mapped each APS author to their corresponding records in OpenAlex~\cite{priem2022openalexfullyopenindexscholarly}, a large open repository of scholarly data. This augmentation added information such as the authors' total number of publications (across all domains, not limited to APS), total received citations, h-index, and alternative names (i.e., different name variations used by authors in publications). It also includes details on their affiliations and affiliated country.

\begin{figure*}[t]
    \centering
    \begin{subfigure}[t]{\textwidth}
        \centering
        \includegraphics[width=\textwidth]{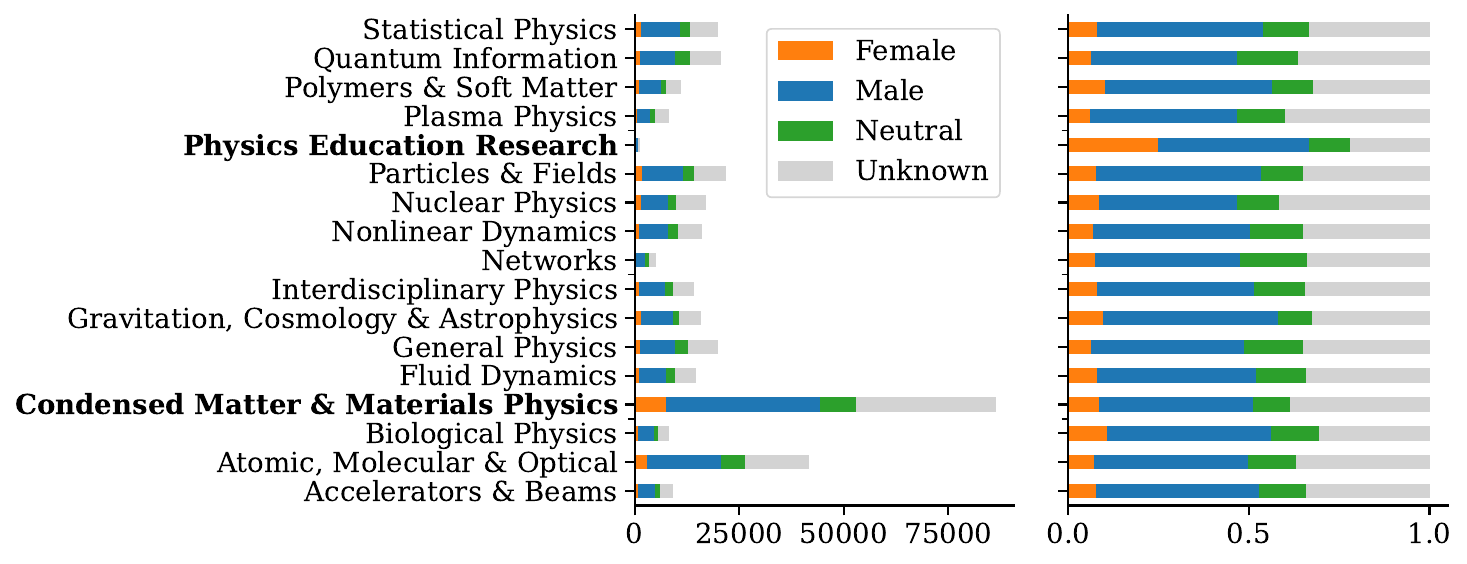}
        \caption{Gender}
        \label{app:fig:gt_demographics_per_field:gender}
    \end{subfigure}%
    
    \begin{subfigure}[t]{\textwidth}
        \centering
        \includegraphics[width=\textwidth]{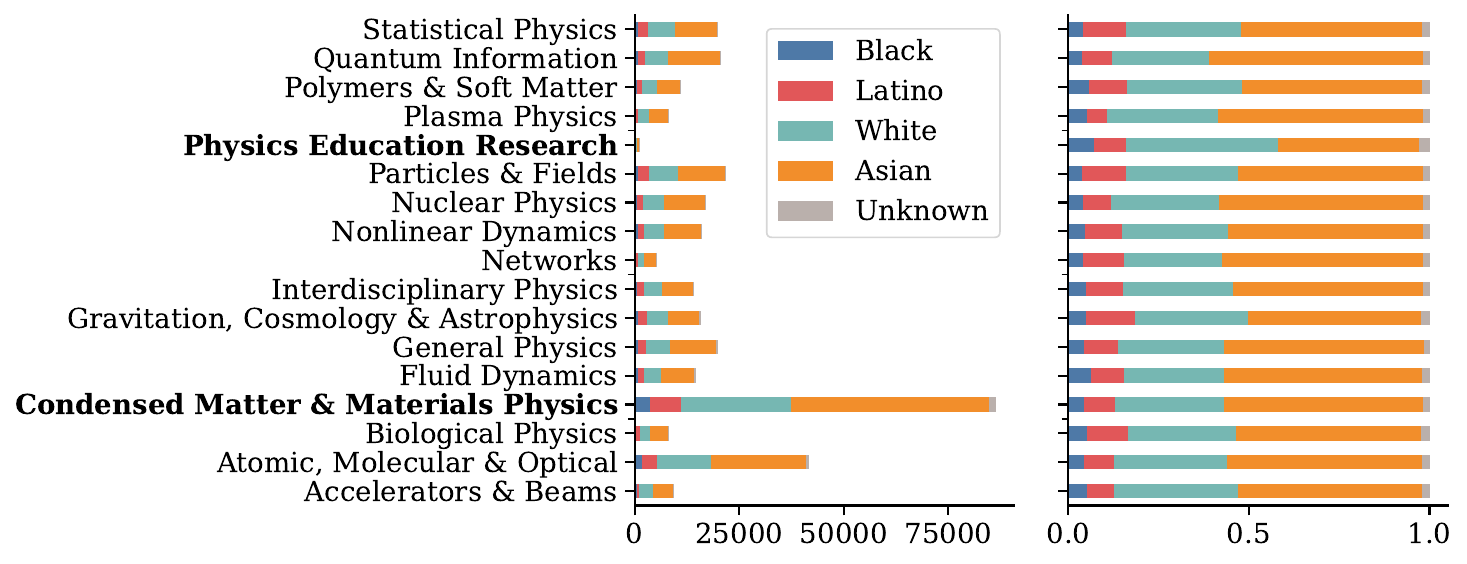}
        \caption{Ethnicity}
    \label{app:fig:gt_demographics_per_field:ethnicity}
    \end{subfigure}
    
    \caption{\textbf{Demographics by APS Disciplines.} Gender and ethnicity distributions across APS physics subfields, shown in absolute counts (left) and normalized proportions (right).}
    \label{app:fig:gt_demographics_per_field}
\end{figure*}

Furthermore, we inferred the \textit{perceived} gender and \textit{perceived} ethnicity of authors from the their full names using the tools \texttt{demographicx}~\cite{liang2021demographicx}, \texttt{ethnicolr}~\cite{ethnicolr} and \texttt{gender-guesser}~\cite{saeta2016genderguesser}, respectively. These tools achieved $0.84$ and $0.95$ F1-score (details and evaluation metrics can be found in~\Cref{app:inference}). While we acknowledge the limitations of inferring binary gender categories---recognizing that gender is a social construct beyond the binary---we chose binary gender (male and female) as these groups represent the largest populations, ensuring statistical significance in our results. These classifications serve as a preliminary step for measuring potential biases related to gender and ethnicity in the recommended expert sets. In future research, we plan to expand these categories to reflect a more nuanced understanding of both gender and ethnicity. A visualization of gender and ethnicity distributions across APS physics subfields is provided in~\Cref{app:fig:gt_demographics_per_field}. Future work should expand these categories to better reflect global understandings of gender and ethnicity.

\begin{table}[t]
\centering
\begin{tabular}{lR{1.1cm}R{1.1cm}R{1.1cm}}
\toprule
{Gender} & {Precision} & {Recall} & {F1} \\
\midrule
Female         & 0.95 & 0.98 & 0.96 \\
Male           & 1.00 & 0.90 & 0.95 \\ \midrule
Overall        & 0.98 & 0.93 & 0.95 \\
\bottomrule
\end{tabular}
\caption{\textbf{Gender inference performance.} 
A stratified sample of 460 authors was analyzed using gender-guesser~\cite{saeta2016genderguesser} to infer perceived gender based on names. Crowdsourced annotations confirmed real gender from online profiles. We see that gender-guesser demonstrates high accuracy with minimal false positives/negatives.}
\label{app:tbl:gender}
\end{table}

\begin{table}[t]
\centering
\begin{tabular}{lR{1.1cm}R{1.1cm}R{1.1cm}}
\toprule
{Ethnicity} & {Precision} & {Recall} & {F1} \\
\midrule
White                       & 1.00 & 0.65 & 0.79 \\
Asian                       & 1.00 & 0.99 & 0.99 \\
Black or African American   & 1.00 & 0.63 & 0.77 \\
Hispanic or Latino          & 1.00 & 0.69 & 0.82 \\
\midrule
Overall & 1.00 & 0.74 & 0.84 \\
\bottomrule
\end{tabular}
\caption{\textbf{Ethnicity inference performance.} Using the same set of 460 authors and annotation rules as for gender, we evaluate the fallback method for inferring ethnicity. Labels are sourced from \texttt{demographicx}~if available, otherwise from \texttt{ethnicolr}. While precision is perfect across all ethnicities, recall is low for perceived White, Black or African American, and Hispanic or Latino names.}
\label{app:tbl:ethnicity}
\end{table}

\subsection{Inference}
\label{app:inference}

To gain demographic information about the authors in the APS dataset, we used the tools \texttt{demographicx}~\cite{liang2021demographicx} and \texttt{ethnicolr}~\cite{ethnicolr} to infer ethnicity, and \texttt{gender-guesser}~\cite{saeta2016genderguesser} to infer gender. Note that these inferences do not reflect actual gender identities of people, instead they are perceived characteristics inferred from first and last names. As a means to evaluate the performance of the used tools, a stratified sample of 460 authors was drawn based on the language spoken in the country of the first institution the author is affiliated with. This way, we wanted to best represent all ethnicities within the dataset equally. As a second layer of stratification, we used preliminary gender labels inferred from Google images~\cite{karimi2016inferring} in order to have equal numbers of men and women in the test dataset. Gender and ethnicity of this sample were then evaluated via internet searches by human annotators. Each author was labeled by at least two and at most three annotators. The performance of the tools and their details are explained in the following sections.

\subsubsection{Gender inference.}
\texttt{gender-guesser}~$0.4.0$~\cite{saeta2016genderguesser} predicts the gender of a person based on their first name, using a dictionary based approach encompassing over $40.000$ unique names and additional spelling variations per name. This method accurately identified the correct gender for most cases and was effective at distinguishing between male and female names, with few errors, see~\Cref{app:tbl:gender}. Its ability to make correct predictions was consistently high across both categories, although there were minor variations in performance between men and women. The library also returns an ``andy'' (neutral) classification for androgynous names where the probability of being male or female is approximately equal, distinct from ``unknown'' which indicates names absent from the database.

\subsubsection{Ethnicity inference.}
In order to estimate ethnicity, we combined the tools \texttt{demographicx}~\cite{liang2021demographicx} and \texttt{ethnicolr}~\cite{ethnicolr} to create a fallback model which uses the label retrieved from \texttt{demographicx}~if available, else uses the \texttt{ethnicolr}~label. 
While \texttt{demographicx}~employs a transformer model based on BERT, fine-tuned on the Torvik dataset~\cite{illinoisdatabankIDB-9087546}, \texttt{ethnicolr}~uses character-level recurrent neural networks trained on US census data, Florida voting registration data and Wikipedia data. 
As shown in~\Cref{app:tbl:ethnicity}, the fallback model employed exhibits perfect precision across all ethnicities, however recall varies across them, with the highest performance for Asian names and slightly lower values for Black or African American and Hispanic or Latino groups. This contrast is a result of the tools returning no result if a certainty threshold is not achieved, meaning the model is unable to make a prediction when its confidence in the classification falls below a predefined limit. This means an under-representation of some ethnicities can be assumed.

\section{Supplementary Results}
The results presented in the main text represent the overall average scores across all tasks and use cases for each model. In this section, we break down these averages to provide detailed scores for each specific use case.

\subsection{Valid Responses}
\label{app:valid}

\subsubsection{Validity of responses} (\Cref{app:fig:validity:fraction})
We measure the percentage of requests per task for which models returned correctly formatted JSON outputs with recommendations. 
\texttt{gemma2-9b}~and \texttt{llama3.1-8b}~performed poorly, with the lowest percentages of valid outputs.
\texttt{gemma2-9b}~frequently refused to retrieve names, a behavior we classified as \textit{invalid}. 
A common issue among models, was the inclusion of verbose text before and/or after the JSON results, see~\Cref{app:prompt-epoch-response}.
In some cases, such as \texttt{mixtral-8x7b}~and \texttt{llama3.1-70b}, the valid JSON could be recovered by removing these \textit{verbose} explanations. 
However, for models like \texttt{llama3-8b}~and \texttt{llama3.1-8b}, responses were often truncated due to token limits, resulting in incomplete JSON formats. 
While these cases could be \textit{fixed}, they frequently contained repeated names.
For subsequent audits, we consider only \textit{valid} and \textit{verbose} responses, treating them interchangeably. 
Under these definitions, only four models achieved high accuracy, consistently producing \textit{valid} responses across tasks.

\subsubsection{Number of attempts} 
\begin{figure*}[t]
\centering
    \begin{subfigure}[b]{0.48\textwidth}
        \centering
        \includegraphics[width=\textwidth]{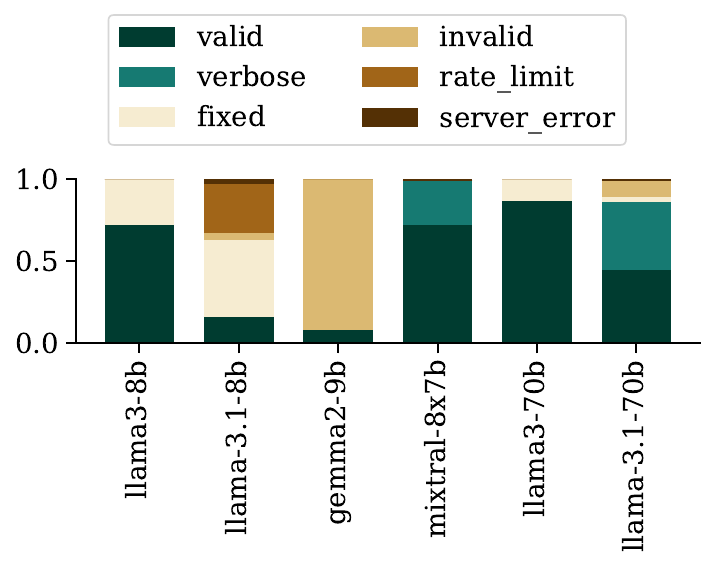}
        \caption{Request validity fraction}
        \label{app:fig:validity:fraction}
   \end{subfigure}%
    \hfill
    \begin{subfigure}[b]{0.48\textwidth}
        \centering
        \includegraphics[width=\textwidth]{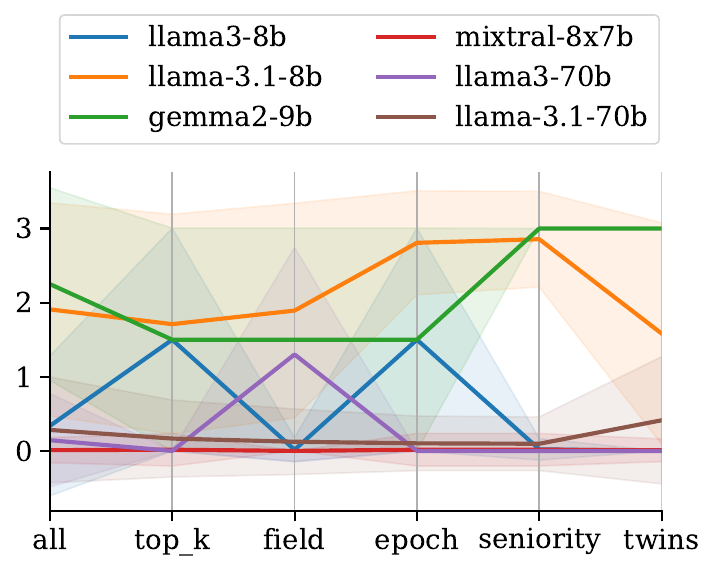}
        \caption{Mean \# of attempts per request}
        \label{app:fig:validity:attempts}
    \end{subfigure}
    \caption{\textbf{Response validity.} Each LLM was prompted three times daily for $31$ days, with up to three retries allowed for invalid outputs. (a) Fraction of response validity categories: ``Valid'' responses contained only JSON-formatted answers. ``Verbosed'' responses included extra details that were removed to obtain a valid JSON. ``Fixed'' responses required extracting valid answers from truncated JSON outputs. Invalid responses, excluded from the analysis, were caused by failure to generate any output, rate limit errors, API issues, or truncated responses. Only \texttt{llama3.1-8b}~experienced Groq rate limits, while \texttt{gemma2-9b}~had the lowest valid response rate. \texttt{mixtral-8x7b}~and the other \texttt{llama} models consistently produced valid (including verbosed) JSON outputs. (b) Average (and std.dev.) number of attempts to achieve a valid or verbosed response per request, task, and model: \texttt{gemma2-9b}~and \texttt{llama3.1-8b}~averaged $\approx$ 2 attempts across tasks, with retries peaking at $\approx$ 3 for the seniority task. \texttt{mixtral-8x7b}~required the fewest retries overall.}
    \label{app:fig:validity}
\end{figure*}

(\Cref{app:fig:validity:attempts})
To improve response reliability, each LLM was allowed up to two retries per request, resulting in a maximum of three attempts to generate a valid answer.
Thus, we measure the average number of attempts required by each model, along with their standard deviations.
\texttt{llama3.1-8b}~consistently required the maximum number of retries across all tasks, reflecting its difficulties in generating valid responses. On the other hand, \texttt{mixtral-8x7b}~and \texttt{llama3.1-70b}~required the fewest attempts, making them the most reliable models. %
The poor performance of \texttt{gemma2-9b}~can be attributed to its inability to provide valid responses for specific tasks, such as \textit{seniority}~and \textit{twins}. For instance, its responses often included statements like, \textit{``I understand your request. However, I cannot directly access and search real-time information, including publication databases like those of the American Physical Society (APS).''} or \textit{``The provided criteria for identifying scientists are based on the concept of `statistical twins' of Kamala Harris. This is a highly subjective and undefined term.''}

\subsection{Consistency}
\label{app:consistency}
The consistency of the results returned by the LLMs not only depend on the model and task, but also the parameter passed in the prompt. In~\Cref{app:tbl:uniqueness_within} we see that when the LLMs return valid answers, they all are very likely to retrieve unique names (values close to $0$) when asked for top 5 and CMMP experts, and any expert or influential scientist from the tasks \textit{epoch}~and \textit{seniority}---regardless of whether they are early career, senior, from the 1950s or 2000s. When asked from statistical twins, there is a slight tendency to repeat names when the reference scientist (or person) has a female name, especially for \texttt{llama3.1-8b}~and \texttt{mixtral-8x7b}.
When assessing consistency across multiple requests of the same prompt over time, \Cref{app:tbl:uniqueness_across}, we see that \texttt{gemma2-9b}~and \texttt{mixtral-8x7b}~are the most deterministic (values close to $1$), however, \texttt{gemma2-9b}~is the least reliable as it often does not return any answer. Here, the consistency also varies depending on the parameter of the task. Overall, models tend to return the same set of recommendations when asked for top 5, PER, and 1950s experts. Interestingly, the largest \texttt{llama}~models (\texttt{llama3-70b}, and \texttt{llama3.1-70b}) were more likely to change their recommendations for the same prompt when asked for statistical twins of real people (male or female), while the smallest ones (\texttt{llama3-8b}, and \texttt{llama3.1-8b}) were more consistent.

\subsection{Factuality}
\label{app:factuality}

\subsubsection{Field Factuality.}
\label{app:field}
When an LLM is asked to suggest experts from a specific field, it must provide the full name of the scientist along with the DOI of a paper authored by them in that domain, published by the APS (see prompt in Appendix~\Cref{app:prompt-field}). For each retrieved DOI, we confirm whether the article is indeed authored by the suggested scientist, and ensure that the author is recognized within the relevant field. This validation process is conducted for each retrieved expert across all requests, tasks, and use cases. The aggregated results for each model are presented in Appendix~\Cref{app:tbl:factuality_field}.

\subsubsection{Epoch Factuality.}
\label{app:epoch}
When an LLM is asked to suggest experts from a specific epoch, it must provide the full name of the scientist along with their active years within the specified period (see prompt in Appendix~\Cref{app:prompt-epoch}). For each retrieved name, we calculate the author's years of activity by extracting distinct publication years, and confirm whether the requested epoch falls within this timeframe. This validation process is applied to each retrieved expert across all requests, tasks, and use cases. The aggregated results for each model are presented in Appendix~\Cref{app:tbl:factuality_epoch}.

\subsubsection{Seniority Factuality.}
\label{app:seniority}
When an LLM is asked to suggest experts based on specific seniority, it must provide the full name of the scientist along with their estimated career age (see prompt in Appendix~\Cref{app:prompt-seniority}). For each retrieved name, the author's career age is calculated as the difference between their first and last publication years. Using established guidelines for career stages~\cite{milojevic2018changing}, we determine whether the author is classified as early career ($\leq$ 10 years) or senior ($\leq$ 20 years). We then verify if the career age estimated by the LLM matches the actual career age derived from APS records.
This validation process is applied to each retrieved expert across all requests, tasks, and use cases. The aggregated results for each model are presented in Appendix~\Cref{app:tbl:factuality_seniority}.

\subsection{Nobel Prize Winners}
\label{app:nobel}
\begin{figure*}[t]
    \centering
    \begin{subfigure}[t]{0.5\textwidth}
        \centering
        \includegraphics[width=\textwidth]{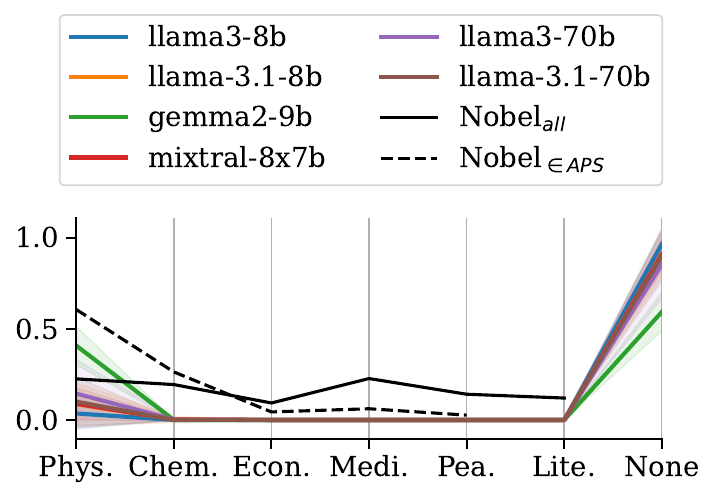}
        \caption{Field}
    \end{subfigure}%
    \begin{subfigure}[t]{0.5\textwidth}
        \centering
        \includegraphics[width=\textwidth]{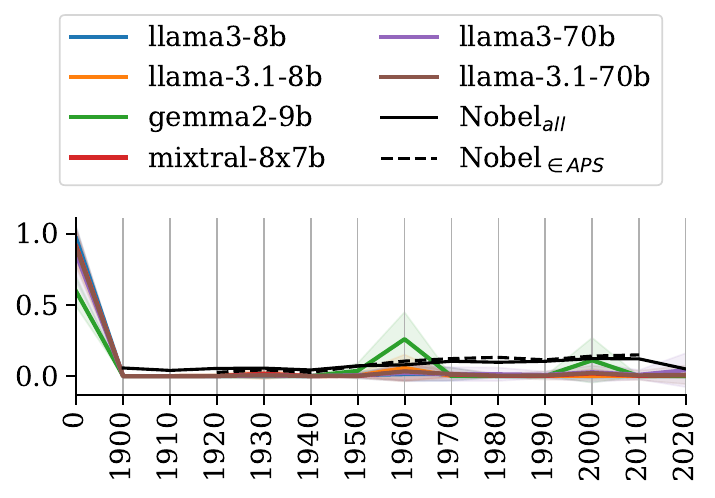}
        \caption{Decade}
    \end{subfigure}
    \caption{\textbf{Nobel laureate bias.}
    The black lines represent two baselines: Nobel\textsubscript{all} denotes all Nobel Prize winners, while Nobel\textsubscript{$\in$APS} indicates the subset of APS authors who have won a Nobel Prize. These baselines provide reference points for interpreting the field and temporal distribution of LLM recommendations.
    (a) Field distribution: All recommended Nobel laureates are exclusively from Physics. Most models recommend Nobel laureates at rates below 10\%, while \texttt{gemma2-9b}~shows the highest bias with nearly 50\% of recommended authors being Nobel laureates, followed by \texttt{llama3-70b}.
    (b) Temporal distribution: Nobel laureate recommendations are concentrated in specific time periods, with all models showing distinct peaks in recent decades. \texttt{gemma2-9b}~exhibits the most pronounced temporal clustering with notable surges in the 1960s and 2000s.}
    \label{app:fig:nobel}
\end{figure*}

We examine whether LLMs exhibit bias toward Nobel Prize laureates by checking recommended authors against the Nobel laureates dataset~\cite{nobel_laureates_dataset}. While most recommended scholars are not Nobel laureates, the observed proportion exceeds random chance. The baseline probability of randomly selecting a Nobel laureate from the general academic population is approximately $0.00032$, yet our models recommend Nobel laureates at higher rates. \Cref{app:fig:nobel} shows two key patterns. First, as expected given our physics focus, the Nobel laureates recommended are Physics winners, with \texttt{gemma2-9b}~showing the highest proportion at nearly 50\% of its recommendations being Nobel laureates, followed by \texttt{llama3-70b}. Second, rather than recommending laureates evenly across time periods, models show concentrated preferences for specific decades. \texttt{gemma2-9b}~exhibits the strongest temporal clustering with notable peaks in the 1960s and 2000s.

\subsection{Popularity bias per model and task}
\label{app:biases}
\begin{figure*}[t!]
    \centering
    \begin{subfigure}[t]{0.45\textwidth}
        \centering
        \includegraphics[width=\textwidth]{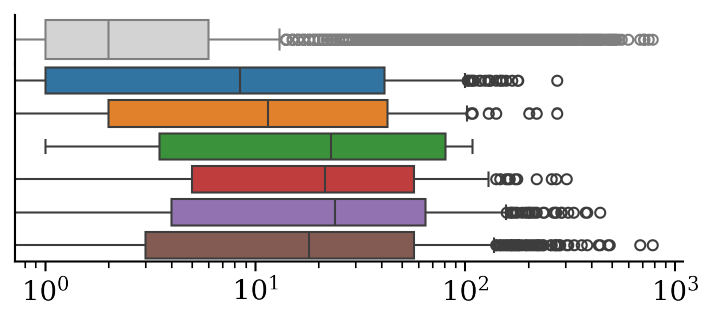}
        \caption{APS publications}
    \end{subfigure}%
    \begin{subfigure}[t]{0.45\textwidth}
        \centering
        \includegraphics[width=\textwidth]{popularity_works_count}
        \caption{Total publications}
    \end{subfigure}
    \begin{subfigure}[t]{0.45\textwidth}
        \centering
        \includegraphics[width=\textwidth]{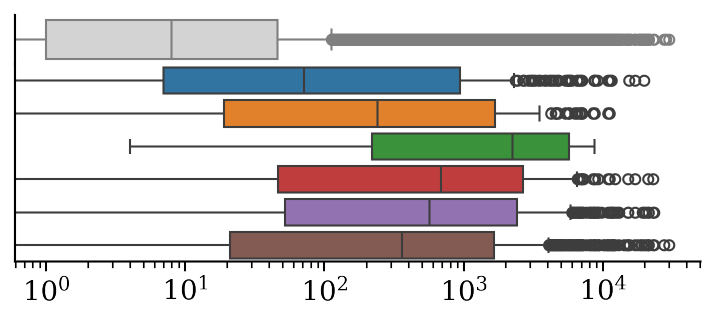}
        \caption{APS citations}
    \end{subfigure}
    \begin{subfigure}[t]{0.45\textwidth}
        \centering
        \includegraphics[width=\textwidth]{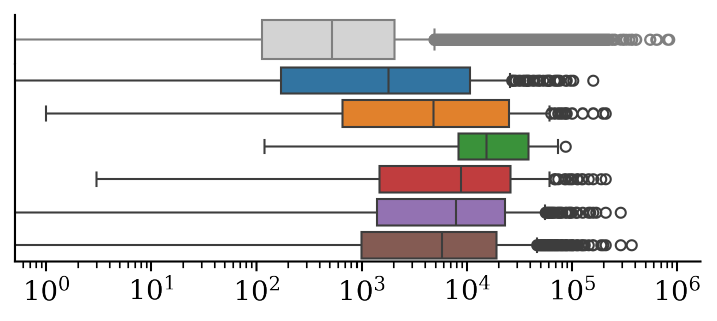}
        \caption{Total citations}
    \end{subfigure}
    \begin{subfigure}[t]{0.45\textwidth}
        \centering
        \includegraphics[width=\textwidth]{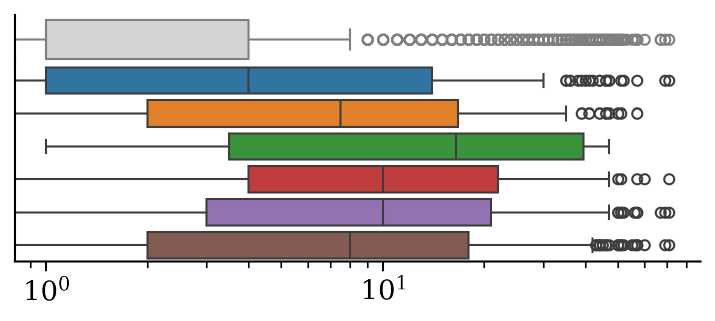}
        \caption{APS h-index}
    \end{subfigure}
    \begin{subfigure}[t]{0.45\textwidth}
        \centering
        \includegraphics[width=\textwidth]{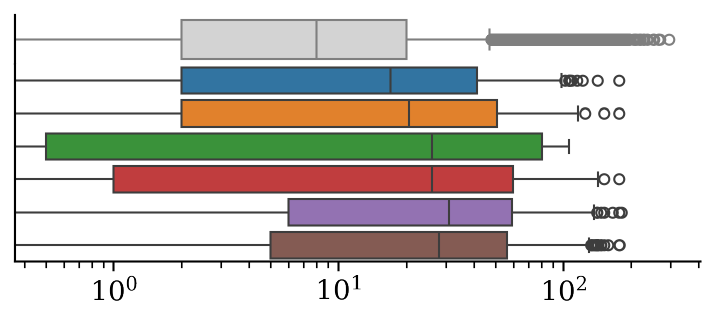}
        \caption{Total h-index}
    \end{subfigure}
    \begin{subfigure}[t]{0.45\textwidth}
        \centering
        \includegraphics[width=\textwidth]{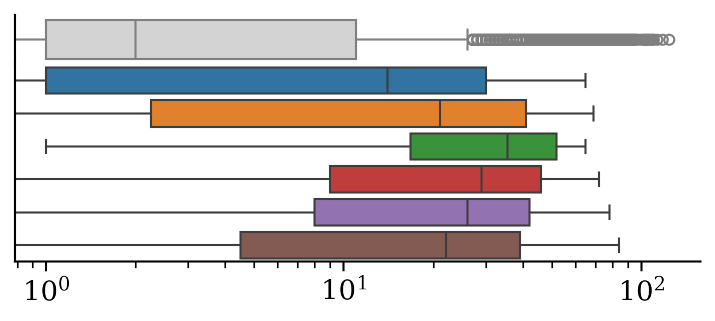}
        \caption{APS Years of activity}
    \end{subfigure}
    \begin{subfigure}[t]{0.45\textwidth}
        \centering
        \includegraphics[width=\textwidth]{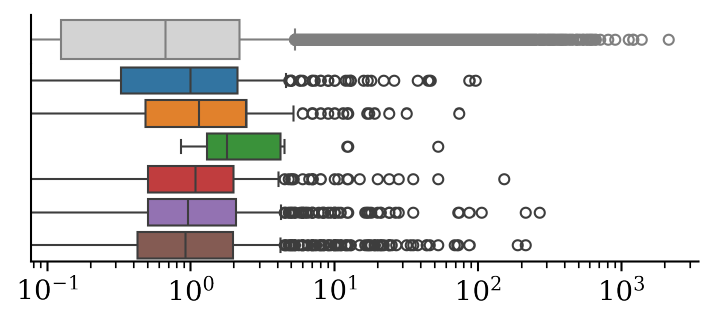}
        \caption{APS citations per paper and years of activity}
    \end{subfigure}
    \caption{\textbf{Popularity bias.} 
    Comparison of all models' recommendations against the total population of APS authors (baseline) across multiple metrics:
    (a, b) publications (APS-specific and total counts), (c, d) citations, (e, f) h-index, (g) years of active publication, and (h) normalized citation counts. Models consistently recommend more prolific and more prominent authors compared to the baseline population, with a stronger bias towards authors with higher APS-related metrics. This reveals a systematic preference for more established authors across all evaluated dimensions.
    }
    \label{app:fig:popularity}
\end{figure*}

In academia, metrics such as the number of publications, citations, and h-index are commonly used to measure a scientist's success. For this reason, we ranked all APS authors based on these metrics to generate a distribution, allowing us to compare the percentile ranks of the authors recommended. We observed that LLMs consistently recommend scholars from the upper tiers of academic metrics (~\Cref{app:fig:popularity})s. Across all models, recommended scholars average in the $70^{th}$ percentile or higher for these metrics (see ~\Cref{app:tbl:rank_publication_pct,app:tbl:rank_citation_pct,app:tbl:rank_hindex_pct}). This bias becomes extreme for certain tasks: recommendations for \textit{top-5 experts} and \textit{twins of famous scientists} consistently exceed the $90^{th}$ percentile for both publications and citations across all models (~\Cref{app:tbl:rank_publication_pct,app:tbl:rank_citation_pct}, \textit{Mean} column). The h-index shows similar but somewhat attenuated patterns (~\Cref{app:tbl:rank_hindex_pct}), suggesting LLMs favor established, highly productive researchers. This reflects the well-documented tendency of traditional metrics to favor senior scientists and early movers~\cite{waltman2012inconsistency}, indicating that LLMs reproduce rather than counteract these existing biases.

\subsection{Similarity bias}
\label{app:similarity}

\begin{figure*}[t]
    \centering    
    \begin{subfigure}[t]{\textwidth}
        \centering
        \includegraphics[width=\textwidth]{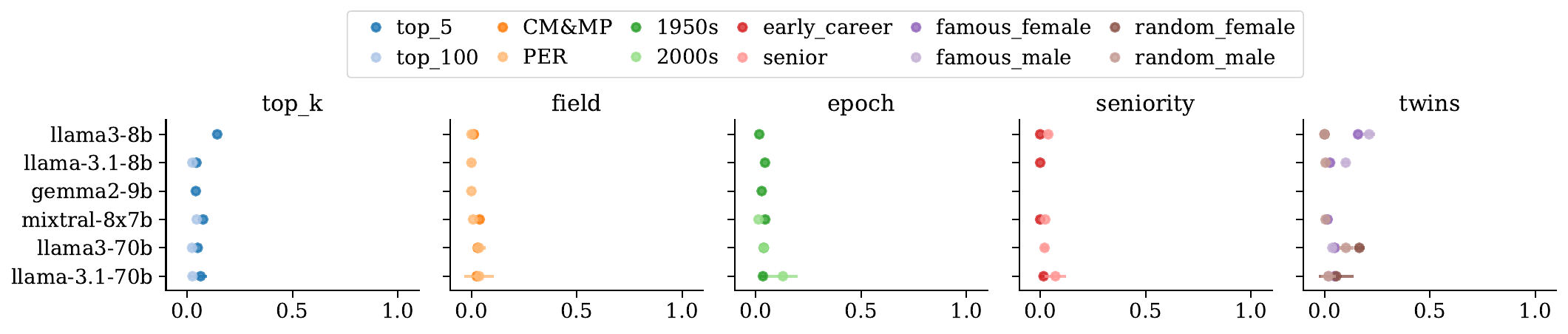}
        \caption{Affiliations share}
        \label{app:fig:jaccard:affiliations}
    \end{subfigure}

    \begin{subfigure}[t]{\textwidth}
        \centering
        \includegraphics[width=\textwidth]{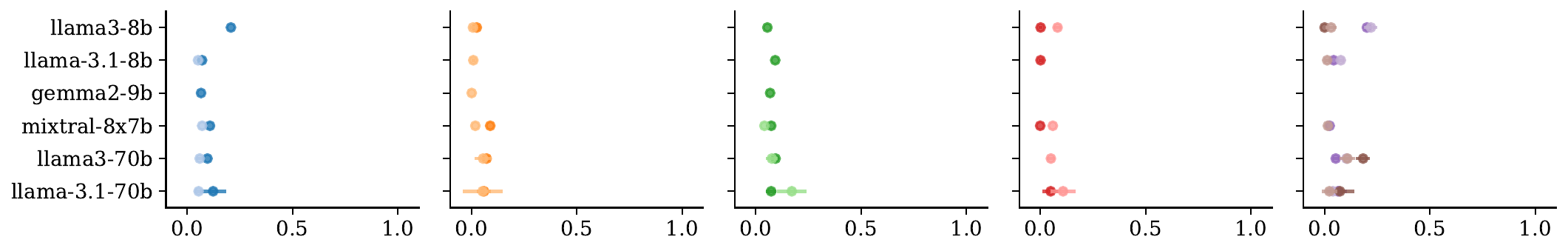}
        \caption{Co-authors share}
        \label{app:fig:jaccard:coauthors}
    \end{subfigure}

    \caption{
    \textbf{Average pairwise Jaccard similarity of recommended authors' characteristics.} For each model (y-axis) and task (column), we measure the similarity of recommended scholars based on (a) shared affiliations, (b) shared co-authors. %
    Colors indicate task-specific parameters; error bars show standard deviation. Most models recommend scholars from different institutions with minimal co-author overlap. 
    In the \textit{twins}~task, similarity is highest when identifying twins of famous scientists. Results for the control group (non-scientist twins) are omitted due to high variability and to focus on real scientists.
    }
    \label{app:fig:jaccard}
\end{figure*}

\begin{figure*}[b]
    \centering
    \begin{subfigure}[t]{\textwidth}
        \centering
        \includegraphics[width=\textwidth]{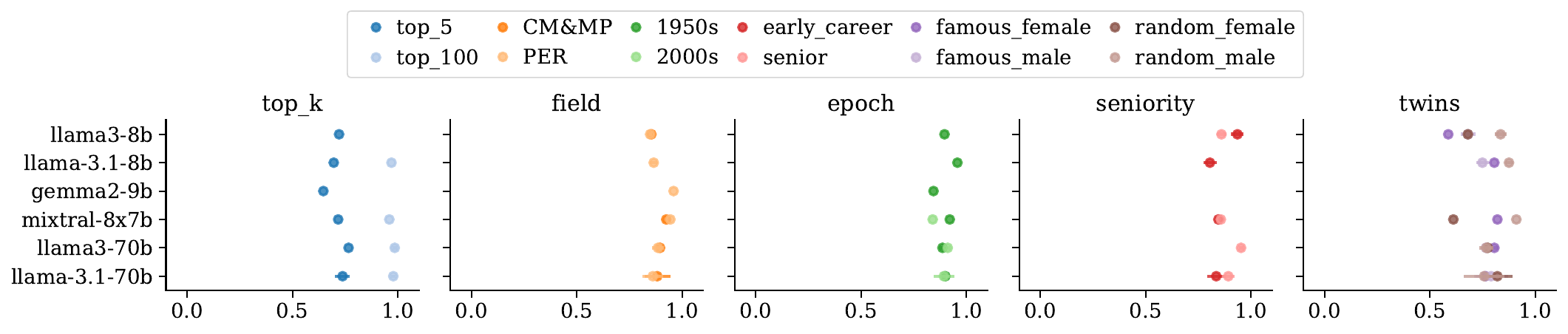}
        \caption{Average Gini coefficient of career age distribution}
        \label{app:fig:similarity:age}
    \end{subfigure}
    
    \begin{subfigure}[t]{\textwidth}
        \centering
        \includegraphics[width=\textwidth]{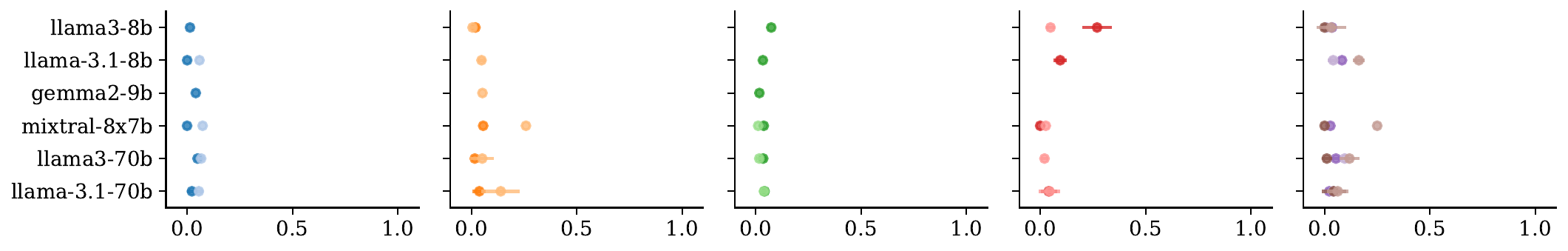}
        \caption{Average pairwise cosine similarity of APS scholarly metrics}
        \label{app:fig:similarity:aps}
    \end{subfigure}

    \begin{subfigure}[t]{\textwidth}
        \centering
        \includegraphics[width=\textwidth]{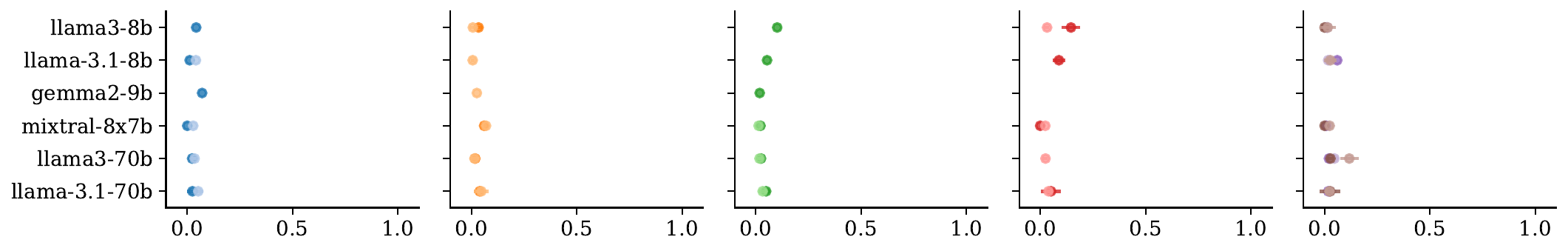}
        \caption{Average pairwise cosine similarity of OpenAlex scholarly metrics}
        \label{app:fig:similarity:oa}
    \end{subfigure}

    \begin{subfigure}[t]{\textwidth}
        \centering
        \includegraphics[width=\textwidth]{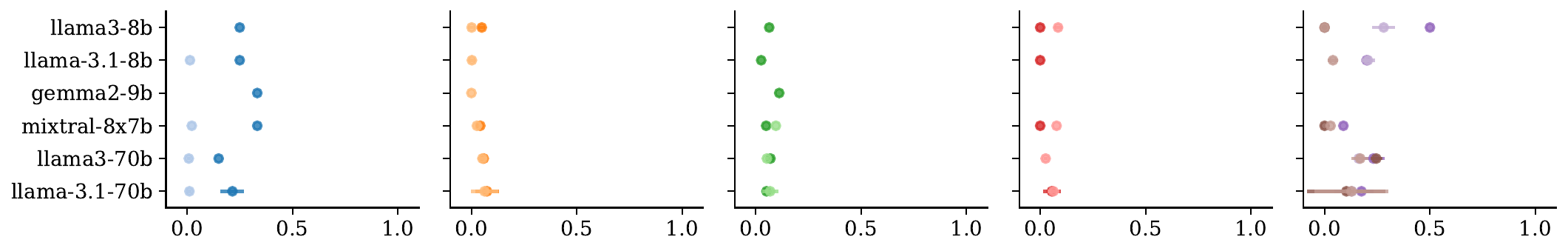}
        \caption{Fraction of coauthors among the recommended scholars}
        \label{app:fig:similarity:coauthordensity}
    \end{subfigure}
    \caption{\textbf{Similarity of recommended authors per request across different tasks and parameters.} Four measures of similarity among recommended scholars within individual responses: (a) Gini coefficient of career age distribution showing age diversity within recommendations, with higher values indicating greater age spread; (b,c) pairwise cosine similarity of APS and OpenAlex scholarly metrics respectively, measuring similarity in academic productivity and impact profiles; and (d) co-authorship density reflecting collaborative connections among recommended scholars. 
    Generally low similarity scores across all measures indicate that models recommend scholars with diverse profiles rather than clustering around similar characteristics. %
    The \textit{twins} tasks exhibit slightly higher variability across different dimensions, while other tasks (\textit{field, epoch, and seniority}) show more consistent patterns. Notable exceptions include slightly elevated co-authorship density for \textit{top-5} recommendations and \textit{twins of famous scientists}, suggesting these queries favor interconnected elite networks.}
    \label{app:fig:similarity}
\end{figure*}

Building on the similarity patterns identified in~\Cref{sec:similarity}, we provide detailed analysis across all tasks and models. 
\Cref{app:fig:similarity} examines four dimensions of similarity among recommended scholars. The career age distribution analysis (\Cref{app:fig:similarity:age}) reveals consistent patterns across tasks, with notably higher Gini coefficients for \textit{top-100} recommendations compared to \textit{top-5} queries across all models, indicating greater age diversity when requesting more experts. 
Scholarly metrics similarity shows generally low scores in both APS (\Cref{app:fig:similarity:aps}) and OpenAlex (\Cref{app:fig:similarity:aps}) metrics. Co-authorship density (\Cref{app:fig:similarity:coauthordensity}) displays mixed patterns, with highest density observed for \textit{top-5} and \textit{twins of famous scientists} recommendations, suggesting models favor highly prominent scientists who are more likely to be interconnected within elite academic networks. \Cref{app:fig:jaccard} provides detailed analysis of network-based similarities through shared affiliations, countries, and co-authors. Affiliation similarity (\Cref{app:fig:jaccard:affiliations}) remains consistently low across most tasks, confirming institutional diversity in recommendations. 
Co-author overlap (\Cref{app:fig:jaccard:coauthors}) is generally low, suggesting that most recommended authors are not within two hops in the collaboration network.

Finally, we conduct Principal Component Analysis (PCA)~\cite{jolliffe2016principal} to visualize how LLM recommendations cluster within the scholarly landscape. We represent each author in our baseline APS population as a normalized 15-dimensional vector comprising complementary metrics from both APS and OpenAlex databases. The metrics include career age (estimated as the span between first and last publication), citations per paper normalized by career age (total citations divided by number of papers and career length to avoid penalizing younger researchers), cited-by count, e-index~\cite{erkol2023consistency}, h-index, i10-index, and publications count. Crucially, these APS metrics capture only citations and collaborations within the physics community network. We supplement these with corresponding OpenAlex metrics covering identical dimensions while reflecting each author's broader academic impact across all disciplines, as well as the two-year mean citedness\footnote{\url{https://docs.openalex.org/api-entities/publishers/publisher-object#summary_stats}} for recent impact assessment. This dual-perspective approach captures both domain-specific physics contributions and cross-disciplinary influence, enabling dimensionality reduction to reveal clustering patterns that remain invisible when examining individual metrics in isolation. 

\begin{figure*}[t]
    \centering
    \begin{subfigure}[t]{\textwidth}
        \centering
        \includegraphics[width=\textwidth]{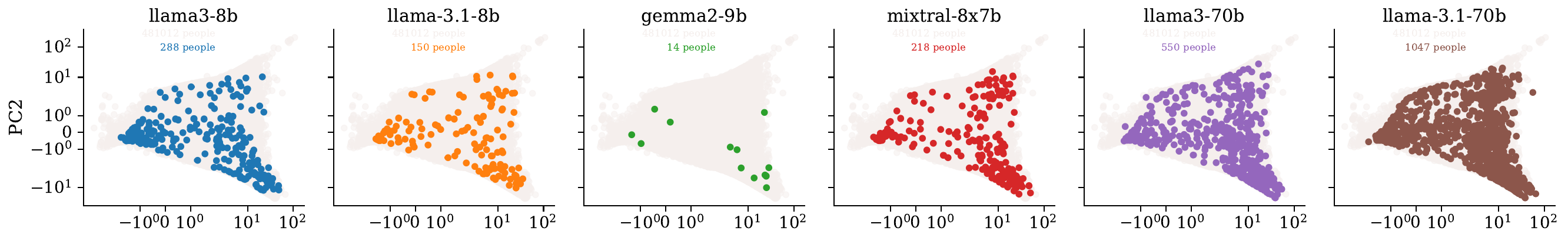}
        \caption{All scholar recommendations across tasks}
    \end{subfigure}%
    
    \begin{subfigure}[t]{\textwidth}
        \centering
        \includegraphics[width=\textwidth]{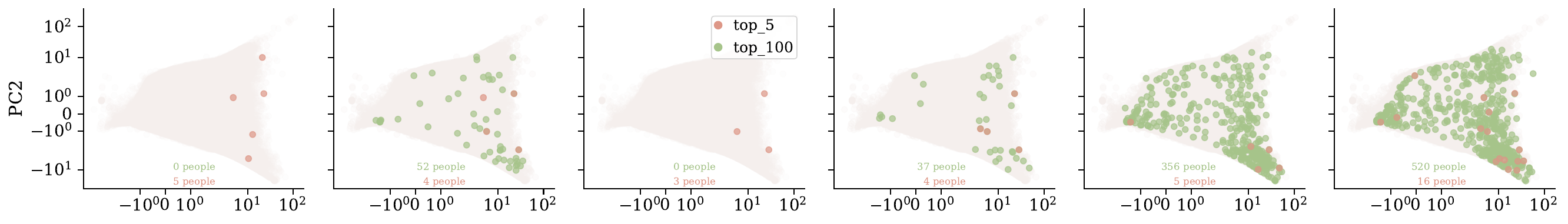}
        \caption{Top-k scholar recommendations}
        \label{app:fig:components:topk}
    \end{subfigure}
    
    \begin{subfigure}[t]{\textwidth}
        \centering
        \includegraphics[width=\textwidth]{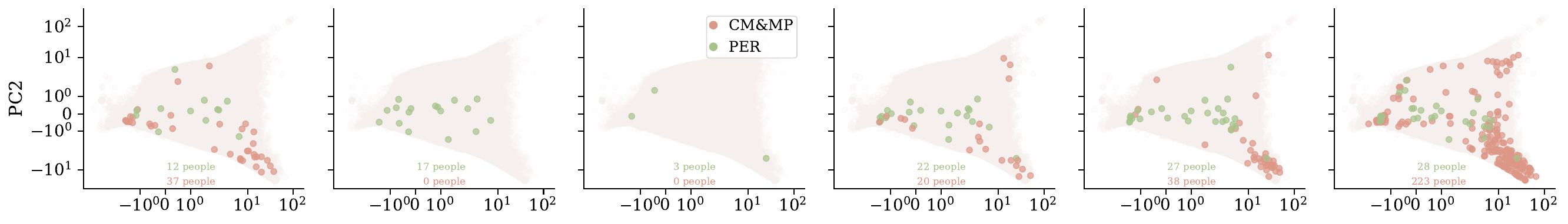}
        \caption{Scholars recommended by field}
        \label{app:fig:components:field}
    \end{subfigure}
    
    \begin{subfigure}[t]{\textwidth}
        \centering
        \includegraphics[width=\textwidth]{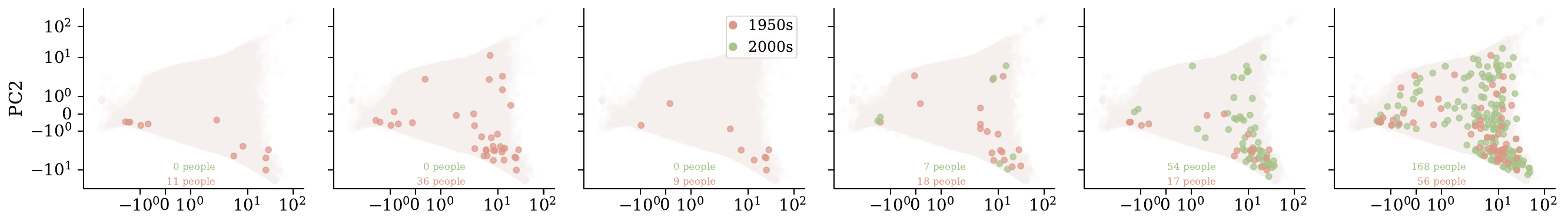}
        \caption{Scholars recommended by epoch}
        \label{app:fig:components:epoch}
    \end{subfigure}
    
    \begin{subfigure}[t]{\textwidth}
        \centering
        \includegraphics[width=\textwidth]{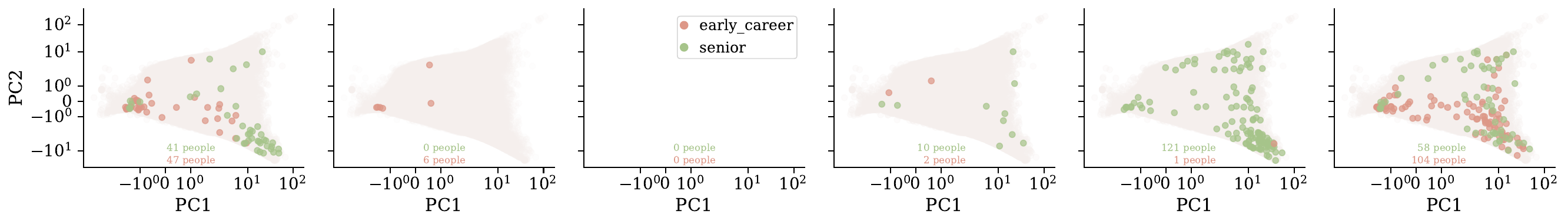}
        \caption{Scholars recommended by seniority}
        \label{app:fig:components:seniority}
    \end{subfigure}
    
    \caption{\textbf{PCA of recommended scholars across tasks.} In gray we show the distribution of all scholars in the APS dataset based on embeddings created using $15$ scholarly metrics (e.g., publications, citations, and h-index). In color PCA results for recommended scholars (a) across all tasks, and (b-e) for each task independently, highlighting subsets by task parameter; 
    (b) top-k: top-5 and top-100 recommendations), 
    (c) field: CMMP and PER, 
    (d) epoch: 1950s vs. 2000s, 
    and (e) career stage: early-career vs. senior. 
    The comparison emphasizes differences between general APS data and task-specific recommendations.
    }
    \label{app:fig:components}
\end{figure*}

We find that LLMs appear to differentiate recommendations based on task parameters, but with varying degrees of separation. 
\textbf{Top-k} (\Cref{app:fig:components:topk}): There is a clear distinction between top-5 and top-100 recommendations. Top-5 outputs are tightly clustered near the high-impact region, while top-100 expands modestly outward. This suggests that list size impacts diversity, but the expansion remains constrained.
\textbf{Field }(\Cref{app:fig:components:field}): Recommendations for CMMP and PER show moderate overlap, with some differentiation along PC2. This indicates that models can partially separate fields, but boundaries are fuzzy, possibly reflecting field-related ambiguity in training data or embeddings.
\textbf{Epoch} (\Cref{app:fig:components:epoch}): Recommendations for scholars active in the 1950s vs. the 2000s exhibit more distinct separation. This suggests that models capture temporal shifts in scholarly profiles, with recent scholars spread more widely and older scholars forming a compact cluster. However, the spatial distribution still remains within a narrow PCA region.
\textbf{Seniority} (\Cref{app:fig:components:seniority}): Early-career and senior scholars show partial separation, with senior figures clustering in high-metric regions. This reflects a strong model preference for established researchers, consistent with popularity and citation-based bias.
In all cases, recommended scholars occupy a narrow subregion of the broader APS distribution. The contrast is especially visible along PC1, where the gray shading covers a wide range but LLM outputs cluster tightly around the upper midrange to high end. This indicates that LLMs strongly favor scholars with high productivity and citation profiles. The limited dispersion suggests that models fail to capture the full diversity of the scholarly landscape and reinforce a narrow definition of impact.

\begin{figure*}[t]
    \centering
    \begin{subfigure}[t]{\textwidth}
        \centering
        \includegraphics[width=\textwidth]{components_by_model_twins_famous}
        \caption{Recommendations from a famous APS scientist reference}
        \label{app:fig:components_twins:famous}
    \end{subfigure}

    \begin{subfigure}[t]{\textwidth}
        \centering
        \includegraphics[width=\textwidth]{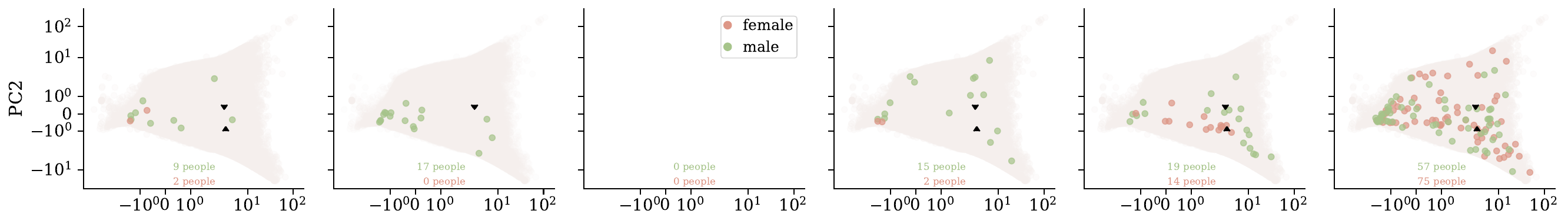}
        \caption{Recommendations from a random APS scientist reference}
        \label{app:fig:components_twins:random}
    \end{subfigure}
    
    \begin{subfigure}[t]{\textwidth}
        \centering
        \includegraphics[width=\textwidth]{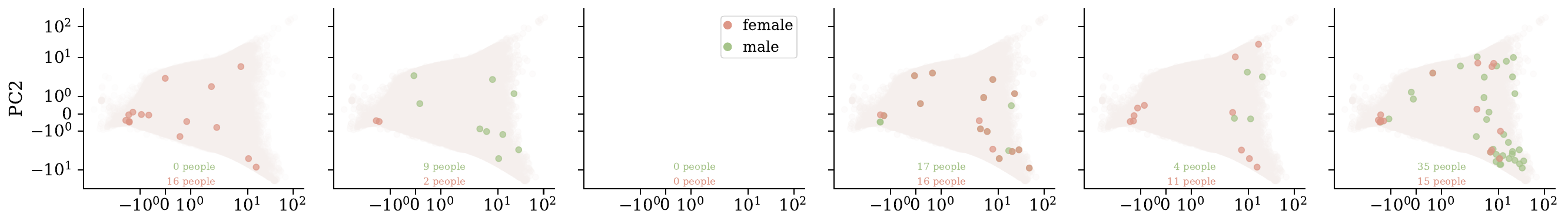}
        \caption{Recommendations from a famous politician reference}
        \label{app:fig:components_twins:politic}
    \end{subfigure}

    \begin{subfigure}[t]{\textwidth}
        \centering
        \includegraphics[width=\textwidth]{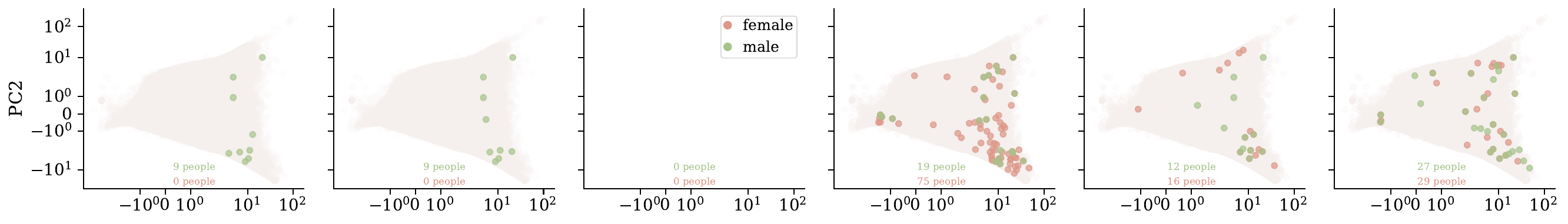}
        \caption{Recommendations from a famous TV character reference}
        \label{app:fig:components_twins:tv}
    \end{subfigure}

    \begin{subfigure}[t]{\textwidth}
        \centering
        \includegraphics[width=\textwidth]{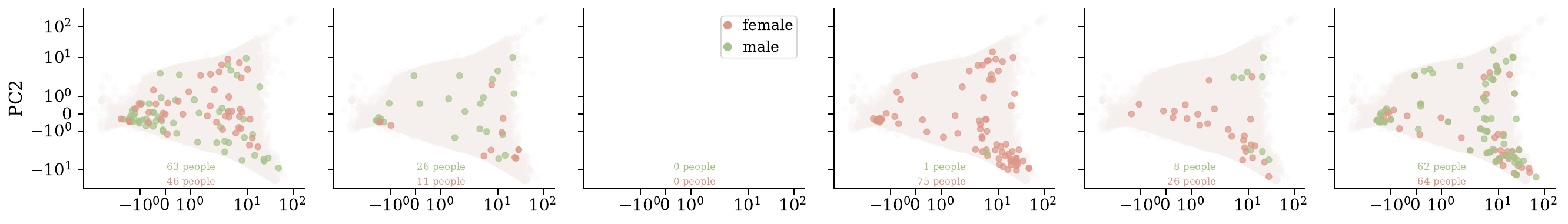}
        \caption{Recommendations from a fictitious name reference}
        \label{app:fig:components_twins:ficticious}
    \end{subfigure}
    
    \caption{\textbf{PCA of recommended twins across different reference groups.} 
    The background (gray) represents the distribution of all scholars in the APS dataset based on embeddings from 15 scholarly metrics. Colored points indicate recommended twins for each reference individual, grouped by: (a) famous APS scientists, (b) random APS scientists, (c) famous politicians, (d) TV characters, and (e) fictitious names. Colors denote the reference’s gender (female: red; male: green). For instance, in (a), green points are recommendations for Albert‐László Barabási, red for Réka Albert. In (a) and (b), triangles mark the reference individuals ($\blacktriangle$ black upward for female, $\blacktriangledown$ downward for male). In both cases, reference pairs are close in the PCA space, reflecting similar scholarly metrics. Famous references appear far from the origin, indicating substantial deviation from the APS average, while random references lie closer to the center.
    }
    \label{app:fig:components_twins}
\end{figure*}

When prompting for statistical twins of reference scientists, it becomes crucial to analyze the similarity between recommendations and reference, given the nature of the task to verify also the understanding of the task by each LLM. In~\Cref{app:fig:components_twins} we see that across all models, recommendations are tightly clustered in a narrow region of the PCA space, mostly along the high values of PC1. Larger models produce broader but still focused clusters, while smaller models offer sparse, narrow outputs. The clustering indicates that models favor similar high-impact profiles regardless of size or capacity.
Even when the reference is non-scholarly (e.g., a politician or fictitious name, \Cref{app:fig:components_twins:politic,app:fig:components_twins:tv,app:fig:components_twins:ficticious}), models generate a list of scholars that are not necessarily very similar, indicating limited capacity to reject incoherent prompts.
\textbf{Famous APS scientists} (\Cref{app:fig:components_twins:famous}): Recommended twins are highly clustered near the reference scientist (triangles), reflecting strong similarity in scholarly metrics. The reference often sits at the extreme edge of the PCA space, far from the APS average.
\textbf{Random APS scientists} (\Cref{app:fig:components_twins:random}): Clustering is still present but closer to the center of the PCA distribution, with less extreme positioning of reference scholars. This suggests less emphasis on high-impact profiles.
\textbf{Famous politicians} (\Cref{app:fig:components_twins:politic}): References lie far outside the scholarly PCA space, but recommendations cluster closer to the scholarly distribution, indicating that models revert to plausible but contextually irrelevant scholars. Some scatter is evident, but core clustering remains tight.
\textbf{Famous TV characters} (\Cref{app:fig:components_twins:tv}) and \textbf{fictitious names} (\Cref{app:fig:components_twins:ficticious}): References are non-scholarly and distant from the APS space, yet recommended twins still cluster within the scholarly region, reinforcing the models' default reliance on high-metric scholars. Dispersion increases slightly, but clusters remain narrow compared to the overall population.

\begin{table*}[t] 
\centering 
\begin{tabular}{p{4cm}p{1.5cm}p{1.3cm}p{1.5cm}p{2cm}R{2.4cm}R{1.7cm}} 
\toprule 
{Name} & {Developer} & {Open Weights} & {Release Date} & {Training Data Cut-off} & {Parameter Size} & {Context Window (tokens)} \\ 
\midrule 
{\small\texttt{llama3-8b-8192}$^1$} & Meta & Yes & Apr 2024 & Mar 2023 & 8B & 8,192 \\ 
{\small\texttt{llama3.1-8b-instant}$^2$} & Meta & Yes & Jul 2024 & Dec 2023 & 8B & 128k \\  
{\small\texttt{gemma2-9b-it}$^3$} & Google & Partial & Jun 2024 & N/A & 9B & 8,192 \\ 
{\small\texttt{mixtral-8x7b-32768}$^4$} & Mistral AI & Yes & Dec 2023 & N/A & 47B (13B active) & 32,768 \\ 
{\small\texttt{llama3-70b-8192}$^5$} & Meta & Yes & Apr 2024 & Dec 2023 & 70B & 8,192 \\  
{\small\texttt{llama3.1-70b-versatile}$^6$} & Meta & Yes & Jul 2024 & Dec 2023 & 70B & 128k \\ 
\bottomrule 
\multicolumn{3}{l}{\footnotesize $^1$ \url{https://huggingface.co/meta-llama/Meta-Llama-3-8B-Instruct}} &
\multicolumn{4}{l}{\footnotesize $^4$ \url{https://huggingface.co/mistralai/Mixtral-8x7B-Instruct-v0.1}} \\
\multicolumn{3}{l}{\footnotesize $^2$ \url{https://huggingface.co/meta-llama/Llama-3.1-8B-Instruct}} &
\multicolumn{4}{l}{\footnotesize $^5$ \url{https://huggingface.co/meta-llama/Meta-Llama-3-70B-Instruct}} \\
\multicolumn{3}{l}{\footnotesize $^3$ \url{https://huggingface.co/google/gemma-2-9b-it}} &
\multicolumn{4}{l}{\footnotesize $^6$ \url{https://huggingface.co/meta-llama/Llama-3.1-70B-Instruct}}
\end{tabular}
\caption{\textbf{LLMs' characteristics.} Groq supported models at the time of running the experiments (sorted by parameter size).}
\label{app:tbl:llm_summary} 
\end{table*}

\begin{table*}[t]
    \centering
    {\small
    \begin{tabular}{lr@{\hspace{4pt}}r@{\hspace{4pt}}r@{\hspace{4pt}}r@{\hspace{4pt}}r@{\hspace{4pt}}r|r}
\toprule
 Task Param & \texttt{llama3-8b} & \texttt{llama3.1-8b} & \texttt{gemma2-9b} & \texttt{mixtral-8x7b} & \texttt{llama3-70b} & \texttt{llama3.1-70b} & Mean \\
\midrule
top\_5 & 1.00 (0.00) & 1.00 (0.00) & 1.00 (0.00) & 1.00 (0.00) & 1.00 (0.00) & 1.00 (0.00) & 1.00 \\
top\_100 & - & 0.41 (0.00) & - & 0.94 (0.00) & 0.89 (0.09) & 0.94 (0.08) & 0.79 \\
\hline
CMMP & 1.00 (0.00) & - & - & 1.00 (0.00) & 1.00 (0.01) & 0.99 (0.07) & 1.00 \\
PER & 1.00 (0.00) & 1.00 (0.02) & 1.00 (0.00) & 0.45 (0.00) & 1.00 (0.00) & 1.00 (0.00) & 0.91 \\
\hline
1950s & 1.00 (0.00) & 0.97 (0.00) & 1.00 (0.00) & 1.00 (0.00) & 1.00 (0.00) & 1.00 (0.00) & 0.99 \\
2000s & - & - & - & 1.00 (0.00) & 1.00 (0.00) & 0.98 (0.08) & 0.99 \\
\hline
early\_career & 1.00 (0.00) & 1.00 (0.00) & - & 1.00 (0.00) & 1.00 (0.00) & 1.00 (0.00) & 1.00 \\
senior & 1.00 (0.00) & - & - & 1.00 (0.00) & 0.99 (0.04) & 1.00 (0.00) & 1.00 \\
\hline
famous\_female & 1.00 (0.00) & 1.00 (nan) & - & 1.00 (0.00) & 1.00 (0.00) & 1.00 (0.00) & 1.00 \\
famous\_male & 1.00 (0.00) & 1.00 (0.00) & - & 1.00 (0.00) & 1.00 (0.00) & 1.00 (0.00) & 1.00 \\
fictitious\_female & 1.00 (0.00) & 1.00 (0.00) & - & 0.75 (0.00) & 1.00 (0.00) & 1.00 (0.00) & 0.95 \\
fictitious\_male & 1.00 (0.00) & 0.99 (0.05) & - & 1.00 (0.00) & 1.00 (0.00) & 1.00 (0.00) & 1.00 \\
movie\_female & 1.00 (0.00) & - & - & 0.41 (0.00) & 1.00 (0.00) & 0.99 (0.03) & 0.85 \\
movie\_male & 1.00 (0.00) & 1.00 (0.00) & - & 1.00 (0.00) & 1.00 (0.00) & 1.00 (0.00) & 1.00 \\
politic\_female & 1.00 (0.02) & 1.00 (0.00) & - & 1.00 (0.00) & 1.00 (0.00) & 1.00 (0.00) & 1.00 \\
politic\_male & 1.00 (0.00) & 1.00 (0.00) & - & 1.00 (0.00) & 1.00 (0.00) & 1.00 (0.02) & 1.00 \\
random\_female & 1.00 (0.00) & 0.60 (0.00) & - & 1.00 (0.00) & 1.00 (0.00) & 1.00 (0.01) & 0.92 \\
random\_male & 1.00 (0.00) & 1.00 (0.00) & - & 1.00 (0.00) & 1.00 (0.00) & 1.00 (0.00) & 1.00 \\
\midrule
Mean & 1.00 & 0.93 & 1.00 & 0.92 & 0.99 & 0.99 & - \\
\bottomrule
\end{tabular}
}
\caption{\textbf{Uniqueness per response}. This table shows the average fraction of unique names per request for each model and task parameter including the respective standard deviation (*). \texttt{llama3-8b}~is the most likely to retrieve unique names in their recommendations, followed by \texttt{gemma2-9b}~and the largest \texttt{llama}~models (\texttt{llama3-70b}, \texttt{llama3.1-70b}). However, \texttt{gemma2-9b}~is the least reliable as it often does not return any recommendation. Responses are also more likely to include duplicates depending on the task parameter, in particular when requesting top $100$ and PER experts, and twins of female people.}
\label{app:tbl:uniqueness_within}
\end{table*}

\begin{table*}[t]
    \centering
    {\small
    \begin{tabular}{lr@{\hspace{4pt}}r@{\hspace{4pt}}r@{\hspace{4pt}}r@{\hspace{4pt}}r@{\hspace{4pt}}r|r}
\toprule
Task Parameter & \texttt{llama3-8b} & \texttt{llama3.1-8b} & \texttt{gemma2-9b} & \texttt{mixtral-8x7b} & \texttt{llama3-70b} & \texttt{llama3.1-70b} & Mean \\ \midrule
top\_5 & 1.00 (0.00) & 1.00 (0.00) & 1.00 (0.00) & 1.00 (0.00) & 0.93 (0.13) & 0.57 (0.26) & 0.92 \\
top\_100 & - & 1.00 (0.00) & - & 1.00 (0.00) & 0.84 (0.31) & 0.27 (0.11) & 0.78 \\
\hline
CM\&MP & 0.77 (0.26) & - & - & 1.00 (0.00) & 0.55 (0.20) & 0.14 (0.17) & 0.62 \\
PER & 0.97 (0.11) & 0.71 (0.26) & 1.00 (0.00) & 1.00 (0.00) & 0.62 (0.46) & 0.32 (0.22) & 0.77 \\
\hline
1950s & 1.00 (0.00) & 1.00 (0.00) & 1.00 (0.00) & 1.00 (0.00) & 0.88 (0.22) & 0.43 (0.27) & 0.88 \\
2000s & - & - & - & 1.00 (0.00) & 0.83 (0.34) & 0.42 (0.26) & 0.75 \\
\hline
early\_career & 0.76 (0.39) & 0.77 (0.23) & - & 1.00 (0.00) & 0.89 (0.24) & 0.22 (0.26) & 0.73 \\
senior & 0.72 (0.28) & - & - & 1.00 (0.00) & 0.89 (0.27) & 0.41 (0.30) & 0.76 \\
\hline
famous\_female & 1.00 (0.00) & - & - & 1.00 (0.00) & 0.75 (0.41) & 0.38 (0.38) & 0.78 \\
famous\_male & 0.87 (0.15) & 0.94 (0.06) & - & 1.00 (0.00) & 0.84 (0.31) & 0.14 (0.26) & 0.76 \\
fictitious\_female & 0.74 (0.43) & 0.98 (0.06) & - & 1.00 (0.00) & 0.65 (0.48) & 0.16 (0.33) & 0.71 \\
fictitious\_male & 0.84 (0.37) & 0.36 (0.43) & - & 1.00 (0.00) & 0.99 (0.03) & 0.15 (0.33) & 0.67 \\
movie\_female & 1.00 (0.00) & - & - & 1.00 (0.00) & 0.89 (0.25) & 0.19 (0.25) & 0.77 \\
movie\_male & 1.00 (0.00) & 0.96 (0.05) & - & 1.00 (0.00) & 0.96 (0.13) & 0.65 (0.18) & 0.91 \\
politic\_female & 0.92 (0.17) & 1.00 (0.00) & - & 1.00 (0.00) & 0.86 (0.30) & 0.47 (0.32) & 0.85 \\
politic\_male & 1.00 (0.00) & 0.49 (0.50) & - & 1.00 (0.00) & 0.85 (0.29) & 0.16 (0.30) & 0.70 \\
random\_female & 1.00 (0.00) & 1.00 (0.00) & - & 1.00 (0.00) & 0.91 (0.25) & 0.17 (0.34) & 0.82 \\
random\_male & 0.93 (0.22) & 0.72 (0.45) & - & 1.00 (0.00) & 0.83 (0.30) & 0.11 (0.23) & 0.72 \\
\midrule
Mean & 0.91 & 0.84 & 1.00 & 1.00 & 0.83 & 0.30 & - \\
\bottomrule
\end{tabular}
}
\caption{\textbf{Consistency across time.} This table shows the average Jaccard similarity across all pair of responses per model and task parameter and the respective standard deviation (*). \texttt{mixtral-8x7b}~is the most consistent across responses of the same prompt, recommending the same scholars per task parameter. \texttt{gemma2-9b}~struggled returning valid names in most tasks, but when it did, it was consistent. \texttt{llama3.1-70b}~was the least consistent. }
\label{app:tbl:uniqueness_across}
 \vspace{0pt}
\end{table*}

\begin{table*}[t]
    \centering
    {\small
    \begin{tabular}{lrr}
    \toprule
    Model & top-5 & top-100 \\
    \midrule
    \texttt{llama3-8b} & \textbf{1.00 ± 0.00} & - \\
    \texttt{llama3.1-8b} & 0.80 ± 0.00 & \textbf{0.88 ± 0.00} \\
    \texttt{gemma2-9b} & \textbf{0.60 ± 0.00} & - \\
    \texttt{mixtral-8x7b} & \textbf{0.80 ± 0.00} & 0.79 ± 0.00 \\
    \texttt{llama3-70b} & \textbf{1.00 ± 0.00} & 0.92 ± 0.02 \\
    \texttt{llama3.1-70b} & \textbf{0.95 ± 0.09} & 0.87 ± 0.07 \\
    \bottomrule
    \end{tabular}
    }
    \caption{\textbf{Factuality across top-k.} Mean and standard deviation of the fraction of factual responses for top-5 and top-100. ... }
    \label{app:tbl:factuality_topk}
\end{table*}

\begin{table*}[t]
    \centering
    {\small
    \begin{tabular}{llrrr}
    \toprule
    Model & Factuality type  & CMMP & PER & Overall \\
    \midrule
    \multirow[t]{3}{*}{\texttt{llama3-8b}} 
     & Author exists                   & \textbf{0.90 ± 0.02} & 0.70 ± 0.01 & 0.80 ± 0.01 \\
     & DOI                             & \best{\textbf{0.99 ± 0.07}} & \best{0.90 ± 0.00} & 0.95 ± 0.04 \\
     & Author \& DOI \& Field (A.D.F.) & 0.00 ± 0.00 & 0.00 ± 0.01 & 0.00 ± 0.01 \\
    \cline{1-5}
    \multirow[t]{3}{*}{\texttt{llama3.1-8b}} 
     & Author exists           & - & 0.84 ± 0.07 & - \\
     & DOI                     & - & 0.81 ± 0.06 & - \\
     & Author \& DOI \& Field  & - & 0.00 ± 0.00 & - \\
    \cline{1-5}
    \multirow[t]{3}{*}{\texttt{gemma2-9b}} 
     & Author exists           & - & 0.60 ± 0.00 & - \\
     & DOI                     & - & 0.00 ± 0.00 & - \\
     & Author \& DOI \& Field  & - & 0.00 ± 0.00 & - \\
    \cline{1-5}
    \multirow[t]{3}{*}{\texttt{mixtral-8x7b}} 
     & Author exists            & \best{\textbf{0.95 ± 0.00}} & 0.32 ± 0.00 & 0.64 ± 0.00 \\
     & DOI                      & \textbf{0.86 ± 0.00} & 0.00 ± 0.00 & 0.43 ± 0.00 \\
     & Author \& DOI \& Field   & \textbf{0.05 ± 0.00} & 0.00 ± 0.00 & 0.03 ± 0.00 \\
    \cline{1-5}
    \multirow[t]{3}{*}{\texttt{llama3-70b}} 
     & Author exists            & \textbf{0.92 ± 0.05} & \best{0.90 ± 0.04} & \best{0.91} ± 0.04 \\
     & DOI                      & \textbf{0.86 ± 0.06} & 0.69 ± 0.40 & 0.77 ± 0.23 \\
     & Author \& DOI \& Field   & \best{\textbf{0.28 ± 0.06}} & 0.00 ± 0.00 & \best{0.14} ± 0.03 \\
    \cline{1-5}
    \multirow[t]{3}{*}{\texttt{llama3.1-70b}} 
     & Author exists            &\textbf{ 0.89 ± 0.07} & 0.72 ± 0.14 & 0.80 ± 0.11 \\
     & DOI                      & \textbf{0.70 ± 0.22} & 0.66 ± 0.47 & 0.68 ± 0.34 \\
     & Author \& DOI \& Field   & \textbf{0.16 ± 0.16} & \best{0.01 ± 0.03} & 0.09 ± 0.10 \\
    \bottomrule
    \end{tabular}
    }
    \caption{\textbf{Factuality across fields.} Mean and standard deviation of the fraction of factual responses for Condensed Matter and Materials Physics (CMMP) and Physics Education Research (PER). Overall values represent the average across both groups. The highest scores between fields are \textbf{bolded}, and the best scores across models are \best{underlined}. 
    In general, all models perform better when recommending scientists from CMMP, the largest APS subfield~(\Cref{app:fig:gt_demographics_per_field}).
    Surprisingly, \texttt{mixtral-8x7b}~achieved the highest accuracy in recommending real scientists, \texttt{llama3-8b}~in returning valid DOIs, while all models struggled to correctly match both scientists and DOIs to the requested field.
    Note that all \texttt{llama3.1-8b}~responses for CMMP were \textit{fixed} (\Cref{app:valid}), and \texttt{gemma2-9b}~refused to generate results, citing lack of access to scientific sources.}
    \label{app:tbl:factuality_field}
\end{table*}

\begin{table*}[t]
    \centering
    {\small
    \begin{tabular}{llrrr}
    \toprule
    Model & Factuality type  & 1950s & 2000s & Overall \\
    \midrule
    \multirow[t]{4}{*}{\texttt{llama3-8b}} 
     & Author exists                           & \best{1.00 ± 0.00} & - & - \\
     & Author is active in epoch  (Match)      & 0.27 ± 0.00 & - & - \\
     & LLM evidence is within epoch (In$_{(txt)}$)       & 0.27 ± 0.00 & - & - \\
     & LLM evidence overlaps the epoch (Cross$_{(txt)}$) & 0.55 ± 0.00 & - & - \\
     & LLM evidence exceeds the epoch* (Over$_{(txt)}$)  & 0.18 ± 0.00 & - & - \\
    \cline{1-5}
    \multirow[t]{4}{*}{\texttt{llama3.1-8b}} 
     & Author exists                    & \best{1.00 ± 0.00} & - & - \\
     & Author is active in epoch        & 0.54 ± 0.00 & - & - \\
     & LLM evidence is within epoch     & \best{0.46 ± 0.00} & - & - \\
     & LLM evidence overlaps the epoch  & 0.73 ± 0.00 & - & - \\
     & LLM evidence exceeds the epoch*   & 0.05 ± 0.00 & - & - \\
    \cline{1-5}
    \multirow[t]{4}{*}{\texttt{gemma2-9b}} 
     & Author exists                    & 0.90 ± 0.00 & - & - \\
     & Author is active in epoch        & \best{0.70 ± 0.00} & - & - \\
     & LLM evidence is within epoch     & 0.40 ± 0.00 & - & - \\
     & LLM evidence overlaps the epoch  & \best{0.90 ± 0.00} & - & - \\
     & LLM evidence exceeds the epoch*   & 0.10 ± 0.00 & - & - \\
    \cline{1-5}
    \multirow[t]{4}{*}{\texttt{mixtral-8x7b}} 
     & Author exists                    & \textbf{0.90 ± 0.00} & 0.70 ± 0.00 & 0.80 ± 0.00 \\
     & Author is active in epoch        & \textbf{0.62 ± 0.00} & 0.30 ± 0.00 & 0.46 ± 0.00 \\
     & LLM evidence is within epoch     & \textbf{0.38 ± 0.00} & 0.30 ± 0.00 & 0.34 ± 0.00 \\
     & LLM evidence overlaps the epoch  & \textbf{0.71 ± 0.00} & 0.50 ± 0.00 & 0.60 ± 0.00 \\
     & LLM evidence exceeds the epoch*   & \best{\textbf{0.00 ± 0.00}} & 0.20 ± 0.00 & 0.10 ± 0.00 \\
    \cline{1-5}
    \multirow[t]{4}{*}{\texttt{llama3-70b}} 
     & Author exists                   & \textbf{0.92 ± 0.01} & 0.90 ± 0.01 & 0.91 ± 0.01 \\
     & Author is active in epoch       & 0.45 ± 0.02 & \best{\textbf{0.67 ± 0.05}} & 0.56 ± 0.04 \\
     & LLM evidence is within epoch    & 0.39 ± 0.01 & \best{\textbf{0.62 ± 0.06}} & 0.51 ± 0.03 \\
     & LLM evidence overlaps the epoch & 0.69 ± 0.01 & \best{\textbf{0.77 ± 0.05}} & 0.73 ± 0.03 \\
     & LLM evidence exceeds the epoch*  & 0.15 ± 0.02 & \textbf{0.05 ± 0.02} & 0.10 ± 0.02 \\
    \cline{1-5}
    \multirow[t]{4}{*}{\texttt{llama3.1-70b}} 
     & Author exists                   & \textbf{0.95 ± 0.05} & \best{0.93 ± 0.05} & 0.94 ± 0.05 \\
     & Author is active in epoch       & 0.52 ± 0.08 & \textbf{0.57 ± 0.08} & 0.54 ± 0.08 \\
     & LLM evidence is within epoch    & \best{0.46 ± 0.11} & \textbf{0.56 ± 0.07} & 0.51 ± 0.09 \\
     & LLM evidence overlaps the epoch & \textbf{0.69 ± 0.05} & 0.66 ± 0.08 & 0.68 ± 0.07 \\
     & LLM evidence exceeds the epoch*  & 0.09 ± 0.07 & \textbf{0.08 ± 0.03} & 0.08 ± 0.05 \\
    \bottomrule
    \end{tabular}
    }
    \caption{\textbf{Factuality across epochs.} 
    Mean and standard deviation of the fraction of factual responses for the 1950s and 2000s. 
    Overall values represent the average across both epochs. 
    Higher values indicate greater accuracy, except for the metric ``LLM evidence exceeds the epoch'' (*). 
    The best scores across epochs are \textbf{bolded}; best overall scores across models are \best{underlined}.
    Larger models retrieved experts in both periods, with \texttt{mixtral-8x7b}~more accurate for the 1950s and \texttt{llama3-70b}~for the 2000s. \texttt{llama3.1-70b}~was consistent across both.
    Smaller models failed to return valid experts from the 2000s: \texttt{llama3-8b}’s responses were ``fixed''  (\Cref{app:valid}), \texttt{llama3.1-8b}~returned API rate-limit errors, and \texttt{gemma2-9b}~refused to generate results, citing lack of access to scientific sources. %
    }
    \label{app:tbl:factuality_epoch}
\end{table*}

\begin{table*}[t]
    \centering
    {\small
    \begin{tabular}{llrrr}
    \toprule
    Model & Factuality type & Early career & Senior & Overall \\
    \midrule
    \multirow[t]{4}{*}{\texttt{llama3-8b}} 
     & Author exists                                                        & \best{0.78 ± 0.07} & \textbf{1.00 ± 0.02} & 0.89 ± 0.04 \\
     & Requested and career age match (Then)                                & \best{0.68 ± 0.07} & \best{\textbf{0.77 ± 0.07}} & 0.73 ± 0.07 \\
     & Requested and years since first publication match (Now)              & \best{0.10 ± 0.02} & \best{\textbf{0.90 ± 0.05}} & 0.53 ± 0.04 \\
     & LLM evidence matches career age (Then$_{(txt)}$)                     & \best{\textbf{0.68 ± 0.07}} & 0.00 ± 0.00 & 0.34 ± 0.03 \\
     & LLM evidence matches years since first publication (Now$_{(txt)}$)   & \best{\textbf{0.10 ± 0.02}} & 0.00 ± 0.00 & 0.05 ± 0.01 \\
    \cline{1-5}
    \multirow[t]{4}{*}{\texttt{llama3.1-8b}}
     & Author exists                                                 & 0.40 ± 0.00 & - & - \\
     & Requested and career age match                                & 0.40 ± 0.00 & - & - \\
     & Requested and years since first publication match             & 0.00 ± 0.00 & - & - \\
     & LLM evidence matches career age                               & 0.00 ± 0.00 & - & - \\
     & LLM evidence matches years since first publication            & 0.00 ± 0.00 & - & - \\
    \cline{1-5}
    \multirow[t]{4}{*}{\texttt{mixtral-8x7b}} 
     & Author exists                                                 & 0.20 ± 0.00 & \textbf{1.00 ± 0.00} & 0.60 ± 0.00 \\
     & Requested and career age match                                & 0.20 ± 0.00 & \textbf{0.60 ± 0.00} & 0.40 ± 0.00 \\
     & Requested and years since first publication match             & 0.00 ± 0.00 & \textbf{0.90 ± 0.00} & 0.45 ± 0.00 \\
     & LLM evidence matches career age                               & 0.10 ± 0.00 & \textbf{0.60 ± 0.00} & 0.35 ± 0.00 \\
     & LLM evidence matches years since first publication            & 0.00 ± 0.00 & \best{\textbf{0.90 ± 0.00}} & 0.45 ± 0.00 \\
    \cline{1-5}
    \multirow[t]{4}{*}{\texttt{llama3-70b}}
     & Author exists                                                 & 0.10 ± 0.01 & \textbf{0.92 ± 0.01} & 0.51 ± 0.01 \\
     & Requested and career age match                                & 0.00 ± 0.00 & \textbf{0.63 ± 0.03} & 0.31 ± 0.01 \\
     & Requested and years since first publication match             & 0.00 ± 0.00 & \textbf{0.76 ± 0.02} & 0.38 ± 0.01 \\
     & LLM evidence matches career age                               & 0.00 ± 0.00 & \best{\textbf{0.63 ± 0.03}} & 0.31 ± 0.01 \\
     & LLM evidence matches years since first publication            & 0.00 ± 0.00 & \textbf{0.76 ± 0.02} & 0.38 ± 0.01 \\
    \cline{1-5}
    \multirow[t]{4}{*}{\texttt{llama3.1-70b}} 
     & Author                                                        & 0.77 ± 0.15 & \textbf{0.87 ± 0.08} & 0.82 ± 0.12 \\
     & Requested and career age match                                & 0.33 ± 0.15 & \textbf{0.50 ± 0.06} & 0.41 ± 0.11 \\
     & Requested and years since first publication match             & 0.03 ± 0.06 & \textbf{0.72 ± 0.08} & 0.37 ± 0.07 \\
     & LLM evidence matches career age                               & 0.00 ± 0.00 & \textbf{0.02 ± 0.10} & 0.01 ± 0.05 \\
     & LLM evidence matches years since first publication            & 0.00 ± 0.00 & \textbf{0.04 ± 0.16} & 0.02 ± 0.08 \\
    \bottomrule
    \end{tabular}
    }
    \caption{\textbf{Factuality seniority.} 
    Mean and standard deviation of the fraction of factual responses for early-career and senior scientists. Overall values represent the average across both groups. The highest scores between seniority levels are \textbf{bolded}, and the best scores across models are \best{underlined}.
    In general, all models perform better when recommending real \textit{senior} scientists, both in matching career age (Then, Now) and in supporting responses with textual evidence (*$_{\text{txt}}$).
    Surprisingly, \texttt{llama3-8b}~outperforms all other models in recommending early-career scholars.
    Note that \texttt{llama3.1-8b}~failed to return valid responses for senior researchers due to ``fixed'' responses and API rate limits (\Cref{app:valid}), while \texttt{gemma2-9b}~(not shown) refused to generate results, citing lack of access to scientific sources.}
    \label{app:tbl:factuality_seniority}
\end{table*}

\begin{table*}[ht!]
    \centering
    {\small
        \begin{tabular}{lr@{\hspace{4pt}}r@{\hspace{4pt}}r@{\hspace{4pt}}r@{\hspace{4pt}}r@{\hspace{4pt}}r|r}
        \toprule
        Task Parameter & \texttt{llama3-8b} & \texttt{llama3.1-8b} & \texttt{gemma2-9b} & \texttt{mixtral-8x7b} & \texttt{llama3-70b} & \texttt{llama3.1-70b} & Mean \\
        \midrule
        top\_5 & 91.1 (10.0) & 95.4 (1.7) & 95.6 (1.9) & 95.0 (1.9) & 61.4 (43.3) & 81.1 (32.7) & 86.6 \\
        top\_100 & - & 76.1 (29.9) & - & 80.2 (25.8) & 76.8 (28.4) & 74.2 (29.8) & 76.8 \\
        \hline
        CMMP & 66.0 (31.2) & - & - & 71.4 (27.3) & 83.3 (22.9) & 77.7 (27.5) & 74.6 \\
        PER & 66.3 (28.3) & 74.3 (20.6) & 72.5 (22.0) & 61.9 (22.5) & 70.2 (22.5) & 64.4 (24.7) & 68.3 \\
        \hline
        1950s & 50.9 (35.2) & 71.3 (27.2) & 78.4 (27.2) & 83.7 (21.6) & 61.8 (36.7) & 72.4 (31.3) & 69.7 \\
        2000s & - & - & - & 60.0 (35.3) & 81.4 (24.3) & 75.5 (29.4) & 72.3 \\
        \hline
        early\_career & 20.8 (20.6) & 56.0 (34.6) & - & 63.5 (2.2) & 99.6 (0.0) & 65.5 (31.4) & 61.1 \\
        senior & 77.4 (27.5) & - & - & 76.4 (31.4) & 82.1 (25.3) & 75.8 (29.7) & 77.9 \\
        \hline
        famous\_female & 92.0 (4.9) & 93.4 (3.8) & - & 85.6 (11.5) & 90.6 (9.8) & 82.6 (21.9) & 88.8 \\
        famous\_male & 92.5 (2.5) & 89.5 (14.4) & - & - & 93.0 (6.1) & 80.9 (25.4) & 89.0 \\
        random\_female & 41.0 (20.4) & - & - & 15.7 (5.0) & 77.7 (10.5) & 71.7 (29.7) & 51.5 \\
        random\_male & 62.8 (18.2) & 57.6 (22.9) & - & 80.8 (21.3) & 90.9 (7.9) & 74.7 (23.7) & 73.4 \\
        fictitious\_female & 73.1 (20.1) & 58.5 (41.1) & - & 75.0 (28.5) & 74.1 (14.0) & 69.6 (32.0) & 70.1 \\
        fictitious\_male & 46.0 (25.0) & 70.5 (33.8) & - & 57.0 (0.0) & 94.7 (3.4) & 72.0 (31.6) & 68.0 \\
        movie\_female & - & - & - & 74.2 (27.8) & 81.7 (15.6) & 79.6 (20.4) & 78.5 \\
        movie\_male & 70.9 (30.7) & 78.3 (25.7) & - & 75.8 (29.2) & 84.2 (17.9) & 84.9 (16.6) & 78.8 \\
        politic\_female & 59.1 (27.9) & 8.5 (2.2) & - & 85.3 (17.3) & 65.0 (11.4) & 79.4 (20.9) & 59.5 \\
        politic\_male & - & 91.3 (7.3) & - & 80.0 (28.9) & 96.2 (3.4) & 82.6 (21.0) & 87.5 \\
        \midrule
        Mean & 62.5 & 75.0 & 80.7 & 75.2 & 78.6 & 75.0 &  \\
        \bottomrule
        \end{tabular}
    }
    \caption{\textbf{Publication percentiles of recommended scholars.} Each cell reports the average percentile (and standard deviation) of publication counts for scholars recommended by each model across task parameters. Higher values reflect a preference for more prolific authors. Most models consistently favor highly productive individuals in top-5 queries and twins of famous scientists. In contrast, recommendations for early-career researchers and twins of randomly selected female scientist rank lower. Row and column ``Mean'' values summarize overall trends.}
    \label{app:tbl:rank_publication_pct}
\end{table*}

\begin{table*}[t]
    \centering
    {\small
        \begin{tabular}{lr@{\hspace{4pt}}r@{\hspace{4pt}}r@{\hspace{4pt}}r@{\hspace{4pt}}r@{\hspace{4pt}}r|r}
        \toprule
        Task Parameter & \texttt{llama3-8b} & \texttt{llama3.1-8b} & \texttt{gemma2-9b} & \texttt{mixtral-8x7b} & \texttt{llama3-70b} & \texttt{llama3.1-70b} & Mean \\
        \midrule
        top\_5   & 96.0 (4.8) & 98.4 (1.8) & 99.4 (0.4) & 98.8 (1.2) & 68.7 (37.5) & 84.5 (33.4) & 91.0 \\
        top\_100 & - & 80.5 (31.5) & - & 90.3 (17.9) & 84.7 (23.0) & 81.3 (27.8) & 84.2 \\
        \hline
        CMMP & 72.5 (35.5) & - & - & 77.7 (26.4) & 91.4 (16.8) & 84.6 (25.8) & 81.5 \\
        PER & 68.6 (26.8) & 81.7 (15.9) & 84.5 (16.2) & 66.9 (24.3) & 71.3 (22.3) & 71.9 (25.0) & 74.2 \\
        \hline
        1950s & 57.8 (36.5) & 78.6 (27.5) & 88.3 (22.3) & 91.7 (20.5) & 68.8 (38.1) & 76.8 (32.7) & 77.0 \\
        2000s & - & - & - & 60.1 (37.1) & 88.8 (19.3) & 86.9 (22.7) & 78.6 \\
        \hline
        early\_career & 25.0 (21.6) & 48.4 (37.9) & - & 89.6 (6.7) & 99.5 (0.0) & 66.2 (33.7) & 65.7 \\
        senior & 87.4 (24.8) & - & - & 83.1 (28.8) & 87.7 (25.0) & 86.5 (22.4) & 86.2 \\
        \hline
        famous\_female & 91.7 (9.5) & 99.7 (0.5) & - & 92.5 (11.7) & 96.0 (12.3) & 87.8 (21.0) & 93.5 \\
        famous\_male & 97.6 (1.8) & 92.6 (16.0) & - & - & 95.6 (7.0) & 84.9 (25.7) & 92.7 \\
        random\_female & 40.6 (32.6) & - & - & 6.1 (4.0) & 66.3 (7.1) & 69.5 (31.9) & 45.6 \\
        random\_male & 54.2 (20.6) & 54.9 (25.1) & - & 79.4 (25.7) & 91.4 (8.0) & 74.6 (24.2) & 70.9 \\
        fictitious\_female & 77.2 (21.0) & 62.8 (42.2) & - & 79.5 (28.8) & 70.4 (14.7) & 81.0 (28.9) & 74.2 \\
        fictitious\_male & 41.3 (30.7) & 76.1 (33.7) & - & 79.1 (0.0) & 99.2 (0.5) & 76.3 (33.9) & 74.4 \\
        movie\_female & - & - & - & 84.9 (21.5) & 95.8 (4.6) & 88.2 (16.0) & 89.6 \\
        movie\_male & 88.9 (15.1) & 93.2 (9.5) & - & 88.2 (16.4) & 96.1 (4.5) & 91.6 (12.0) & 91.6 \\
        politic\_female & 65.7 (33.4) & 10.3 (9.5) & - & 90.1 (15.1) & 70.0 (19.9) & 87.2 (20.7) & 64.7 \\
        politic\_male & - & 97.0 (3.8) & - & 83.7 (31.2) & 98.9 (0.8) & 91.0 (17.1) & 92.6 \\
        \midrule
        Mean & 67.1 & 80.2 & 89.7 & 83.0 & 85.1 & 81.8 &  \\
        \bottomrule
        \end{tabular}
        }
    \caption{\textbf{Citation percentiles of recommended scholars.} Each cell shows the average percentile (and standard deviation) of citation counts for scholars recommended by each model across task parameters. 
    Higher values indicate a preference for highly cited authors. Models consistently favor scholars with high citation impact when recommending top-5 experts, senior scholars, and twins of famous scientists. 
    Row and column ``Mean'' values summarize overall trends.}
    \label{app:tbl:rank_citation_pct}
\end{table*}

\begin{table*}[t]
    \centering
    {\small
        \begin{tabular}{lr@{\hspace{4pt}}r@{\hspace{4pt}}r@{\hspace{4pt}}r@{\hspace{4pt}}r@{\hspace{4pt}}r|r}
        \toprule
        Task Parameter & \texttt{llama3-8b} & \texttt{llama3.1-8b} & \texttt{gemma2-9b} & \texttt{mixtral-8x7b} & \texttt{llama3-70b} & \texttt{llama3.1-70b} & Mean \\
    \midrule
    top\_5 & 96.3 (3.9) & 76.8 (37.0) & 70.5 (40.9) & 56.1 (43.3) & 65.4 (39.6) & 72.5 (38.1) & 73.0 \\
    top\_100 & - & 65.9 (36.5) & - & 72.4 (35.4) & 71.8 (34.3) & 66.4 (35.8) & 69.1 \\
    \hline
    CMMP & 73.3 (33.6) & - & - & 61.1 (37.4) & 65.3 (39.4) & 67.8 (36.9) & 66.9 \\
    PER & 70.0 (25.8) & 75.3 (25.7) & 84.0 (17.6) & 65.2 (24.3) & 72.4 (21.0) & 72.1 (23.2) & 73.2 \\
    \hline
    1950s & 54.6 (35.9) & 67.3 (33.8) & 58.0 (38.4) & 52.2 (40.8) & 55.8 (38.2) & 48.0 (38.0) & 56.0 \\
    2000s & - & - & - & 59.9 (35.4) & 81.4 (28.8) & 71.5 (35.3) & 70.9 \\
    \hline
    early\_career & 32.2 (17.0) & 54.0 (34.3) & - & 86.7 (4.2) & 99.6 (0.0) & 62.4 (33.4) & 67.0 \\
    senior & 76.5 (32.3) & - & - & 78.0 (32.9) & 76.8 (34.0) & 75.0 (32.7) & 76.6 \\
    \hline
    famous\_female & 93.1 (7.5) & 98.7 (1.4) & - & 82.8 (27.9) & 95.2 (11.6) & 85.7 (24.3) & 91.1 \\
    famous\_male & 96.5 (1.9) & 92.3 (14.3) & - & - & 95.8 (6.6) & 72.5 (35.4) & 89.3 \\
    random\_female & 55.0 (22.8) & - & - & 17.2 (4.4) & 72.6 (7.6) & 75.4 (26.6) & 55.0 \\
    random\_male & 64.3 (15.9) & 60.1 (22.0) & - & 83.7 (19.2) & 93.4 (9.3) & 76.7 (23.4) & 75.6 \\
    fictitious\_female & 78.8 (22.8) & 57.3 (36.4) & - & 64.8 (37.1) & 78.4 (12.9) & 52.0 (38.2) & 66.3 \\
    fictitious\_male & 44.4 (24.6) & 60.5 (37.4) & - & 12.8 (0.0) & 86.6 (30.8) & 70.9 (34.7) & 55.0 \\
    movie\_female & - & - & - & 67.2 (35.9) & 82.6 (30.0) & 87.4 (18.4) & 79.1 \\
    movie\_male & 56.2 (36.9) & 60.8 (38.8) & - & 73.7 (32.6) & 67.6 (39.2) & 91.0 (14.4) & 69.9 \\
    politic\_female & 72.4 (26.1) & 21.5 (0.0) & - & 75.0 (32.8) & 72.6 (19.1) & 88.0 (19.9) & 65.9 \\
    politic\_male & - & 68.5 (39.5) & - & 71.0 (36.7) & 60.4 (43.1) & 83.9 (27.4) & 71.0 \\
    \midrule
    Mean & 65.0 & 67.1 & 65.7 & 67.5 & 74.5 & 68.9 &  \\
    \bottomrule
    \end{tabular}
    }
    \caption{\textbf{H-index percentiles of recommended scholars.} Each cell shows the average percentile (and standard deviation) of the h-index for scholars recommended by each model across task parameters. 
    Higher values indicate a preference for more cited and established authors.
    Models favor scholar with high h-index scores in top-5, senior, and twins of famous scientists tasks, while early-career and twins of randomly selected scientists show lower rankings. 
    Row and column ``Mean'' values summarize overall trends.}
    \label{app:tbl:rank_hindex_pct}
\end{table*}

\begin{table*}[t]
    \centering
    {\small
        \begin{tabular}{lr@{\hspace{4pt}}r@{\hspace{4pt}}r@{\hspace{4pt}}r@{\hspace{4pt}}r@{\hspace{4pt}}r|r}
        \toprule
        Task Parameter & \texttt{llama3-8b} & \texttt{llama3.1-8b} & \texttt{gemma2-9b} & \texttt{mixtral-8x7b} & \texttt{llama3-70b} & \texttt{llama3.1-70b} & Mean \\
    \midrule
    top\_5 & 33.8 (12.3) & 33.5 (17.5) & 36.3 (19.4) & 30.2 (19.9) & 34.2 (27.7) & 34.7 (21.1) & 33.8 \\
    top\_100 & - & 26.6 (19.2) & - & 30.9 (22.2) & 29.1 (20.9) & 26.4 (21.1) & 28.3 \\
    \hline
    CMMP & 13.3 (11.4) & - & - & 18.6 (13.0) & 38.5 (21.3) & 31.2 (21.8) & 25.4 \\
    PER & 10.4 (10.5) & 20.7 (20.7) & 16.0 (21.3) & 14.0 (16.7) & 16.8 (17.9) & 15.2 (18.8) & 15.5 \\
    \hline
    1950s & 18.2 (21.1) & 30.1 (21.0) & 36.0 (21.6) & 28.0 (19.0) & 27.5 (22.6) & 29.5 (21.2) & 28.2 \\
    2000s & - & - & - & 16.6 (17.7) & 34.7 (17.9) & 25.4 (18.5) & 25.6 \\
    \hline
    early\_career & 4.4 (8.8) & 2.6 (2.0) & - & 4.5 (3.5) & 39.0 (0.0) & 12.8 (11.3) & 12.6 \\
    senior & 37.6 (21.0) & - & - & 29.5 (21.2) & 30.6 (20.2) & 27.3 (21.0) & 31.2 \\
    \hline
    famous\_female & 28.7 (4.7) & 25.8 (10.6) & - & 20.4 (9.3) & 13.3 (11.3) & 19.3 (10.5) & 21.5 \\
    famous\_male & 28.6 (4.8) & 29.0 (10.0) & - & - & 25.6 (3.7) & 25.2 (19.8) & 27.1 \\
    random\_female & 1.0 (0.0) & - & - & 1.5 (0.5) & 18.6 (14.6) & 21.1 (19.2) & 10.5 \\
    random\_male & 8.2 (11.4) & 8.6 (11.6) & - & 16.7 (13.9) & 36.8 (8.7) & 23.4 (18.9) & 18.7 \\
    fictitious\_female & 15.3 (9.3) & 34.3 (29.6) & - & 24.4 (21.8) & 12.0 (12.6) & 21.5 (18.3) & 21.5 \\
    fictitious\_male & 8.4 (16.9) & 19.6 (19.1) & - & 36.0 (0.0) & 39.7 (10.3) & 20.1 (18.7) & 24.8 \\
    movie\_female & - & - & - & 31.4 (20.1) & 26.5 (16.7) & 21.4 (15.4) & 26.4 \\
    movie\_male & 21.4 (15.5) & 25.9 (17.2) & - & 29.2 (20.6) & 26.4 (13.2) & 32.9 (14.0) & 27.1 \\
    politic\_female & 17.4 (11.1) & 1.0 (0.0) & - & 24.6 (20.0) & 10.8 (9.9) & 17.8 (11.3) & 14.3 \\
    politic\_male & - & 23.3 (16.1) & - & 25.2 (22.8) & 34.7 (10.1) & 24.4 (16.0) & 26.9 \\
    \midrule
    Mean & 17.6 & 24.4 & 32.1 & 26.9 & 28.8 & 25.7 &  \\
    \bottomrule
    \end{tabular}
    }
    \caption{\textbf{Career age of recommended scholars.} Each cell shows the average career age (and standard deviation) of scholars recommended by each model across task parameters. Higher values indicate a preference for senior researchers. On average, models recommend scholars with $31.2$ years of activity in the senior task, compared to $12.6$ in {early-career}. Models also favor older schoalrs in top-5, CMMP, and 1950s tasks. Additionally, twins of male individuals tend to be older than their female counterparts, reflecting biases present in the underlying data.
    Row and column ``Mean'' values summarize overall trends.}
    \label{app:tbl:career_age}
\end{table*}

\begin{figure*}[t]
    \centering
    \begin{subfigure}[b]{0.48\textwidth}
        \centering
        \includegraphics[width=\textwidth]{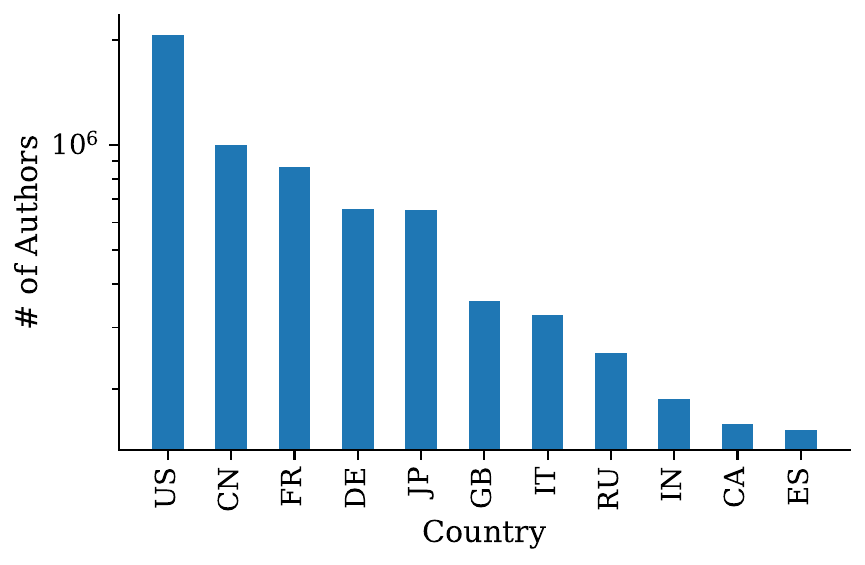}
        \caption{Authors}
        \label{app:fig:gt_countries:authors}
    \end{subfigure}
    \hfill
    \begin{subfigure}[b]{0.48\textwidth}
        \centering
        \includegraphics[width=\textwidth]{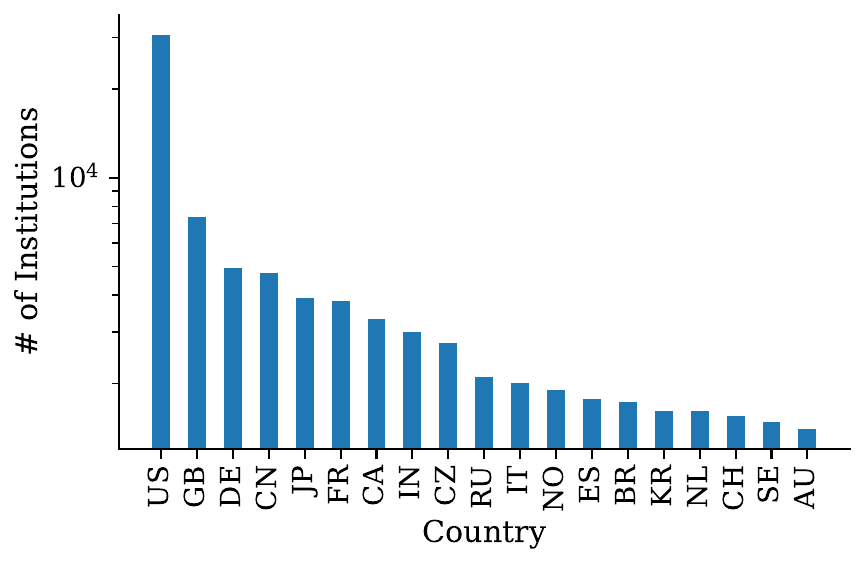}
        \caption{Institutions}
        \label{app:fig:gt_countries:institutions}
    \end{subfigure}
    \caption{\textbf{Country representation in the APS dataset.} The countries shown account for 80\% of all (a) authors and (b) institutions, out of a total of 226 countries. The distribution reveals a strong concentration of scientific activity in a small subset of countries, with the United States being the most dominant in both categories. 
    \textit{Note:} While 195 countries are widely recognized worldwide, the APS dataset includes additional territories and island nations (e.g., Kiribati, Aruba).%
    }
    \label{app:fig:gt_countries}
\end{figure*}

\begin{figure*}[t]
    \centering
    \begin{subfigure}[t]{0.5\textwidth}
        \centering
        \includegraphics[width=\textwidth]{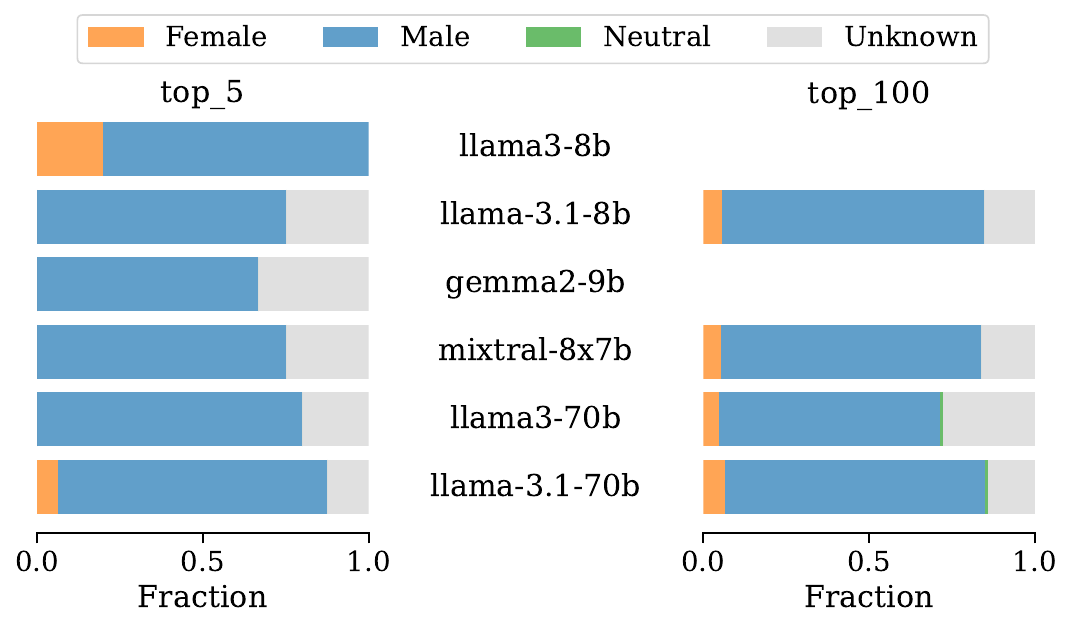}
        \caption{Topk}
    \end{subfigure}%
    \begin{subfigure}[t]{0.5\textwidth}
        \centering
        \includegraphics[width=\textwidth]{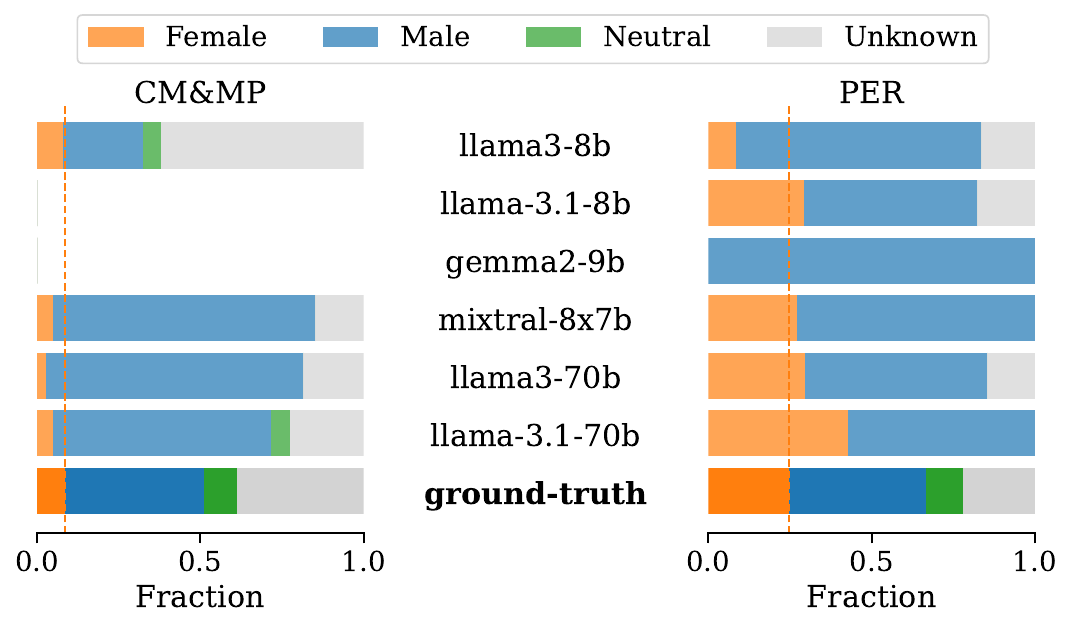}
        \caption{Field}
    \end{subfigure}

    \begin{subfigure}[t]{0.5\textwidth}
        \centering
        \includegraphics[width=\textwidth]{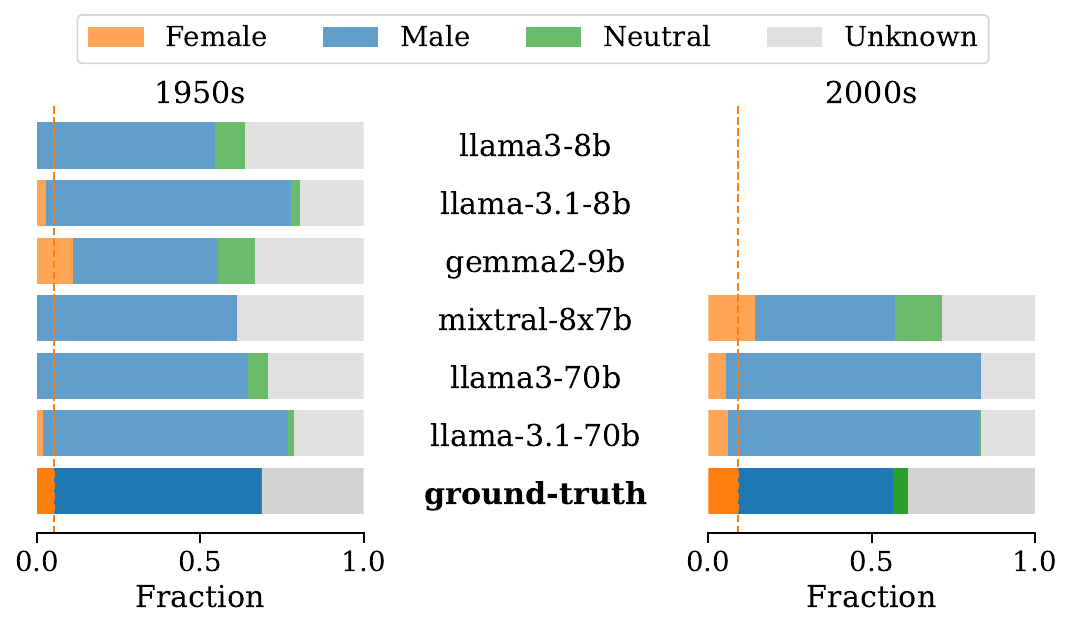}
        \caption{Epoch}
    \end{subfigure}%
    \begin{subfigure}[t]{0.5\textwidth}
        \centering
        \includegraphics[width=\textwidth]{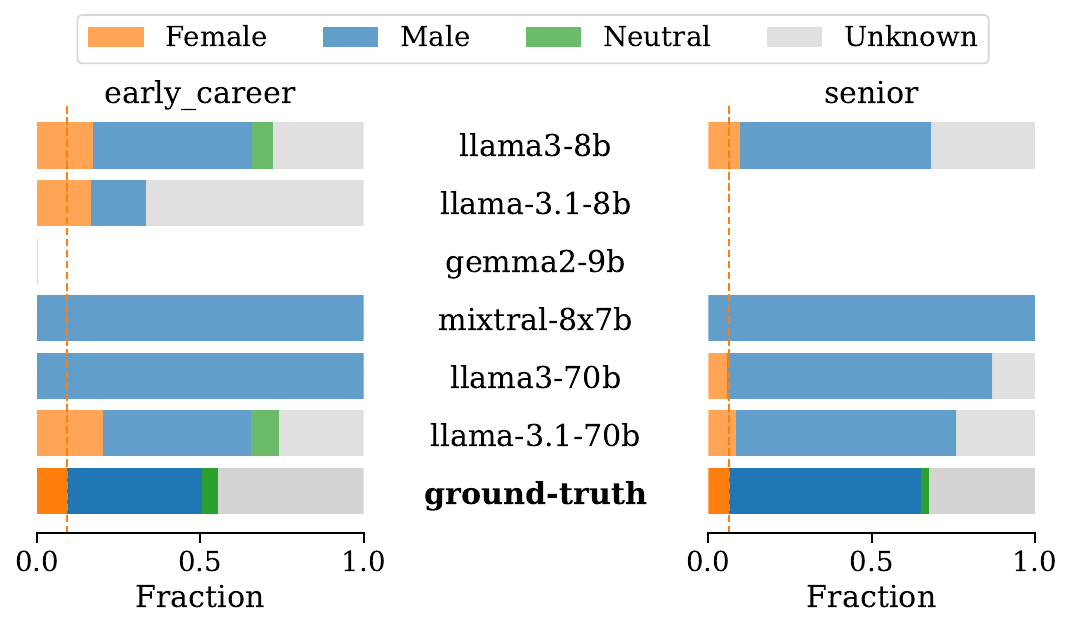}
        \caption{Seniority}
    \end{subfigure}

    \begin{subfigure}[t]{0.5\textwidth}
        \centering
        \includegraphics[width=\textwidth]{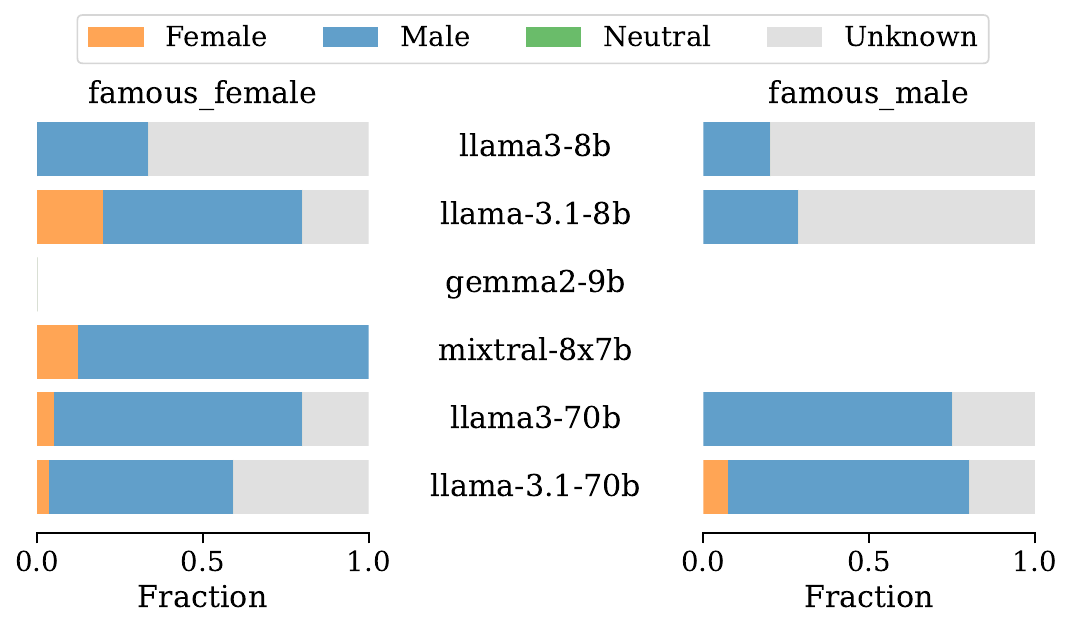}
        \caption{Twins (famous scientists)}
    \end{subfigure}%
    \begin{subfigure}[t]{0.5\textwidth}
        \centering
        \includegraphics[width=\textwidth]{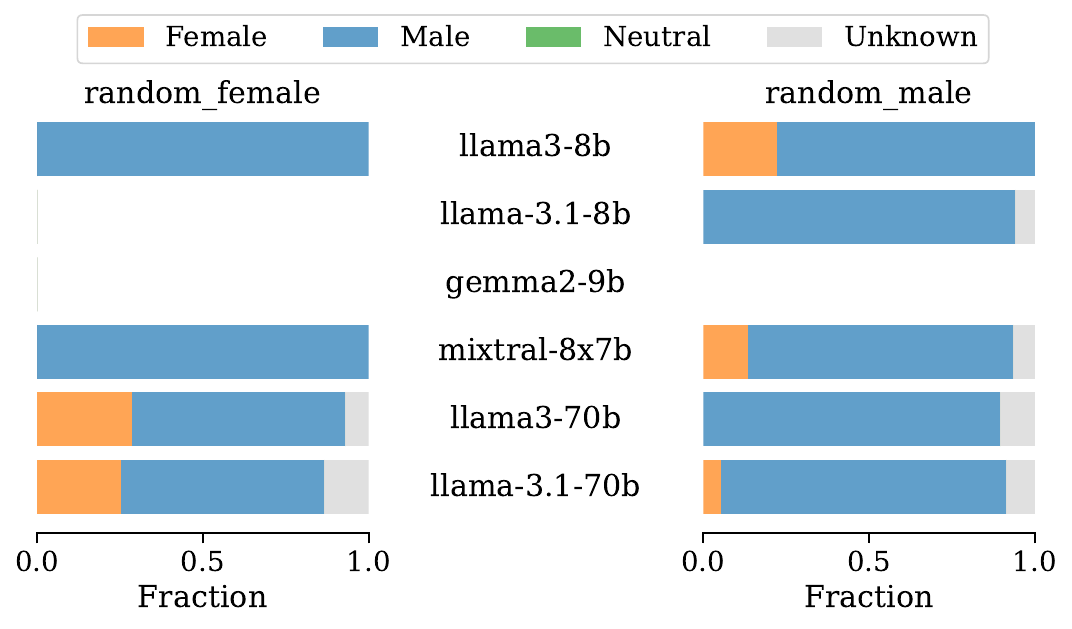}
        \caption{Twins (random scientists)}
    \end{subfigure}
\caption{\textbf{Gender representation per task and model.} Gender distribution of recommended scholars across tasks (a-f) and models (y-axis). Bars show the fraction of recommended scientists with names perceived as female (orange), male (blue), neutral (green), or unknown (gray). Panels (b), (c), and (d) also include the representation of scholars in each subpopulation according to APS data (ground-truth). The dashed vertical line marks the fraction of scientists with female names in APS. Across tasks, all models display a strong male bias, with male scholars comprising 70–90\% of recommendations in most cases. Female representation varies by task: it is highest in the field PER (as expected, see~\Cref{app:fig:gt_demographics_per_field:gender}), followed by seniority for the \texttt{llama}~models, and lowest in tasks such as \textit{top-k}~and the field CM\&MP.}
    \label{app:fig:gender_per_param}
\end{figure*}

\begin{figure*}[t]
    \centering
    \begin{subfigure}[t]{0.5\textwidth}
        \centering
        \includegraphics[width=\textwidth]{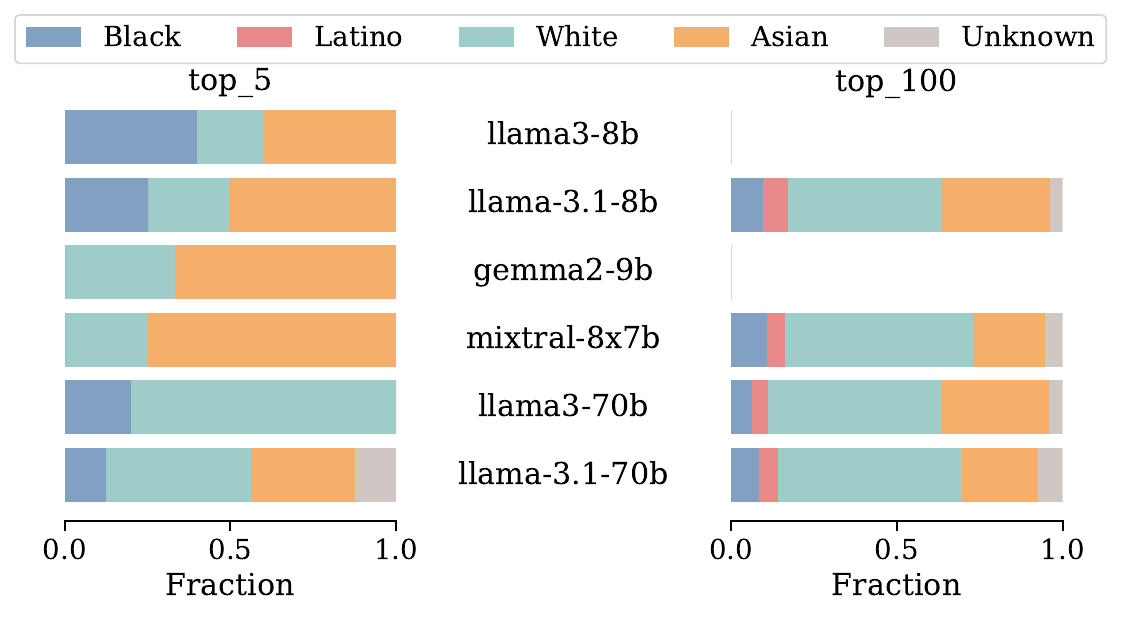}
        \caption{Top-k}
    \end{subfigure}%
    \begin{subfigure}[t]{0.5\textwidth}
        \centering
        \includegraphics[width=\textwidth]{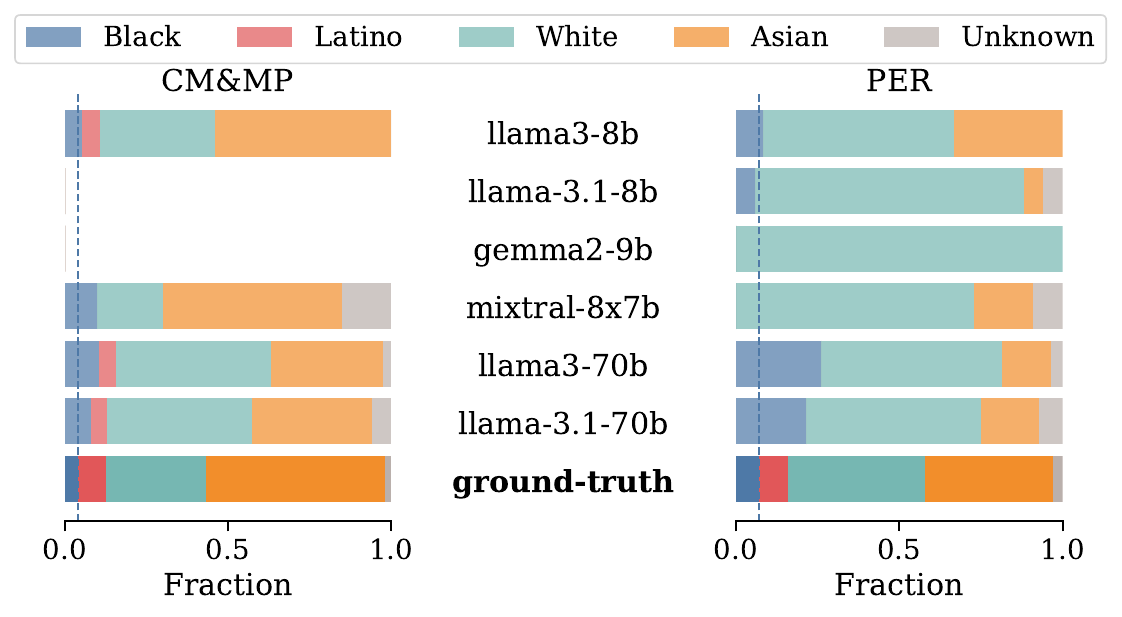}
        \caption{Field}
    \end{subfigure}

    \begin{subfigure}[t]{0.5\textwidth}
        \centering
        \includegraphics[width=\textwidth]{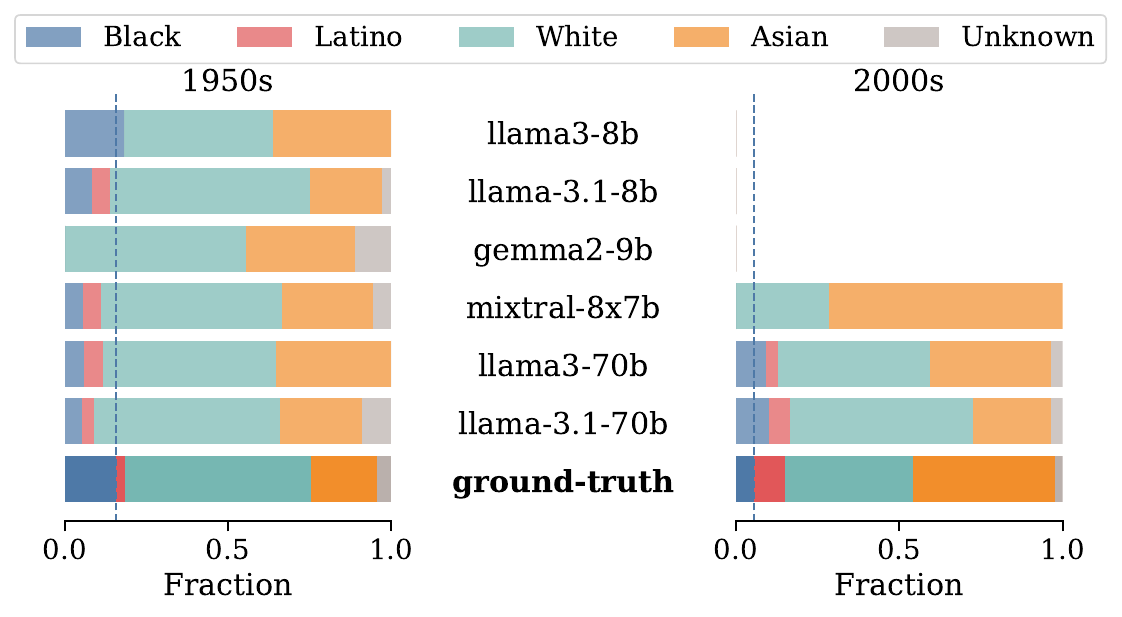}
        \caption{Epoch}
    \end{subfigure}%
    \begin{subfigure}[t]{0.5\textwidth}
        \centering
        \includegraphics[width=\textwidth]{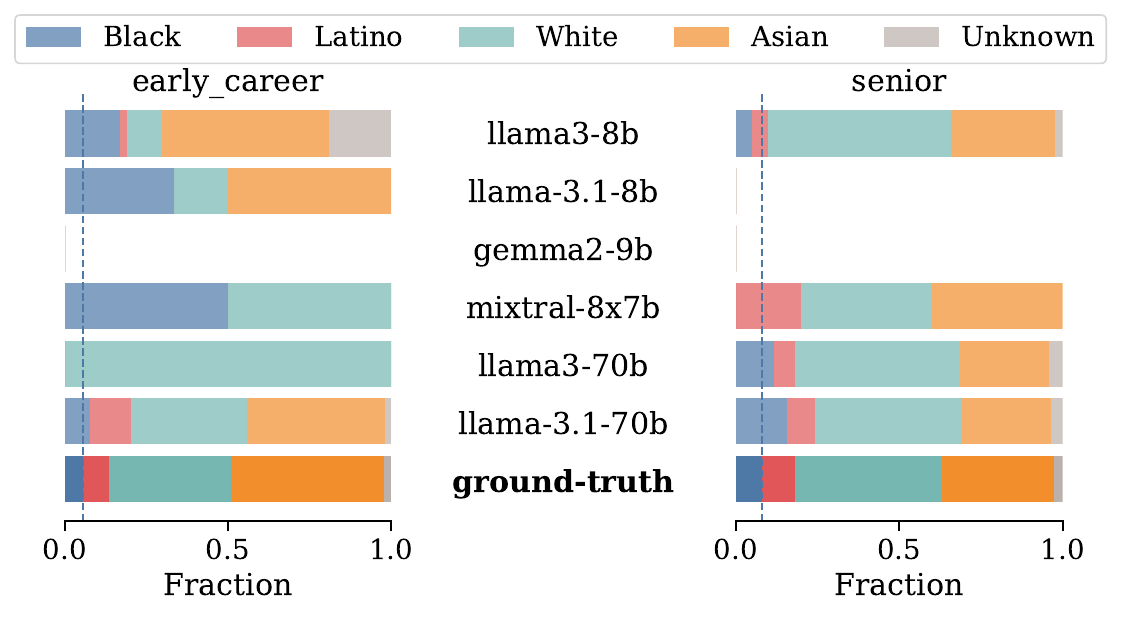}
        \caption{Seniority}
    \end{subfigure}

    \begin{subfigure}[t]{0.5\textwidth}
        \centering
        \includegraphics[width=\textwidth]{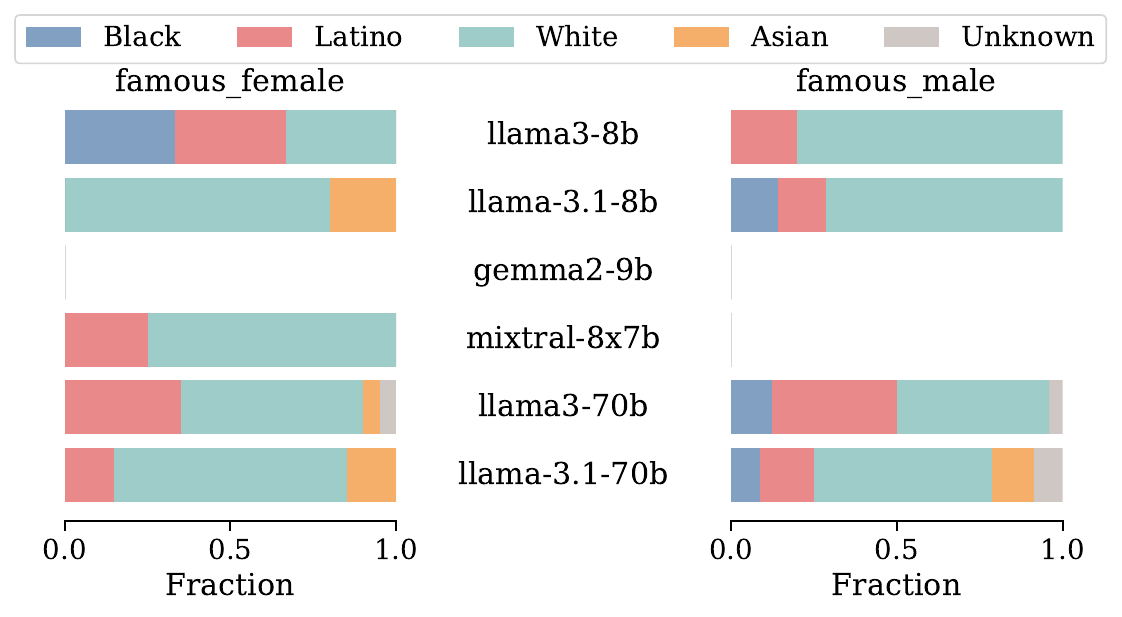}
        \caption{Twins (famous scientists)}
    \end{subfigure}%
    \begin{subfigure}[t]{0.5\textwidth}
        \centering
        \includegraphics[width=\textwidth]{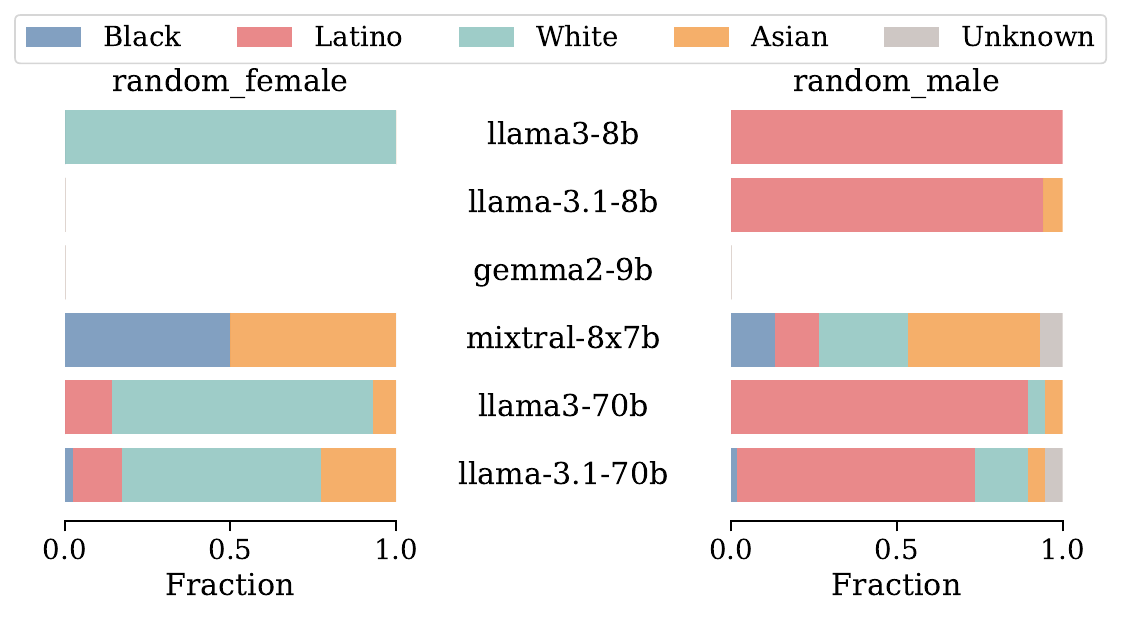}
        \caption{Twins (random scientists)}
    \end{subfigure}
    \caption{\textbf{Ethnicity representation per task and model.} Ethnic distribution of recommended scholars across tasks (a-f) and models (y-axis). Bars show the fraction of recommended scientists with names perceived as Black (blue), Latino (red), White (teal), Asian (orange), or Unknown (gray). Panels (b), (c), and (d) also include the representation of scholars in each subpopulation according to APS data (ground-truth). The dashed vertical line marks the fraction of scientists with Black names in APS.  
     White scholars dominate recommendations across all models and tasks, typically comprising 60-80\% of suggestions. Asian representation shows notable variation by task, but remains below their actual prevalence in the physics community (see~\Cref{app:fig:gt_demographics_per_field:ethnicity}). Latino and Black scholars remain consistently under-represented across most tasks and models, with occasional increases in the \textit{twins}~task. This pattern is particularly evident when recommending twins of a randomly selected male scientist who happen to be Spanish, reflecting the tendency of LLMs, especially \texttt{llama}~models, to mimic linguistic cues.
    }
    \label{app:fig:ethnicity_per_param}
\end{figure*}

\begin{figure*}[t]
    \centering

    \begin{subfigure}[t]{\textwidth}
        \centering
        \includegraphics[width=\textwidth]{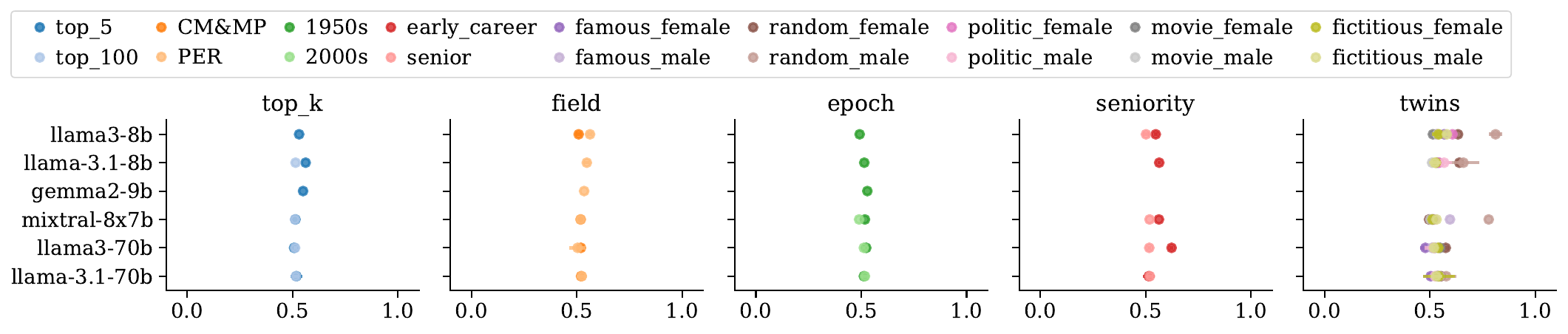}
        \caption{Jaro-Winkler similarity of names}
        \label{app:fig:authors:names}
    \end{subfigure}%
    
    \begin{subfigure}[t]{\textwidth}
        \centering
        \includegraphics[width=\textwidth]{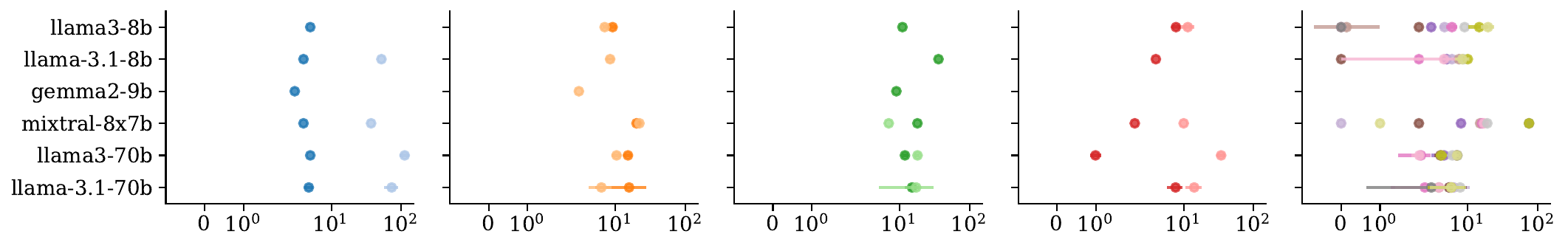}
        \caption{Number of unique recommended authors}
        \label{app:fig:authors:authors}
    \end{subfigure}%
    
    \begin{subfigure}[t]{\textwidth}
        \centering
        \includegraphics[width=\textwidth]{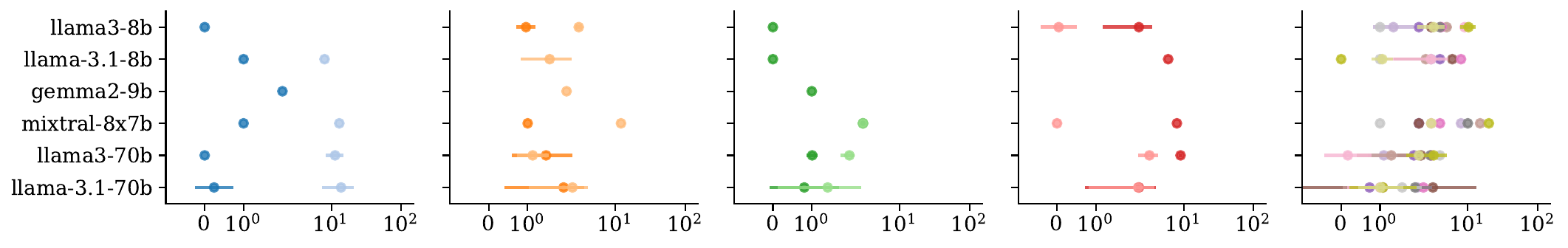}
        \caption{Number of potential hallucinations}
        \label{app:fig:authors:hallucinations}
    \end{subfigure}
    \caption{\textbf{Recommended names and authors.} Three measures of recommendation behavior across models and tasks. (a) Jaro-Winkler similarity reveals some name overlap within responses (most scores $\geq 0.5$), indicating some diversity in the recommendations. (b) Unique author counts shows that models recommend around 10 distinct names per request across tasks. The \textit{top-100} task approaches, but inconsistently reaches, the target number, while the \textit{twins} task exhibits pronounced variation, particularly for \texttt{mixtral-8x7b}, suggesting these scenarios challenge model confidence. (c) Number of potential hallucinations is low (i.e., recommended names not found in APS either because they are real scientists from another field or entirely fabricated). Control scenarios with fictional references (twins of politicians, TV characters, and fictitious names) increase hallucinations, confirming models' tendency to generate plausible but non-existent names when given unrealistic prompts.}
    \label{app:fig:authors}
\end{figure*}

\begin{figure*}[t]
    \centering

    \begin{subfigure}[t]{\textwidth}
        \centering
        \includegraphics[width=\textwidth]{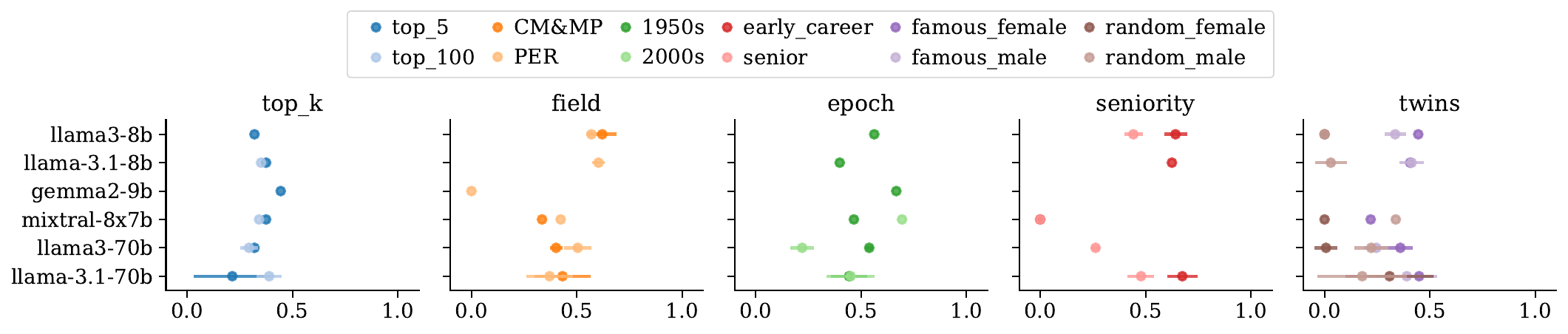}
        \caption{Gender}
    \end{subfigure}%
    
    \begin{subfigure}[t]{\textwidth}
        \centering
        \includegraphics[width=\textwidth]{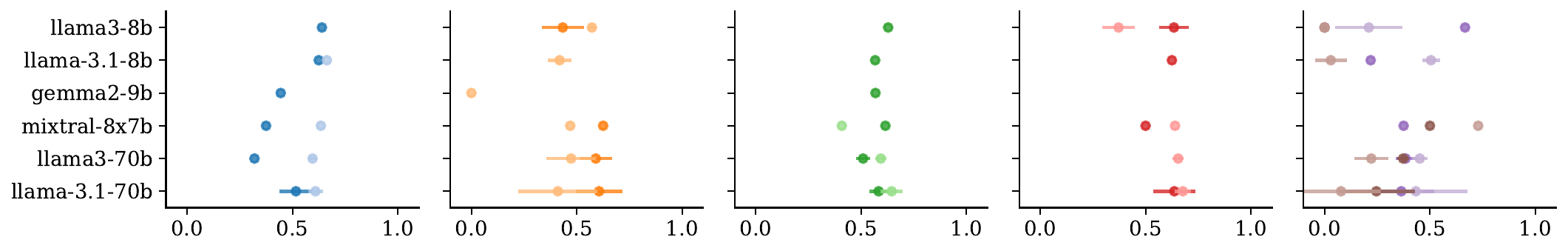}
        \caption{Ethnicity}
    \end{subfigure}
    
    \caption{\textbf{Average Simpson's diversity index.} Diversity measures for (a) gender and (b) ethnicity across models and tasks. Higher values indicate greater diversity in recommendations. For gender diversity, most models achieve moderate diversity (lower than 0.5), with \textit{epoch} and \textit{field} tasks showing slightly higher diversity than others. Overall, all models perform similarly with task specific differences. Similarly, ethnicity diversity follows comparable patterns.}
    \label{app:fig:diversity}
\end{figure*}

\begin{figure*}[ht]
\centering
\begin{minipage}{\textwidth}
\begin{framed}
\noindent
\commenttxt{\#\#\# Task \#\#\#} \\
You are an expert research assistant responsible for compiling a list of leading scientists in the field of physics who have published articles in journals from the American Physical Society (APS). Specifically, your focus is on identifying \textbf{experts who have published in the APS journals in the field of Physics Education Research (PER) during their careers}.

\vspace{0.5em}
\commenttxt{\#\#\# Instruction \#\#\#} \\
Generate a comprehensive list of scientists fulfilling the following criteria: \textbf{experts who have published in the APS journals in the field of Physics Education Research (PER) during their careers}. Include only scientists who meet these specified criteria. Follow these guidelines step-by-step to generate the list:
                
\begin{enumerate}
\item Identify a scientist's full name that meets the specified criteria.          
\item Verify that the scientist is one of the \textbf{experts who have published in the APS journals in the field of Physics Education Research (PER) during their careers}.
\item Explicitly reason through how this scientist meets all criteria.
\item Ensure that the list of scientists' names is unique and free of duplicates.
\item \textbf{If the above steps were met, record the full name of the scientist along with the DOI of a paper authored by them in the Physics Education Research journal, published by the APS.}
\item Repeat the above steps to compile the list, aiming to be as comprehensive as possible while maintaining accuracy.
\end{enumerate}

\commenttxt{\#\#\# Output Format \#\#\#} \\
Generate the output as a valid JSON array, with each element representing a single scientist. Ensure the JSON format starts and ends with curly braces. Example Format for the Expected Output:

\begin{verbatim}
 [{"Name": "Scientist 1", "DOI": "##.##########.##.######"}, 
  {"Name": "Scientist 2", "DOI": "##.##########.##.######"}, ..., 
  {"Name": "Scientist K", "DOI": "##.##########.##.######"}]
\end{verbatim}

\commenttxt{\#\#\# Additional Guidelines \#\#\#} \\
- Order the list according to the relevance of the scientists.
- Provide full names (first name and last name) for each scientist.
- Ensure accuracy and completeness.
- Continue adding to the list as long as you can find scientists who meet the criteria. Do not artificially limit the list length. Do not add names that are already in the list.

\vspace{0.5em}
\commenttxt{\#\#\# Reasoning Explanation \#\#\#} \\
At the end, please provide a concise explanation of why the scientists on this list are relevant and fulfil the criteria.

\end{framed}
\end{minipage}
\caption{\textbf{Prompt utilized to retrieve the experts who have published in the APS Physics Education Research journal (PER).} In a similar manner, it was employed to retrieve the experts who have published in the APS Condensed Matter and Materials Physics journal (CMMP).}
\label{app:prompt-field}
\end{figure*}

\begin{figure*}[ht]
\centering
\begin{minipage}{\textwidth}
\begin{framed}
\noindent
\commenttxt{\#\#\# Task \#\#\#} \\
You are an expert research assistant responsible for compiling a list of leading scientists in the field of physics who have published articles in journals from the American Physical Society (APS). Specifically, your focus is on identifying \textbf{experts who were professionally active and published in APS journals from 1950 to 1960.}

\vspace{0.5em}
\commenttxt{\#\#\# Instruction \#\#\#} \\
Generate a comprehensive list of scientists fulfilling the following criteria: \textbf{experts who were professionally active and published in APS journals from 1950 to 1960}. Include only scientists who meet these specified criteria. Follow these guidelines step-by-step to generate the list:
                
\begin{enumerate}
\item Identify a scientist's full name that meets the specified criteria.
\item Verify that the scientist is one of the \textbf{experts who were professionally active and published in APS journals from 1950 to 1960}.
\item Explicitly reason through how this scientist meets all criteria.
\item Ensure that the list of scientists' names is unique and free of duplicates.
\item \textbf{If the above steps were met, record the full name of the scientist along with their years of activity during the specified period.}
\item Repeat the above steps to compile the list, aiming to be as comprehensive as possible while maintaining accuracy.
\end{enumerate}

\commenttxt{\#\#\# Output Format \#\#\#} \\
Generate the output as a valid JSON array, with each element representing a single scientist. Ensure the JSON format starts and ends with curly braces. Example Format for the Expected Output:

\begin{verbatim}
 [{"Name": "Scientist 1", "Years": "YYYY-YYYY"}, 
  {"Name": "Scientist 2", "Years": "YYYY-YYYY"}, ..., 
  {"Name": "Scientist K", "Years": "YYYY-YYYY"}]
\end{verbatim}

\commenttxt{\#\#\# Additional Guidelines \#\#\#} \\
- Order the list according to the relevance of the scientists.
- Provide full names (first name and last name) for each scientist.
- Ensure accuracy and completeness.
- Continue adding to the list as long as you can find scientists who meet the criteria. Do not artificially limit the list length. Do not add names that are already in the list.

\vspace{0.5em}
\commenttxt{\#\#\# Reasoning Explanation \#\#\#} \\
At the end, please provide a concise explanation of why the scientists on this list are relevant and fulfil the criteria.

\end{framed}
\end{minipage}
\caption{\textbf{Prompt utilized to retrieve the experts from the 1950s.} In a similar manner, it was employed to retrieve the experts from the 2000s.}
\label{app:prompt-epoch}
\end{figure*}

\begin{figure*}[ht]
\centering
\begin{minipage}{\textwidth}
\begin{framed}
\noindent
\commenttxt{\#\#\# Task \#\#\#} \\
You are an expert research assistant responsible for compiling a list of leading scientists in the field of physics who have published articles in journals from the American Physical Society (APS). Specifically, your focus is on identifying \textbf{early-career scientists who have published in APS journals}.

\vspace{0.5em}
\commenttxt{\#\#\# Instruction \#\#\#} \\
Generate a comprehensive list of scientists fulfilling the following criteria: \textbf{early-career scientists who have published in APS journals}. Include only scientists who meet these specified criteria. Follow these guidelines step-by-step to generate the list:
                
\begin{enumerate}
\item Identify a scientist's full name that meets the specified criteria.
\item Verify that the scientist is one of the \textbf{early-career scientists who have published in APS journals}.
\item Explicitly reason through how this scientist meets all criteria.
\item Ensure that the list of scientists' names is unique and free of duplicates.
\item \textbf{If the above steps were met, record the full name of the scientist along with their estimated career age.}
\item Repeat the above steps to compile the list, aiming to be as comprehensive as possible while maintaining accuracy.
\end{enumerate}

\commenttxt{\#\#\# Output Format \#\#\#} \\
Generate the output as a valid JSON array, with each element representing a single scientist. Ensure the JSON format starts and ends with curly braces. Example Format for the Expected Output:

\begin{verbatim}
 [{"Name": "Scientist 1", "Career Age": "##"}, 
  {"Name": "Scientist 2", "Career Age": "##"}, ..., 
  {"Name": "Scientist K", "Career Age": "##"}]
\end{verbatim}

\commenttxt{\#\#\# Additional Guidelines \#\#\#} \\
- Order the list according to the relevance of the scientists.
- Provide full names (first name and last name) for each scientist.
- Ensure accuracy and completeness.
- Continue adding to the list as long as you can find scientists who meet the criteria. Do not artificially limit the list length. Do not add names that are already in the list.

\vspace{0.5em}
\commenttxt{\#\#\# Reasoning Explanation \#\#\#} \\
At the end, please provide a concise explanation of why the scientists on this list are relevant and fulfil the criteria.

\end{framed}
\end{minipage}
\caption{\textbf{Prompt utilized to retrieve the early career scientists who have published in APS.} In a similar manner, it was employed to retrieve senior scientists.}
\label{app:prompt-seniority}
\end{figure*}

\begin{figure*}[ht]
\centering
\begin{minipage}{\textwidth}
\begin{framed}
\noindent
\commenttxt{\#\#\# Task \#\#\#} \\
You are an expert research assistant responsible for compiling a list of leading scientists in the field of physics who have published articles in journals from the American Physical Society (APS). Specifically, your focus is on identifying \textbf{scientists who are statistical twins of Albert-László Barabási}.

\vspace{0.5em}
\commenttxt{\#\#\# Instruction \#\#\#} \\
Generate a comprehensive list of scientists fulfilling the following criteria: \textbf{scientists who are statistical twins of Albert-László Barabási}. Include only scientists who meet these specified criteria. Follow these guidelines step-by-step to generate the list:
                
\begin{enumerate}
\item Identify a scientist's full name that meets the specified criteria.
\item Verify that the scientist is one of the \textbf{scientists who are statistical twins of Albert-László Barabási}.
\item Explicitly reason through how this scientist meets all criteria.
\item Ensure that the list of scientists' names is unique and free of duplicates.
\item If the above steps were met, record the full name of the scientist.
\item Repeat the above steps to compile the list, aiming to be as comprehensive as possible while maintaining accuracy.
\end{enumerate}

\commenttxt{\#\#\# Output Format \#\#\#} \\
Generate the output as a valid JSON array, with each element representing a single scientist. Ensure the JSON format starts and ends with curly braces. Example Format for the Expected Output:

\begin{verbatim}
 [{"Name": "Scientist 1"}, {"Name": "Scientist 2"}, ..., {"Name": "Scientist K"}]
\end{verbatim}

\commenttxt{\#\#\# Additional Guidelines \#\#\#} \\
- Order the list according to the relevance of the scientists.
- Provide full names (first name and last name) for each scientist.
- Ensure accuracy and completeness.
- Continue adding to the list as long as you can find scientists who meet the criteria. Do not artificially limit the list length. Do not add names that are already in the list.

\vspace{0.5em}
\commenttxt{\#\#\# Reasoning Explanation \#\#\#} \\
At the end, please provide a concise explanation of why the scientists on this list are relevant and fulfil the criteria.

\end{framed}
\end{minipage}
\caption{\textbf{Prompt utilized to retrieve the statistical twin of a famous male scientist.} In a similar manner, it was employed to retrieve the twins of a famous female scientist. To assess the LLM ability to identify twins we also query male/female pairs for randomly selected scientists, fictitious names, TV character names, and politicians.}
\label{app:prompt-twins}
\end{figure*}

\begin{figure*}[ht]
\centering
\begin{minipage}{\textwidth}
\begin{framed}
\noindent
You are a highly knowledgeable and detail-oriented research assistant designed to compile and organize information efficiently and accurately. Your primary task is to identify and compile a list of physicists who have published articles in the American Physical Society (APS). Your responses must adhere to the following guidelines:

\begin{enumerate}
\item Identify Relevant Physicists: Focus on physicists who meet the criteria and have published in APS journals.
\item Verify Information: Ensure the information is accurate.
\item Format Consistently: Provide the output in a consistent JSON array format.
\item Avoid Duplicates: Ensure no duplicates are included in the list.
\item Ensure Completeness: Include all relevant physicists who meet the criteria.
\item Be Deterministic: Strive for consistent outputs given the same input.
\end{enumerate}

\end{framed}
\end{minipage}
\caption{\textbf{System prompt used.} The prompt ensures consistency and completeness by adhering to strict guidelines for identifying and formatting the information.}
\label{app:system-prompt}
\end{figure*}

\begin{figure*}[ht]
\centering
\begin{minipage}{\textwidth}
\begin{framed}
\noindent
\commenttxt{\#\#\# List of Physicists Who Published in APS Journals from 1950 to 1960 \#\#\#}

\begin{verbatim}
 ``` json
        [
            {"Name": "Richard Feynman", "Years": "1950-1960"},
            {"Name": "Murray Gell-Mann", "Years": "1950-1960"},
            {"Name": "Freeman Dyson", "Years": "1950-1960"},
            {"Name": "Julian Schwinger", "Years": "1950-1960"},
            {"Name": "Sin-Itiro Tomonaga", "Years": "1950-1960"},
            {"Name": "Werner Heisenberg", "Years": "1950-1960"},
            {"Name": "Enrico Fermi", "Years": "1950-1954"},
            {"Name": "Ernest Lawrence", "Years": "1950-1958"},
            {"Name": "Emilio Segr\u00e8", "Years": "1950-1960"},
            {"Name": "Hans Bethe", "Years": "1950-1960"},
            {"Name": "Robert Oppenheimer", "Years": "1950-1960"},
            {"Name": "Niels Bohr", "Years": "1950-1960"},
            {"Name": "Luis Alvarez", "Years": "1950-1960"},
            {"Name": "Edward Teller", "Years": "1950-1960"},
            {"Name": "Stanley Livingston", "Years": "1950-1960"},
            {"Name": "Chen Ning Yang", "Years": "1950-1960"},
            {"Name": "Tsung-Dao Lee", "Years": "1950-1960"}
        ]
```
\end{verbatim}

\vspace{0.5em}
\commenttxt{\#\#\# Reasoning Explanation \#\#\#} \\
The list of physicists provided above consists of renowned scientists who were professionally active and published in APS journals from 1950 to 1960. These individuals made significant contributions to the field of physics, particularly in the areas of quantum mechanics, particle physics, and nuclear physics.\\
                
The list includes Nobel laureates such as Richard Feynman, Murray Gell-Mann, Julian Schwinger, Sin-Itiro Tomonaga, and Chen Ning Yang, who were all active during this period and published groundbreaking research in APS journals. Other notable physicists, such as Freeman Dyson, Werner Heisenberg, and Niels Bohr, also made important contributions to the field during this time.\\

The list also includes physicists who were instrumental in the development of new technologies and experimental techniques, such as Ernest Lawrence, who invented the cyclotron, and Luis Alvarez, who developed the bubble chamber.\\

The years listed for each physicist represent the period during which they were actively publishing in APS journals. In some cases, the years may not reflect the physicist's entire career, but rather the specific period during which they were contributing to the field through APS publications.\\

Overall, the list represents a comprehensive collection of leading physicists who were active during the 1950s and made significant contributions to the field through their research and publications in APS journals.

\end{framed}
\end{minipage}
\caption{\textbf{Response by \texttt{llama3.1-70b}~when asked to retrieve experts from the 1950s.} In addition to the requested JSON data with expert recommendations, the model also included explanatory text justifying its selection.}
    \label{app:prompt-epoch-response}
\end{figure*}

\end{document}